\documentclass[]{siamart1116}

\floatstyle{plaintop}
\restylefloat{table}

\usepackage{amsfonts}
\usepackage{amsmath}
\usepackage{graphicx}
\usepackage{epstopdf}
\usepackage{algorithm,algpseudocode}
\usepackage{mathtools}
\usepackage{multirow}
\usepackage{booktabs}
\usepackage{bm}
\usepackage[caption=false]{subfig}
\usepackage{float}

\ifpdf
  \DeclareGraphicsExtensions{.eps,.pdf,.png,.jpg}
\else
  \DeclareGraphicsExtensions{.eps}
\fi

\numberwithin{theorem}{section}

\newcommand{\TheTitle}{{A Thermo-Electro-Mechanical Model for Long-Term Reliability of Aging Transmission Lines}} 
\newcommand{\TheAuthors}{E. A. Barros de Moraes, Prakash KC, and M. Zayernouri}

\headers{}{}

\author{
  Eduardo A. Barros de Moraes\thanks{Department of Mechanical Engineering \& Department of Computational Mathematics, Science and Engineering, Michigan State University, 428 S Shaw Ln, East Lansing, MI 48824, USA.}
  \and
  Prakash KC\thanks{Department of Mechanical Engineering, Michigan State University, 428 S Shaw Ln, East Lansing, MI 48824, USA.}
  \and
  Mohsen Zayernouri \thanks{Department of Mechanical Engineering \& Department of Statistics and Probability, Michigan State University, 428 S Shaw Ln, East Lansing, MI 48824, USA, Corresponding Author; \email{zayern@msu.edu}}
}

\title{{\TheTitle}\thanks{This work was supported by the National Science Foundation Award (DMS-1923201). The HPC resources and services were provided by the Institute for Cyber-Enabled Research (ICER) at Michigan State University.}}

\usepackage{amsopn}

\newtheorem{remark}{Remark}

\ifpdf
\hypersetup{
	pdftitle={\TheTitle},
	pdfauthor={\TheAuthors}
}
\fi

\begin{document}
	\maketitle
	
	\begin{abstract}
	Integrity and reliability of a national power grid system are essential to society's development and security. Among the power grid components, transmission lines are critical due to exposure and vulnerability to severe external conditions, including high winds, ice, and extreme temperatures. The combined effects of external agents with high electrical load and presence of damage precursors greatly affects the conducting material’s properties due to a thermal runaway cycle that accelerates the aging process. In this paper, we develop a thermo-electro-mechanical model for long-term failure analysis of overhead transmission lines. A phase-field model of damage and fatigue, coupled with electrical and thermal modules, provides a detailed description of the conductor's temperature evolution. We define a limit state function based on maximum operating temperature to avoid excessive overheating and sagging. We study four representative scenarios deterministically, and propose the Probabilistic Collocation Method (PCM) as a tool to understand the stochastic behavior of the system. We use PCM in forward parametric uncertainty quantification, global sensitivity analysis, and computation of failure probability curves in a straightforward and computationally efficient fashion, and we quantify the most influential parameters that affect the failure predictability from a physics-based perspective.
    \end{abstract}

\section{Introduction}
\label{sec:introduction}
The power grid network in the United States is a complex and interconnected system. Despite its robustness, loss of single components can potentially lead to cascading failures \cite{albert2004structural,crucitti2004model,kinney2005modeling} where the majority of disturbances is caused by natural
events, such as storms, hurricanes, tornadoes, earthquakes, wildfire, and warming global temperatures \cite{ward2013effect,gao2018potential}. Given such complexity and apparent vulnerability, reliable prediction of
system failure is of major importance. Identifying sensitive regions in the American power grid while predicting cascade failure is a challenging task, with significant national interest due to energy distribution and security concerns. Such large-scale networks are dynamically influenced by a substantial number of stochastic
inputs, such as weather, damage, and aging.

Reliability of transmission lines due to environmental effects has been studied in detail. Thermal failure in the form of overheating due to wildfires was studied in \cite{koufakis2010wildfire,guo2018determination}. In \cite{sarajlic2018identification}, the authors used differential evolution algorithm to solve the inverse problem of parameter identification of heat equation parameters for overhead conductor temperature evolution. Damage in transmission line and towers due to ice and wind loads was studied in \cite{yang2013probability} based on estimated generalized Pareto distributions of precipitation and wind loads. Other models studied the impact of severe weather on the reliability of transmission lines \cite{rezaei2016analysis,yang2019failure,hou2021damage}. The effect of ambient temperature, wind speed, and current on conductor temperature has been studied in \cite{bockarjova2007transmission}. They showed that the conductor temperature increased with ambient temperature and current. However, the conductor temperature decreased with wind speed.
\begin{figure}[t]
	\centering
	\subfloat[Ambient Temperature.]{\includegraphics[width=0.33\textwidth]{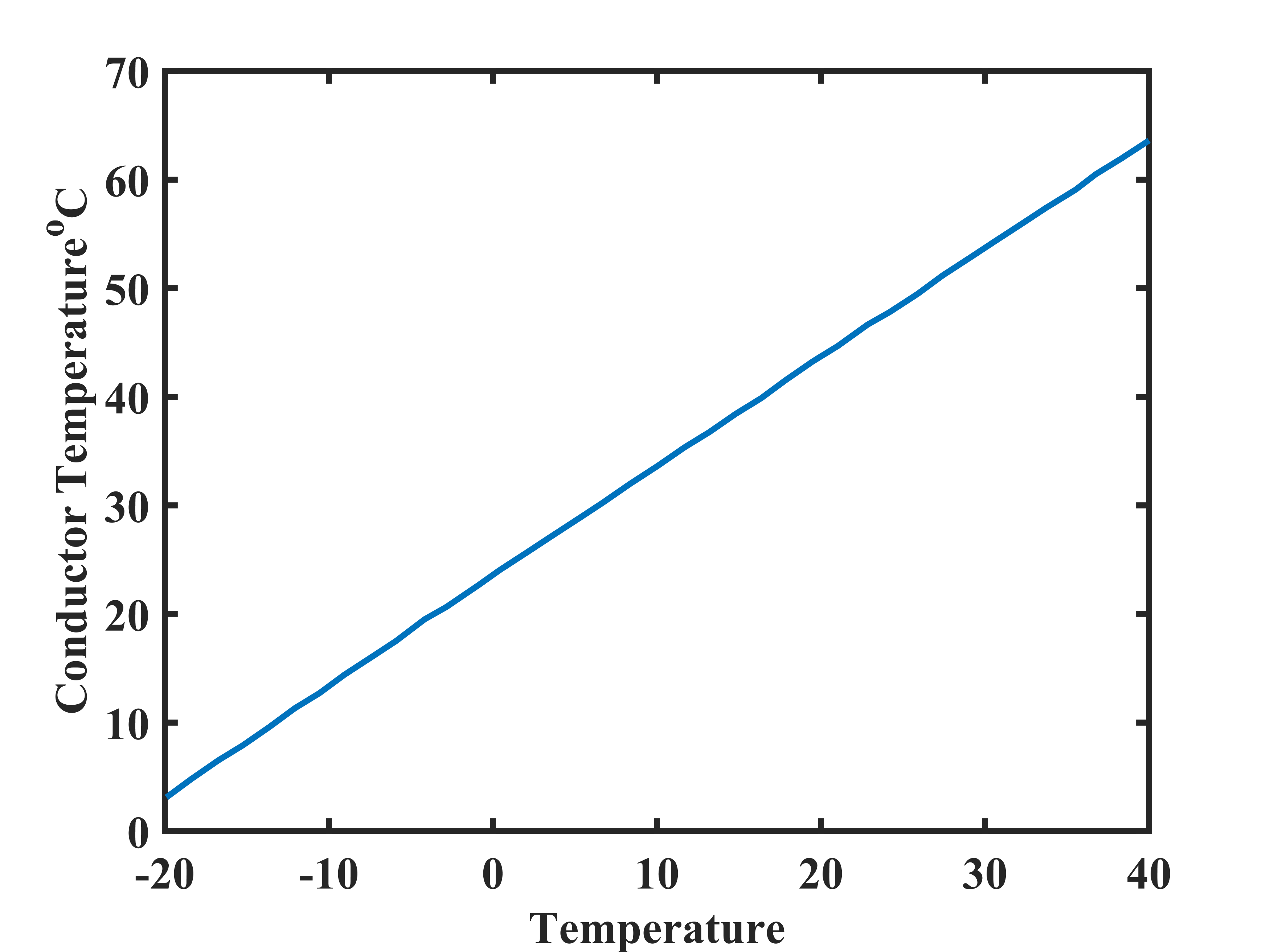}}
	\subfloat[Wind speed.]{\includegraphics[width=0.33\textwidth]{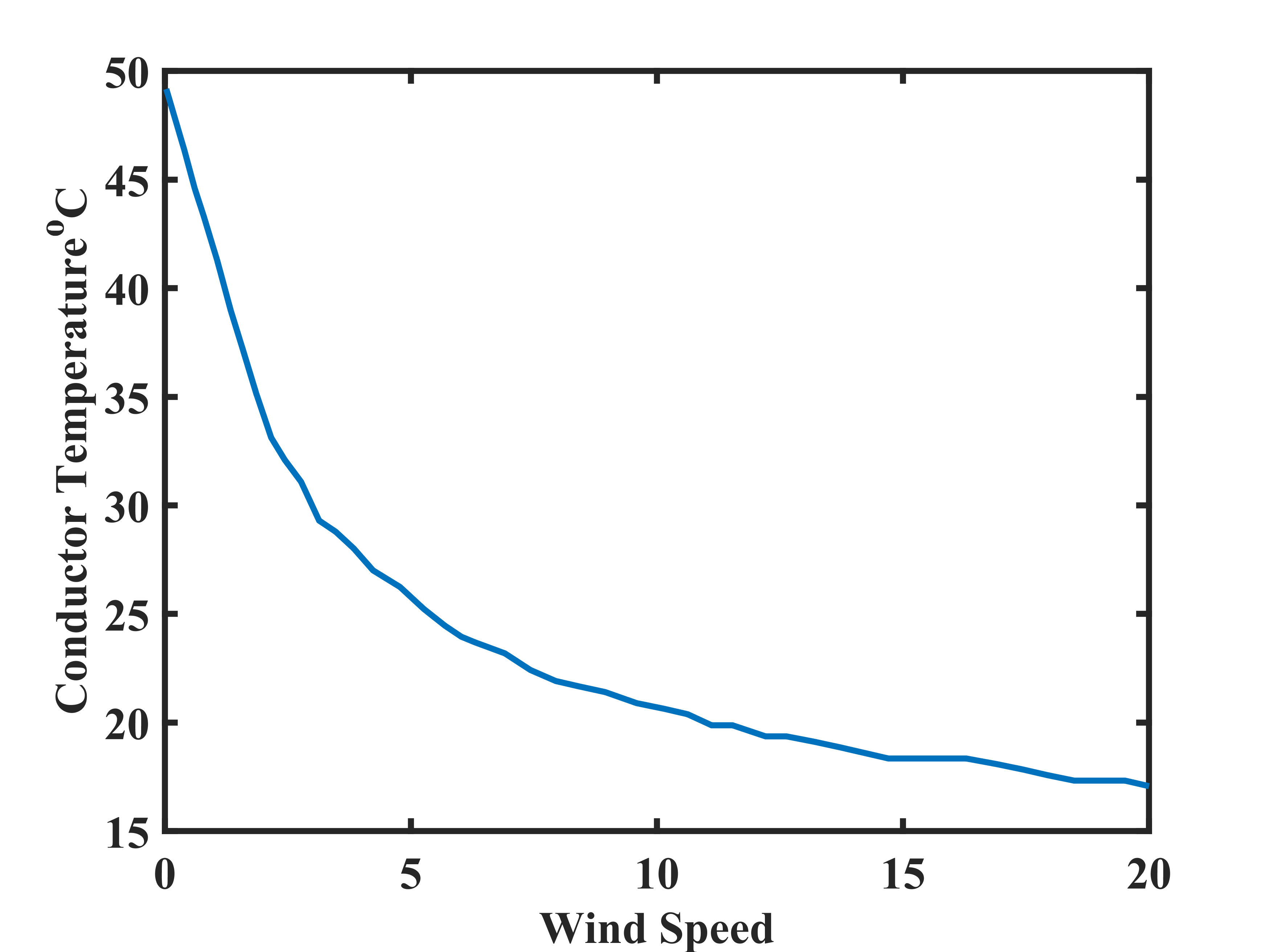}}
	\subfloat[Current.]{\includegraphics[width=0.33\textwidth]{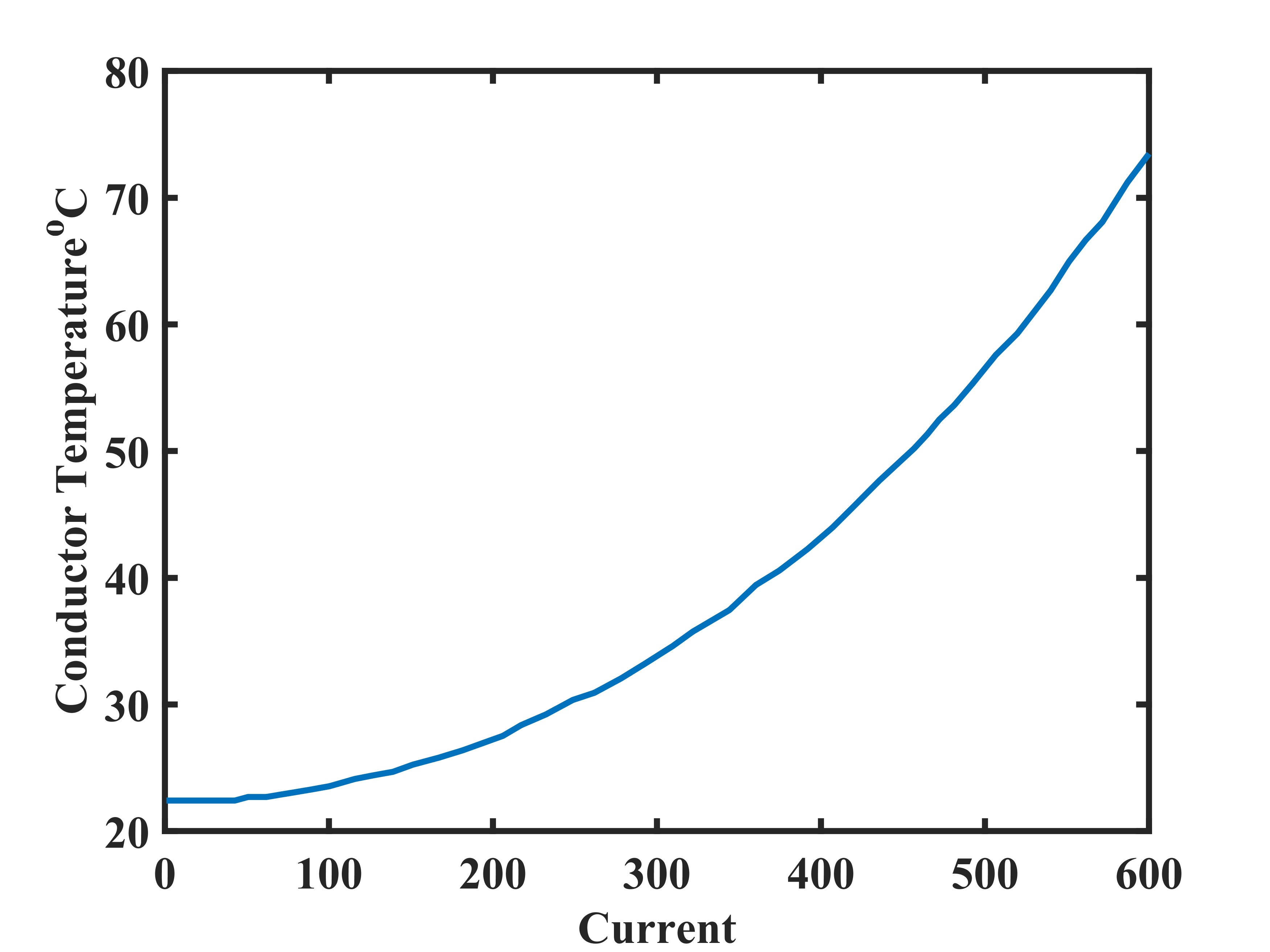}}
	\caption{Effect of Ambient Temperature, Windspeed, and Current on Conductor Temperature.}\cite{bockarjova2007transmission}
	\label{fig:effect}
\end{figure}

All those contributions are fundamental to a better understanding of the failure mechanisms in power transmission systems, however they are limited in the specific scope of each work or are generally short-term simulations. Development of models that couple the electrical and thermal properties of overhead conductors with state-of-the-art damage and fatigue mechanics formulations could improve the predictability of transmission line failure for long-term operations. 

Phase-field models have become a solid research field in damage and fatigue modeling, ranging from problems in brittle \cite{miehe_phase_2010,miehe_thermodynamically_2010,borden_higher-order_2014,de2021data}, ductile \cite{ambati_phase-field_2015,ambati_phase-field_2016}, dynamic \cite{borden_phase-field_2012,hofacker_phase_2013}, and non-isothermal fatigue fracture mechanisms \cite{boldrini_non-isothermal_2016}. The smooth crack representation in phase-field formulation naturally captures crack initiation, propagation, branching and coalescence without explicit tracking of the crack geometry. More recently, damage phase-field models have incorporated electrical effects, yet they are restricted to piezoelectric applications \cite{miehe2010phase_el,tan2022phase}. Other continuum damage approaches have been used in the context of electrical conductors \cite{kaiser2021fundamentals}, self-sensing materials \cite{nayak2019microstructure}, solder joints \cite{benabou2015continuum}, and thermo-electro-mechanical wear of electrical contacts \cite{shen2021numerical}. Yet, damage and fatigue effects on thermo-electrical behavior of transmission line conductors and their effect on life-cycle prediction is still missing, therefore efficient and robust Uncertainty Quantification (UQ) and Sensitivity Analysis (SA) methods are essential.  

The reliability analysis of the transmission line requires a comprehensive understanding of the interconnection between mechanical, thermal, and electrical aspects, as well as the different sources of uncertainties. While the phase field model is used widely for modeling the evolution of damage, dislocation dynamics can provide detailed insight into the microstructural mechanism \cite{de2021atomistic}. Studies have shown that dislocation interaction significantly affects material behavior and can lead to failure under certain conditions \cite{chhetri2023comparative, de2023machine}. Further, at high temperatures, all materials deviate from Hook law exhibiting viscous and elastic behavior. Recent advances in fractional visco-elasto-plastic models \cite{suzuki2022general} provide a comprehensive framework to capture these behaviors for structural analysis \cite{suzuki2016fractional}, accurate damage evolution \cite{suzuki2021thermodynamically}, and large-scale behavior \cite{suzuki2023fractional}. Integrating these fractional-order models enhances the transmission line failure prediction, capturing more complex behavior. Additionally, fractional order can be treated as an uncertain variable to propagate their associated randomness to the system response \cite{kharazmi2019operator}.

In complex systems such as the one proposed in this paper cannot benefit from analytical integration during forward UQ/SA or when computing the probability of failure, as it is typically done in simpler problems, nor rely on traditional sampling mechanisms in reliability engineering, such as the Monte Carlo method (MC) \cite{machado2015reliability}. Different approaches have been proposed for the stochastic solution of PDEs under parametric uncertainty. Methods such as Polynomial Chaos \cite{xiu2002modeling,xiu2002wiener,knio2006uncertainty}, or its generalization via Galerkin projections \cite{babuska2004galerkin,babuvska2005solving,stefanou2009stochastic} suffer from being intrusive, as they directly modify the governing equations, thus have limitations in complex systems. The solution is to use non-intrusive methods, which use the forward deterministic solver as a black-box. Traditional sampling methods such as MC \cite{fishman2013monte,smith2013uncertainty} compute moments of the Quantity of Interest (QoI) in a straight-forward fashion, yet is limited by slow convergence rates. Probabilistic Collocation Methods (PCM) \cite{xiu2005high,babuvska2007stochastic}, promote an efficient computation of moments by direct numerical integration of QoI computed at the collocation points, and yields faster convergence rates. In high-dimensional parametric spaces, one could tackle the curse of dimensionality from full tensorial products through Sparse Grids \cite{smolyak1963quadrature}, or dimensionality reduction techniques, such as active subspace methods \cite{constantine2014active,constantine2015exploiting,constantine2017global}. Uncertainty quantification in the context of power grid systems have been discussed using PCM \cite{hockenberry2004evaluation,lin2014uncertainty} and Gaussian Processes \cite{tartakovsky2019physics}, yet their focus is solely on the electrical short-term behavior. In damage phase-field applications PCM has been effective in solving forward UQ and SA \cite{barros2021integrated}. 

In this work, we propose a coupled thermo-electro-mechanical system with governing equations for displacements, material damage, material fatigue, temperature, and voltage. Initially damaged conductor materials lead to overheating, affecting the mechanical load, subsequent aging and increased resistivity, culminating in the premature failure of the transmission line. Rather than proposing probability distribution for failure parameters, we use PCM to propagate parametric uncertainty to the output temperature solutions of a physics-based material model. We compute the relative influence of each stochastic parameter under a global analysis of variance, and evaluate the probability of failure over time under different representative scenarios also through the PCM building-block. 

The main contributions of this work are:

\begin{itemize}
	\item We propose a coupling between a phase-field damage and fatigue model with thermal and electrical effects that further drive the aging mechanisms.
	\item We use the PCM method beyond the uncertainty and sensitivity analyzes, adopting the collocation method to compute directly the probability of failure when we transform the limit state function into a corresponding failure-indicator Bernoulli random variable.
	\item We study the long-term behavior of the coupled physical system and identify critical elements responsible for premature failure in different representative scenarios of damaged conductors, from normal operating conditions, to seasonal high winds and increasing electric current and air temperature.
\end{itemize}

This work provides a robust connection between the detailed physical principles that govern mechanical, thermal, and electrical properties in a transmission line conductor and the probabilistic nature associated with uncertainties in parameters and loading conditions. For the first time, high-fidelity simulations of material aging are being used to provide predictive estimations of thermal failure probability.

This paper is organized as follows: in Section~\ref{sec:reliability}, we present the thermo-electro-mechanical model for failure of transmission lines, and discuss the temperature-based failure criterion. Then, we discuss the deterministic solution through the Finite-Element Method in Section~\ref{sec:deterministic}, along with the definition of representative scenarios. The stochastic methods are discussed in Section~\ref{sec:stochastic}, where we present the PCM building blocks for the uncertainty, sensitivity, and probability of failure analyzes. Finally, we address concluding remarks in Section~\ref{sec:conclusions}.
\section{Reliability of Transmission Lines}
\label{sec:reliability} 

\subsection{Problem Statement}
Operating temperature is a critical indicator for safe operation of transmission lines due to risk of annealing of the material components, and excessive sagging during high temperature conditions. Both outcomes pose serious consequences to the safety of surrounding areas, and compromise the expected efficiency of electric energy transmission. 

Although complete cable rupture also represents a measure of transmission line failure, incurring inefficiency and repair costs, high operating temperatures is potentially more dangerous since it could go unnoticed. Several conditions may contribute to an excess temperature operation, including an initial damaged state, aging of transmission line, increase in electric energy demand, and higher-than-normal air temperature conditions. Under these circumstances, material temperature may exceed accepted values at specific points of the cable, significantly altering the prospect of useful life-time.

Therefore, operating conductor temperature becomes the focal point of our analysis, and the interplay between different physical effects, loading conditions, and material parameters will define the failure state of the transmission line in long-term life of overhead cables. Our objective is to study the persistent multi-physics effect on the operating temperature of a transmission line with the presence of an initial damage. To achieve such long-term coupled system, several simplifying assumptions need to be made, which we will discuss in more detail in the next section.

In general terms, reliability can be defined through a limit state function $g(R,S;t)$ based on the load $S$ and the capacity of the system $R$, and it can be a function of time $t$. In structural engineering, $S$ and $R$ can be defined as the operating stress and material's ultimate strength, for example. Based on maximum temperature failure criterion discussed above, we define $S$ as the transmission line's maximum temperature $\theta_{max}$ at time $t$, and $R$ as a temperature limit before excessive sag or annealing occur, denoted $\theta_{lim}$. Then, the limit state function is defined as

\begin{equation}
\label{eq:g}
g(\theta_{lim},\theta_{max};t) = \theta_{lim} - \theta_{max}(t).
\end{equation}

In general, both $R$ ans $S$ can be defined as random variables, yet here we will fix $\theta_{lim}$, and only let $\theta_{max}(t)$, which is also a function of space, be a random variable. Then, the probability of failure $p_f(t)$ for the system can be defined as

\begin{equation}
\label{eq:pf}
p_f(t) = P\{g(\theta_{lim},\theta_{max};t) < 0\}.
\end{equation}

\begin{figure}[t]
	\centering
	\includegraphics[width=\textwidth]{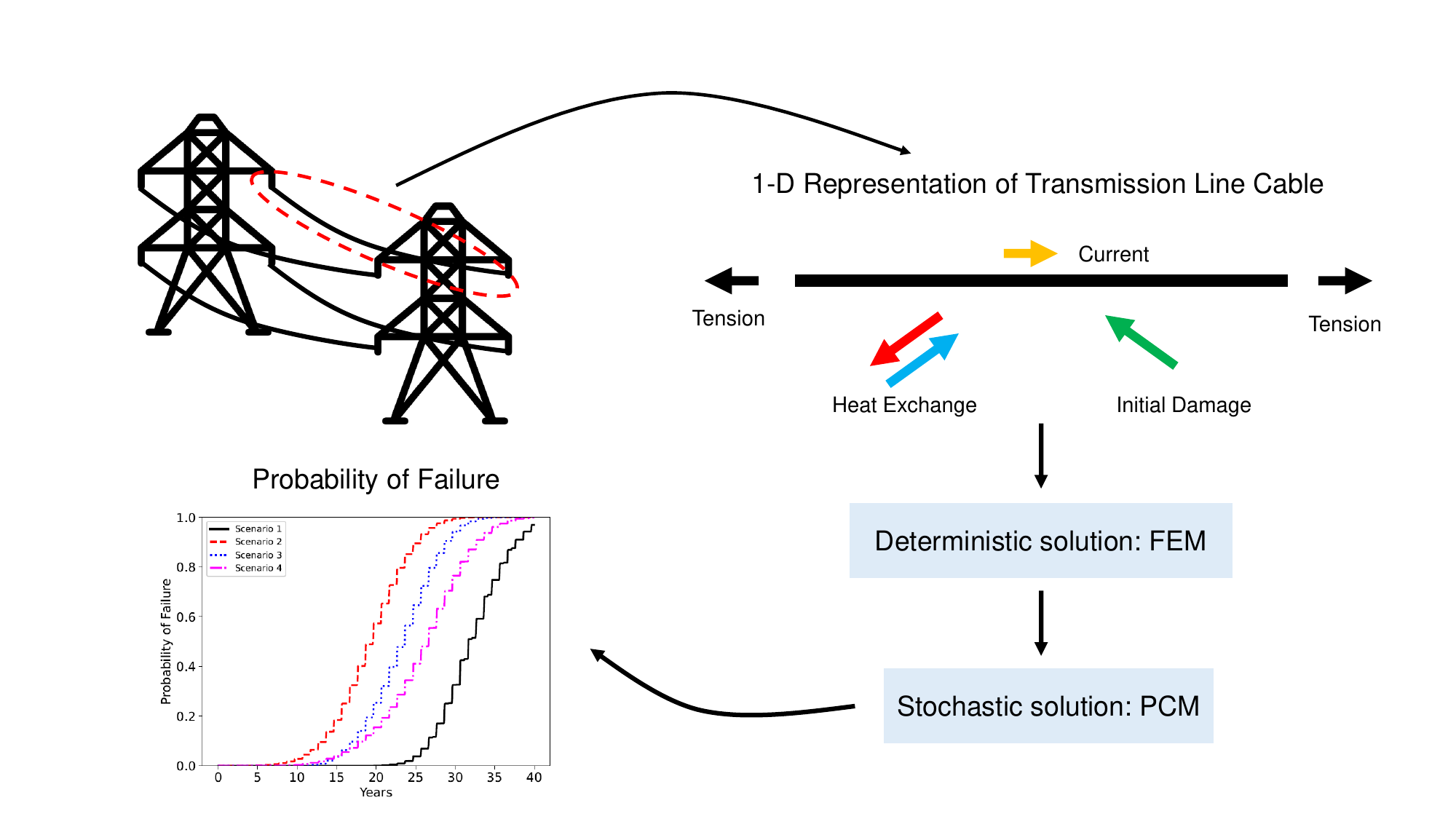}
	\caption{Schematic representation of the thermo-electro-mechanical framework for transmission line failure. A one-dimensional representation of an initially damaged conductor cable is subject to mechanical tension, heat exchange with the environment, and an electric current, which combined lead to thermal failure from overheating. The deterministic solution of the physical system through Finite-Element Method enters the PCM stochastic solver as a black-box, which generates physics-based predictions of probability of failure.}
	\label{fig:scheme}
\end{figure}

The thermo-electrical-mechanical model acts as the source of computation of $\theta_{max}$, providing a high-fidelity measure of the maximum operating temperature across the cable through detailed simulations. Through the stochastic framework built on PCM, we obtain the probability of failure based simply on the physics at the material level. The overall flow of the proposed framework is illustrated in Fig.~\ref{fig:scheme}.

\subsection{Thermo-Electro-Mechanical Damage Phase-Field Model}

We consider one segment of transmission cable showed in Fig.~\ref{fig:scheme}. The material is supported by two transmission towers and tensioned such as to maintain a sag below the maximum allowed threshold. From the mechanical perspective, we look at this system as a one-dimensional body under tension, and we consider the projected span as the operating length in order to keep the one-dimensional assumption. Sag is extremely important and we take it into account when computing the tensile load due to temperature effects, however we refrain from explicit sag modeling in this framework and focus solely on the tensile effects on the cable as drivers to material damage and fatigue.

The tension in the cable is driven primarily by its operating temperature. As it heats, there is an increment in the sag due to thermal elongation, which leads to a decrease in the horizontal tension. Conversely, as the material cools, extra tension is induced by contraction of the material. 

Material temperature is a result of combined Joule heating from electric current, with convective cooling from the ambient air with a given wind speed. In turn, higher temperatures are assumed to further degrade the material and enhance its aging process, while affecting electrical conductivity, increasing the resistivity.

With a focused look at the material behavior, the electrical component of the model is simplified and not related to the electrical design of transmission lines. Instead, the model is concerned with the effective electric current that passes through a single cable of the conductor and affects its material properties, regardless of the voltage level that the line provides to consumers. The ultimate effect of this consideration is that resistive losses along the line will heat the material and cause a voltage drop between its endpoints.

Finally, we consider that any damage in the material is responsible for increasing the material's resistivity, adding degradation to the already present temperature effect on resistivity, therefore leading to further dissipation from Joule heating. All the combined effects make damage a precursor of a positive feedback loop of heating in the transmission line, causing early temperature failure due to compounding thermal runaway. In all physical components, we adopt a quasi-static perspective to allow for long-term simulations, such that in the proposed time-scale, the system is assumed to reach equilibrium faster than changes in loading conditions.

\subsubsection{Mechanical model} 

We consider a non-isothermal phase-field framework for damage and fatigue modeling following the principles in \cite{boldrini_non-isothermal_2016}, consisting in two PDEs for the evolution of displacement $u$ and damage $\varphi$ fields, and a separate ODE for evolution of fatigue $\mathcal{F}$. The damage phase field is representative of volumetric fraction of degraded
material, taking values of $\varphi = 0$ for virgin material, $\varphi = 1$ for a complete fracture, and takes intermediate damaged states with $0 < \varphi < 1$. Damage evolves following an Allen–Cahn type equation, derived along with the equilibrium equation for $u$ through the principle of virtual power and entropy inequalities with thermodynamic consistency. The fatigue field $\mathcal{F}$ is treated as an internal variable whose evolution equation is obtained through constitutive relations that must satisfy the entropy inequality for all admissible processes. 

\begin{remark}
	The fatigue variable was originally introduced to model the evolution of material degradation associated to the presence of microcracks, in the context of high-cycle fatigue of brittle materials. Here, we cast an alternative interpretation of the fatigue $\mathcal{F}$ as a measure of material aging, which is more consistent with the long-term process of interest to this work.
\end{remark}

We adopt a 1-D representation for the mechanical body such that it occupies the domain $\Omega \in \mathbb{R}$ at time $t \in (0,T]$. From the general governing equations, specific forms of material evolution can be obtained from a choice in the free-energy potentials. One alternative is to consider the free-energy function:

\begin{equation}
\label{eq:free_energy}
\Psi (\nabla u, \varphi, \nabla \varphi ,\mathcal{F}) 
= d(\varphi) Y \left(\nabla u\right)^2 + 
g_c \frac{\gamma}{2} \left(\nabla \varphi\right)^2  
+ \mathcal{K} (\varphi, \mathcal{F}) ,
\end{equation}

\noindent where $Y$ denotes the Young modulus, $g_c$ is the fracture energy release rate, and $\gamma > 0$ represents the phase-field layer width parameter. The degradation function is taken to be $d(\varphi) = (1-\varphi)^2$, and beyond the original degradation of elastic response, here it will also degrade the electric conductivity. Finally, $\mathcal{K} (\varphi, \mathcal{F})$ models the evolution of material damage when fatigue, or aging, occurs. 

The original model in \cite{boldrini_non-isothermal_2016} is time-dependent. Here, however, we are interested in long-term behavior and assume that the material fully reaches an equilibrium stage between consecutive time-steps, turning the governing equations to a quasi-static representation for $u$ and $\varphi$. Yet, evolution of $\mathcal{F}$ is still an ODE representative of long-term aging.

\begin{equation}
\label{eq:mechanical}
\nabla \cdot \,\left( (1-\varphi)^2 Y \nabla u \right) 
-  \gamma g_c \, \nabla \cdot \, ( \nabla \varphi \otimes \nabla \varphi  )+  f = 0,
\end{equation}

\begin{equation}
\label{eq:damage}
\gamma g_c \Delta \varphi  
+  (1- \varphi) (\nabla u)^T Y (\nabla u)
- \frac{1}{\gamma}  [ g_c \mathcal{H}' (\varphi) + \mathcal{F} \mathcal{H}_f' (\varphi) ] = 0,
\end{equation}

\begin{equation}
\label{eq:fatigue}
\dot{\mathcal{F}} = -  \frac{ \hat{F}}{\gamma}   \mathcal{H}_f (\varphi),
\end{equation}

\noindent subjected to appropriate boundary conditions. For mechanical equilibrium, either the stress or displacement are known at the boundaries, we assume $\nabla \varphi = 0$ at the $\partial \Omega$, and the $\otimes$ operator denotes outer product.

The potentials $\mathcal{H}(\varphi)$ and $\mathcal{H}_f(\varphi)$ describe the damage evolution from $0$ to $1$ as fatigue changes from zero to $g_c$. By taking their derivatives with respect to $\varphi$, we obtain potentials $\mathcal{H}'(\varphi)$ and $\mathcal{H}_f'(\varphi)$. Choosing the transition to be continuous and monotonically increasing, suitable choices for the potentials are:

\noindent\begin{minipage}{.55\linewidth}
	\begin{equation}
	\label{eq:potentials}
	\mathcal{H} (\varphi) =
	\begin{cases}
	0.5 \varphi^2   & \mbox{for} \; 0 \leq \varphi \leq 1 ,
	\vspace{0.1cm} 
	\\
	0.5 + \delta (\varphi -1)  &  \mbox{for} \; \varphi > 1 ,
	\vspace{0.1cm}
	\\
	- \delta \varphi &   \mbox{for} \; \varphi < 0 .
	\end{cases}
	\end{equation} 	
\end{minipage}%
\begin{minipage}{.45\linewidth}
	\begin{equation}
	\label{eq:potentials2}
	\mathcal{H}_f (\varphi) =
	\begin{cases}
	- \varphi & \mbox{for} \; 0 \leq \varphi \leq 1,
	\vspace{0.1cm} 
	\\
	-1 & \mbox{for} \; \varphi > 1,
	\vspace{0.1cm}
	\\
	\hspace{0.25cm} 0 & \mbox{for} \;  \varphi < 0.
	\end{cases}
	\end{equation}
\end{minipage}

We describe the evolution $\mathcal{F}$ through $\hat{F}$, related to the formation and growth of micro-cracks that occur in cyclic loadings, and influenced by temperature. We consider a representation of $\hat{F}$ that depends on the value of stress associated to the virgin material, defined through a linear relation:

\begin{equation} 
\label{eq:fhat}
\hat{F} =  \rho a \left(\frac{\theta}{\theta_0}\right) (1-\varphi) \left|Y \nabla u\right|,
\end{equation}

\noindent where the parameter $a$ represents the rate of aging, modulated by the ratio of current temperature $\theta$ to a reference temperature $\theta_0$, and $\rho$ is the material density.

The mechanical system as described above allows for damage healing when there is a reduction in the tensile stress. In order to avoid healing mechanisms and simulate an irreversible damaging process, we take an approach similar to \cite{miehe_phase_2010} and define a variable that represents the local maximum strain energy history, $\mathbb{H}$, defined as

\begin{equation}
\label{eq:history}
\mathbb{H}(x,t) = \max((\nabla u(x,t))^T Y (\nabla u(x,t)),\mathcal{H}(x,t)).
\end{equation}

We introduce the history field variable in the damage equation, which becomes

\begin{equation}
\label{eq:damage_H}
\gamma g_c \Delta \varphi  
+  (1- \varphi) \mathbb{H}
- \frac{1}{\gamma}  [ g_c \mathcal{H}' (\varphi) + \mathcal{F} \mathcal{H}_f' (\varphi) ] = 0.
\end{equation}

\subsubsection{Thermal model}

The original model from \cite{boldrini_non-isothermal_2016} contains a fatigue-driven term in the temperature evolution equation, associated with temperature increase due to repetitive, yet fast, loading. In the quasi-static regime, temperature is computed from thermal equilibrium, and heat due to cyclic load becomes negligible.

Therefore, we adopt the steady-state heat equation 

\begin{equation}
\label{eq:temperature}
\nabla \cdot (\kappa \nabla \theta) + q = 0.
\end{equation}

Thermal conductivity is represented by $\kappa$, while $q$ denotes the heat sources/sinks that will depend on the physical mechanisms. Specifically, we consider the Joule heating $q_J$ from electrical current as a source, and convective cooling $q_C$ due to wind as a sink such that

\begin{equation}
q = q_J - q_C.
\end{equation}

Specific form of $q_C$ for a flow past a cylinder can be derived, yet here we adopt a formulation from the Electric Power Engineering Handbook \cite{grigsby2006electric} 

\begin{equation}
q_C = \frac{0.0128 \sqrt{(p v)}}{\theta_{air}^{0.123}\sqrt{d}} (\theta_c - \theta_{air}),
\end{equation} 

\noindent $p$ represents the atmospheric pressure in $atm$, $v$ denotes the wind velocity in $ft/s$, $\theta_{air}$ and $\theta_c$ being the air and conductor's temperatures in $K$, respectively, and $d$ is the conductor diameter in $in$. We convert all parameters and the final $q_C$ (in $W/sq\ in$) to their respective SI units. 

The specific form of $q_J$ will be discussed next.

\subsubsection{Electrical model}

Transmission line design takes into account different effects such as AC frequency, inductance, reactance, and elctro-magnetic interactions with the air, ground, and nearby conductors, and electric load. In the end, those effects are crucial to determine the passing current $I$ across the conductor, yet they are out of the scope of the present study. With the focus on modeling the material's response to combined multi-physics effects, we are not concerned about the design of the transmission line as an electric power component, rather we are interested in understanding how the electricity-driven Joule heating affects a damaged material. In other words, we simply take the current $I$ as an input of the multi-physics system, that could be obtained from other calculations. 

Furthermore, the long-term analysis does not benefit from detailed time simulations of transient effects of AC currents. The major focus is on the effect of damage-induced Joule heating, therefore the model refers to electric current as the DC-equivalent mean current that remain constant between consecutive time-steps. This greatly simplifies the calculations, allowing for efficient measures of the quantities of interest, in this case the heat source term. We adopt prescribed values for the current, yet they can be obtained from more sophisticated methods or from real-life data, and used in this framework with no issues.

Under that perspective, the problem consists in solving the conservation of current model for bulk materials:

\begin{equation}
\nabla \cdot J = 0,
\end{equation}

\begin{equation}
J = \sigma_{E} E,
\end{equation}

\begin{equation}
E = -\nabla V,
\end{equation}

\noindent where $J$ represents the electric current density per cross-section area, $E$ is the corresponding electric field due to the voltage $V$. Also, $\sigma_E$ denotes the degraded electrical conductivity at the operating temperature, and it is a function of the non-degraded conductivity through the degradation function $d(\varphi)$

\begin{equation}
\sigma_E = (1 - \varphi)^2 \sigma_{E,T}.
\end{equation}

In turn, the non-degraded conductivity $\sigma_{E,T}$ at the operating temperature can be obtained from the conductivity at the reference temperature $\sigma_{E,0}$ by

\begin{equation}
\sigma_{E,T} = \frac{\sigma_{E,0}}{1 + \alpha(\theta - \theta_0)},
\end{equation}

\noindent where $\alpha$ is the temperature coefficient of resistivity, which is a positive for metals.

Combining the equations for conservation of electric current, we obtain a single PDE for the evolution of voltage field:

\begin{equation}
\label{eq:voltage}
\nabla \cdot (-\sigma_{E} \nabla V) = 0,
\end{equation} 

\noindent along with either $V$ or $J$ prescribed at the boundaries. The voltage field will be disrupted by the presence of damage and will result in a different voltage drop $\Delta V$ between the extremities of the conductor.

The Joule heating heat source is then defined as

\begin{equation}
q_J = J \cdot E.
\end{equation}

In essence, damage will increase the voltage drop, incurring losses in electric power due to increased resistance of the conductor, further aggravating the thermal load to Joule heating.

\subsection{Further Considerations for the Multi-Physics Model}

As mentioned before in the previous section, the horizontal tension acting on the cable structure is driven by temperature. We now discuss the process of obtaining the horizontal load acting on the transmission line cable.

When supported by two points, cables sag in the shape of a catenary. There are diverse examples in the literature on simulation of their pure mechanical behavior \cite{karoumi1999some,stengel2014finite}, yet here we adopt the simplifying assumption that our damage phase-field model is strictly 1-D. This means that we assume that the length of the cable $L$ is sufficiently close to the span distance $S_c$ such that we adopt them to be equivalent. This does not mean that sag $D$ is not present, rather that the driver of mechanical damage and fatigue is solely the horizontal component of the cable tension, $H$. 

However, the effects of sag still need to be considered, specially under variable temperature conditions. Heat causes the length to increase, and the cable accommodates the elongation by increasing the sag, relieving some tension. On the other hand, colder conditions contract the cable, sag decreases and results in increased tension. Therefore, mechanical load at one end of the cable is a result of a pre-tension in conjunction to tension changes due to operating temperature. The model itself determines the appropriate mechanical loading condition.

We follow the formulations presented in \cite{grigsby2006electric}. Let $W_b$ be the weight per unit length, and $H_0$ the initial pre-tension (usually prescribed to be around $20\%$ of material's ultimate strength). We compute an initial sag $D$ as being:

\begin{equation}
D_0 = \frac{W_b S_c^2}{8 H_0},
\end{equation}

\noindent so the theoretical length (even though we use $L = S$ in the simulation for simplicity) necessary to accommodate the cable with sag $D$ over span $S_c$ is

\begin{equation}
L_0 = S_c + \frac{8 D_0^2}{3 S_c}.
\end{equation}

Changes in temperature alter the length of the cable through the classical relation

\begin{equation}
L = L_0 (1 + \alpha_L \Delta \theta),
\end{equation}

\noindent where $\alpha_L$ is the coefficient of thermal expansion. The change of length is followed by the adjustment of sag

\begin{equation}
D = \sqrt{\frac{3 S_c (L - S_c)}{8}},
\end{equation}

\noindent finally providing the appropriate value of horizontal tension

\begin{equation}
\label{eq:load}
H = \frac{W S^2}{8D},
\end{equation}

\noindent where $W$ is a measure of total weight that could include ice and wind components to the base weight. 

In this work, we consider only a wind component to the total weight relation, given by \cite{grigsby2006electric}

\begin{equation}
W = \sqrt{W_b^2 + W_w^2}.
\end{equation}

The wind component $W_w$ in $lb/ft$ is derived from wind pressure $P_w$ expressed in $lb/ft^2$, considering the wind velocity $v$ in $mph$. We again transform all parameter expression units to SI units before further computations.

\begin{equation}
P_w = 0.0025v^2.
\end{equation}

\begin{equation}
W_w = \frac{P_w d}{12}.
\end{equation}

\subsection{Multi-physics framework}

\begin{figure}[t]
	\centering
	\includegraphics[width=\textwidth]{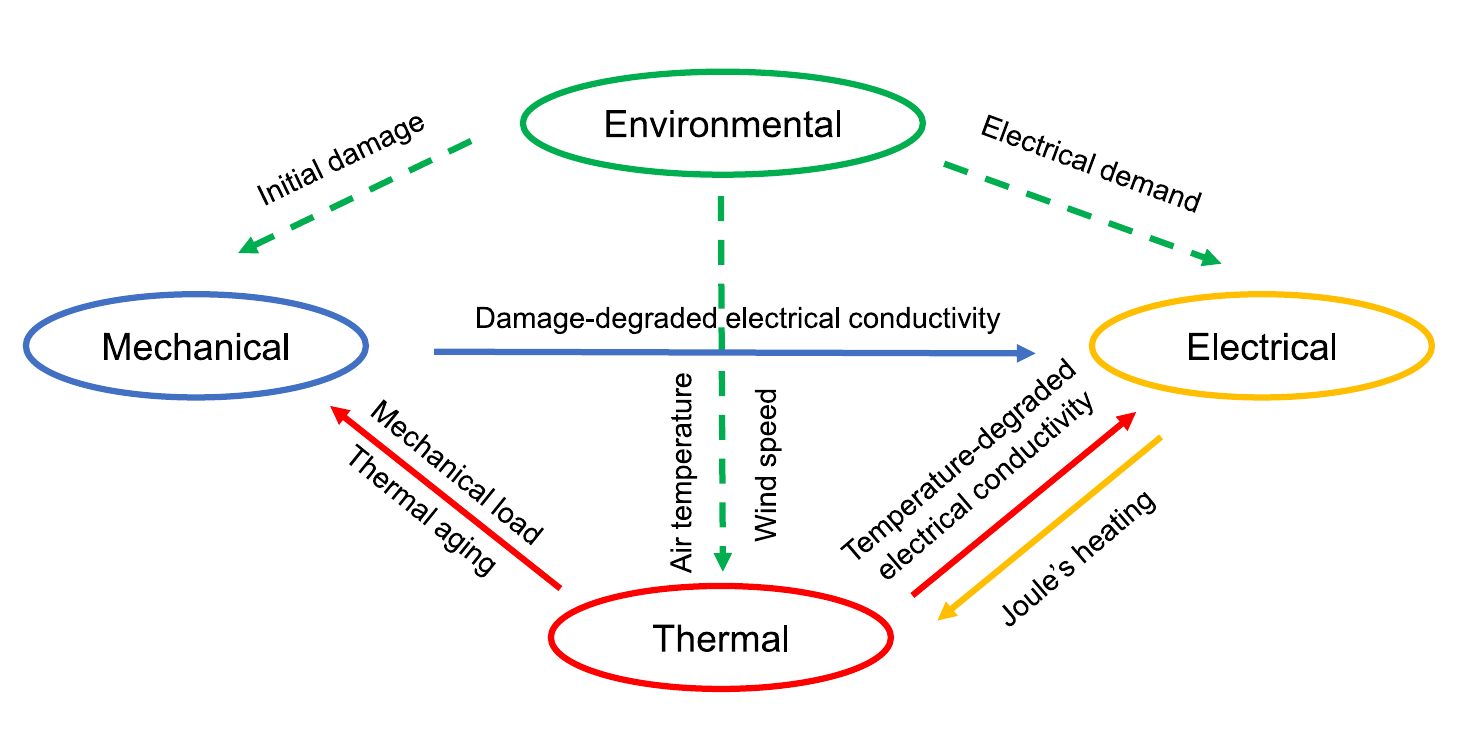}
	\caption{Schematic diagram of the relationship between the thermal, electrical, and mechanical models based on their governing equations, in addition to an abstract, scenario-dependent environmental module that provides initial conditions and input parameters.}
	\label{fig:relations}
\end{figure}

We illustrate the framework for transmission line failure modeling in Fig~\ref{fig:relations}. Eqs.(\ref{eq:mechanical}), (\ref{eq:damage_H}), (\ref{eq:fatigue}), (\ref{eq:temperature}), and (\ref{eq:voltage}) form the main governing equations for the coupled system, in addition to equations for the fatigue potentials, thermal, and mechanical load, and degradation of electrical conductivity. The diagram provides a more comprehensive view of the thermal, electrical, and mechanical components, and their connection with one another. We further introduce an abstract representation of an Environmental module, which determines inputs and initial conditions for all the other modules based on the representative scenario being considered and its corresponding parameters.

\section{Deterministic Solution}
\label{sec:deterministic}

In this section, we discuss the Finite-Element discretization of the proposed multi-physics framework, and describe the solution procedure for the deterministic case. The deterministic solution serves two main purposes: first, its implementation is later used as a black-box for non-intrusive stochastic methods for uncertainty quantification. Second, analysis of deterministic results serves as an interpretable guide to further evaluations of multi-dimensional uncertainty propagation.

\subsection{Finite-Element Discretization}

In 1-D, all variables become scalar fields, and we drop the multi-dimensional representation of differential operators. We adopt standard linear finite-element method for spatial discretization. We multiply Eqs.(\ref{eq:mechanical}), (\ref{eq:damage_H}), (\ref{eq:fatigue}), (\ref{eq:temperature}), and (\ref{eq:voltage}) by test function $w$, and perform integration by parts, and recall that the volume differential can be expressed with cross-section area $A$, and obtain their corresponding weak form

\begin{equation}
\int_0^L -(1 - \varphi)^2 Y A(x) \frac{du}{dx}\frac{dw}{dx} dx + \int_0^L \gamma g_c A(x) \left(\frac{d\varphi}{dx}\right)^2 \frac{dw}{dx} dx + \int_0^L f A(x) w dx = 0,
\end{equation}

\begin{equation}
\begin{split}
\int_0^L - \gamma g_c A(x) \frac{d\varphi}{dx}\frac{dw}{dx} dx + \int_0^L A(x) \mathbb{H} w dx - \int_0^L A \mathbb{H} \varphi w dx \\ - \int_0^L \frac{g_c A(x)}{\gamma} \varphi w dx + \int_0^L \frac{A(x)}{\gamma}\mathcal{F}w dx = 0,
\end{split}
\end{equation}

\begin{equation}
\int_0^L \dot{\mathcal{F}} w A(x) dx = \int_0^L \frac{-\rho a (1 - \varphi) Y \vert\frac{du}{dx}\vert (-\varphi)}{\gamma}\frac{\theta}{\theta_0} w A(x) dx, 
\end{equation}

\begin{equation}
\begin{split}
\int_0^L - \kappa A(x) \frac{d\theta}{dx}\frac{dw}{dx} dx + \int_0^L \sigma_E A(x)\left(\frac{dV}{dx}\right)^2 w dx \\- \int_0^L c \theta A_s(x) w dx + \int_0^L c \theta_{air} A_s(x) w dx = 0,
\end{split}
\end{equation}

\begin{equation}
\int_0^L \sigma_E \frac{dV}{dx} \frac{dw}{dx} A(x) w dx = 0,
\end{equation}

\noindent where we consider $c =  \frac{0.0128 \sqrt{(p v)}}{\theta_{air}^{0.123}\sqrt{d}}$, and $A_s(x)$ as the variable superficial area over which the convective wind-based heat transfer occurs.

We adopt a linear approximation over each element $k$, such that the approximation of field variables is a linear combination of nodal basis functions:

\begin{align}
u^k &= N \hat{u}^k,\\
\varphi^k &= N \hat{\varphi}^k ,\\
\mathcal{F}^k &= N \hat{\mathcal{F}}^k ,\\ 
\theta^k &= N \hat{\theta}^k ,\\
V^k &= N \hat{V}^k.
\end{align}

Finite-element interpolations for spatial derivatives are computed through linear combinations of shape function derivatives:

\begin{align}
\left(\frac{du}{dx}\right)^k &= B \hat{u}^k,\\
\left(\frac{d\varphi}{dx}\right)^k &= B \hat{\varphi}^k,\\
\left(\frac{d\theta}{dx}\right)^k &= B \hat{\theta}^k,\\
\left(\frac{dV}{dx}\right)^k &= B \hat{V}^k\\,
\end{align}

\noindent where we define $N$, $B$, $\hat{u}^k$, $\hat{\varphi}^k$, $\hat{\mathcal{F}}^k$, $\hat{\theta}^k$, $\hat{V}^k$ as

\begin{align}
N &= \begin{bmatrix}
N_1 & N_2
\end{bmatrix},\\
B &= \begin{bmatrix}
N_{1,x} & N_{2,x}
\end{bmatrix},\\
\hat{u}^k &= \begin{bmatrix}
u_1^k & u_2^k
\end{bmatrix},\\
\hat{\varphi}^k &= \begin{bmatrix}
\varphi_1^k & \varphi_2^k
\end{bmatrix},\\
\hat{\mathcal{F}}^k &= \begin{bmatrix}
\mathcal{F}_1^k & \mathcal{F}_2^k
\end{bmatrix},\\
\hat{\theta}^k &= \begin{bmatrix}
\theta_1^k & \theta_2^k
\end{bmatrix},\\
\hat{V}^k &= \begin{bmatrix}
V_1^k & V_2^k
\end{bmatrix},
\end{align}

\noindent where $N_1$ and $N_2$ are linear interpolation functions for a one-dimensional finite element.

We replace the previous approximations into the weak form, and adopt a forward Euler scheme for evolution of $\mathcal{F}$, obtaining the following discretization for each $k-$th element:

\begin{align}
K_u \hat{u}^k &= w_u + M \hat{f}^k,\\
K_{\varphi} \hat{\varphi}^k &= w_\varphi,\\
M \hat{\mathcal{F}^{n+1}}^k &= M \hat{\mathcal{F}^{n}}^k  + \Delta t w_{\mathcal{F}},\\
K_\theta \hat{\theta}^k  &= w_\theta,\\
K_V \hat{V}^k &= 0,
\end{align}

\noindent where superscripts $n$ and $n+1$ represent the current and next time-steps. The discrete form uses the operator definitions

\begin{align}
K_u &= \int_k (1 - N\hat{\varphi}^k)^2 Y A(x) B^T B dx,\\
w_u &= \int_k \gamma g_c A(x) (B \hat{\varphi}^k)^2 B dx,\\
M &= \int_k A(x) N^T N dx,\\
K_\varphi &= \int_k \gamma g_c A(x) B^T B dx + \int_k \mathbb{H} A(x) N^T N dx + \int_k \frac{g_c A(x)}{\gamma} N^T N dx,\\
w_\varphi &= \int_k \mathbb{H} A(x) N dx + \int_k \frac{A(x)}{\gamma} N^T \hat{\mathcal{F}^n}^k N dx,\\
K_\theta &= \int_k \kappa A(x) B^T B dx + \int_k c A_s(x) N^T N dx,\\
w_\theta &= \int_k \sigma_E A(x) \left(B \hat{V}^k\right)^2 N dx  + \int_k c A_s(x) \theta_{air} N dx,\\
K_v &= \int_k (1 - N\hat{\varphi}^k)^2 \sigma_{E,T} A(x) B^T B dx.
\end{align}

By applying the standard assembly operator in the above matrices and vectors, we obtain their respective global forms, dropping the superscript $k$, and we obtain the final form of global system of equations

\begin{align}
K_u \hat{u} &= w_u + M \hat{f}, \label{eq:u}\\
K_{\varphi} \hat{\varphi} &= w_\varphi, \label{eq:phi}\\
M \hat{\mathcal{F}}^{n+1} &= M \hat{\mathcal{F}}^n  + \Delta t w_{\mathcal{F}}, \label{eq:F}\\
K_\theta \hat{\theta}  &= w_\theta, \label{eq:theta}\\
K_V \hat{V} &= 0 \label{eq:V}.
\end{align}

We adopt a staggered solution scheme, in which we evaluate the equations in sequence at each time-step. We summarize the procedure for deterministic solution in Algorithm~\ref{algo:deterministic}

\begin{algorithm}[t]
	\caption{Solution of Thermo-Electro-Mechanical Model.}
	\label{algo:deterministic}
	\begin{algorithmic}[1]
		\State Choose initial pre-tension $H_0$.
		\For{Each time-step}
		\State Compute the current tensile load from Eq.~(\ref{eq:load}).
		\State Solve for displacements, Eq.~(\ref{eq:u}).
		\State Update strain energy history, Eq~(\ref{eq:history}).
		\State Solve damage field Eq.~(\ref{eq:phi}).
		\State Update fatigue using Eq.~(\ref{eq:F}).
		\State Solve the temperature field from Eq.~(\ref{eq:theta}).
		\State Solve voltage field through Eq.~(\ref{eq:V}).
		\EndFor
	\end{algorithmic}
\end{algorithm}

\subsection{Representative Scenarios}

We aim to study the coupled thermo-electro-mechanical system under four different representative scenarios:

\begin{itemize}
	\item \textbf{Scenario 1 - Normal Operating Conditions:} In this scenario, all loading conditions will follow an unchanged pattern throughout the years, and the only driver of material failure, and subsequently of temperature increase is an initial damage precursor.
	\item \textbf{Scenario 2 - High Seasonal Winds:} In this scenario, we simulate the aging and degradation of an initially damaged transmission line located in a region where one expects high winds during a specific season. For example, it could represent expected average winds over a few days during hurricane season, or a known windy condition particular to a certain time of the year.
	\item \textbf{Scenario 3 - Increasing Electric Demand:} Here, we simulate an increase of electric power demand in the region, which is driven by external factors that we do not address here. We model the effect of such factors in a linearly increasing average current over the years.
	\item \textbf{Scenario 4 - Increasing Air Temperature:} The last scenario is concerned with the role of increasing ambient temperature over the course of several years, for example as a result of climate change mechanisms, such that it decreases the rate of convective cooling and leads to early failure.
\end{itemize}

Note that all scenarios are long-term simulations of either slowly changing conditions or specific seasonal extreme events, such that there is no immediate effect in a time-scale of minutes, hours, or even days, yet they lead to accumulation of damage in the long run, making the life expectancy of the conductor to expire sooner than designed for normal operating conditions, whether with or without material damage.
 
\subsection{Numerical Results}

We consider a conductor made of aluminum undergoing cyclic air temperature, wind, and electric current loading conditions. We parameterize the cyclic loadings following the relations

\begin{align}
\theta_{air} (t) &= \theta_b + \theta_A \sin(2 \pi t),\\
w_s (t) &= w_b + w_A \sin(2 \pi t),\\
I(t) &= -I_b - I_A (\sin 4 \pi t),
\end{align}

\noindent where $t$ is measured in years, subscript $b$ indicates a base value, and subscript $A$ denotes an amplitude measure. 

We defined the current cycle to correspond to higher demand in Winter and Summer, and less demand during Spring and Fall. Definition of wind speed $w_s$ followed an arbitrary convention, with the possibility of it being replaced by a time-series data that best represents a region of interest. We represent the horizontal tension by adopting $u=0$ at $x=0$, and introducing the value of $H$ at $x=L$. For damage we adopt $\frac{d \varphi}{dx} = 0$ at the boundaries. For the current conservation equation, the boundary condition is similar to the mechanical case, and we take $V = 0$ at $x=0$ and apply a current density $J$ at $x=L$.

\begin{figure}[t]
	\centering
	\includegraphics[width=0.45\textwidth]{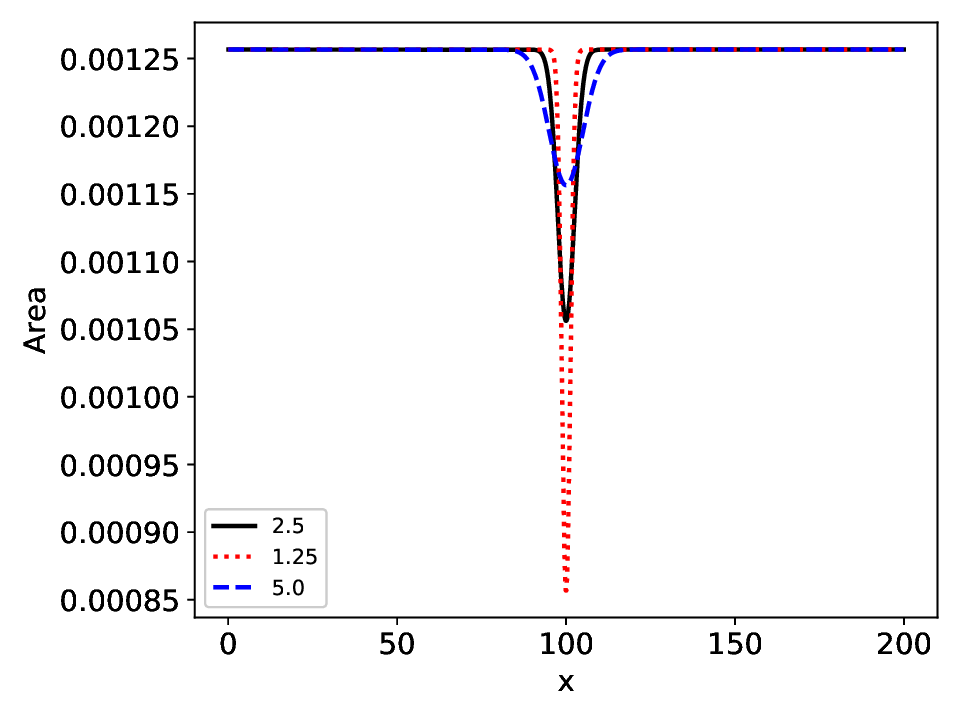}
	\caption{Variable cross-section area as a function of parameter $A_\sigma$.}
	\label{fig:A}
\end{figure}

Furthermore, we represent the damage precursor as a variable cross-section area throughout the elements. In reality, all materials are manufactured with imperfections, along with wear that is spread out across the geometry until a critical spot initiates significant damage. Here we adopt a reduced cross-section area at the center of the cable as a representation of the effects of multiple defects and damage precursors that could initiate a fracture. In that sense, we adopt the following relation for the cross-section area:

\begin{equation}
A(x) = A_0 \left(1 - \frac{1}{A_\sigma \sqrt{2 \pi}} \exp\left( \frac{- (x - L/2)^2 }{2 A_\sigma^2}\right)\right),
\end{equation} 

\noindent where $A_0$ is the undamaged cross-section area, and $A_\sigma$ represents the ratio of spread to depth of the representative area variation. Fig~\ref{fig:A}. illustrates different area profiles based on $A_\sigma$.

Typical values for all parameters associated to a reference, normal condition (Scenario 1) simulation case are presented in Table~\ref{tab:param}. All simulations run up to 4000 time-steps, representing a life-cycle of 40 years. As the Aluminum conductor starts to anneal at temperature greater than 366\ K \cite{vasquez2017end,hathout2018impact} and rapture if the temperature exceeds 373\ K \cite{cimini2013temperature} , so we consider the maximum limit temperature in the conductor at which all simulations are stopped to be $\theta_{lim} = 373\ K (100C)$.

\begin{table}[t]
	\centering
	\caption{Geometry, material, and loading parameters for reference Scenario 1 simulation.}
	\label{tab:param}
	\begin{tabular}{lll}
		Parameter                              & Value                & Unit         \\ \toprule
		Length of Cable $L$                    & 200                  & $m$          \\
		\# Elements $N$                        & 1000                 &              \\
		Diameter $d$                           & 0.04                 & $m$          \\
		Initial Damage Intensity$A_{sigma}$       & 2.5                 &     -         \\
		Time-step for fatigue $\Delta t$       & 0.01                 & $y$           \\ \midrule
		Young modulus $Y$                      & 69                   & $Gpa$        \\
		Damage layer width $\gamma$            & 0.02                  & $m$          \\
		Fracture energy $g_c$                  & 10                   & $kN/m$       \\
		Density $\rho$                         & 2700                 & $kg/m^3$     \\
		Aging coefficient $a$                  & $1\times 10^{-10}$    & $m^5/(y kg)$ \\
		Thermal conductivity $\kappa$          & 237                  & $W/(m K)$    \\
		Electrical conductivity $\sigma_{E,0}$ & $3.77 \times 10^7$   & $S/m$        \\
		Temperature coefficient $\alpha$       & $3.9 \times 10^{-3}$ & $K^{-1}$      \\ \midrule
		Pre-tension $H_0$                     & 40                   & $kN$         \\
		Base Air Temp $\theta_b$         & 288                  & $K$          \\
		Air Temp Amplitude $\theta_A$    & 10                   & $K$          \\
		Base wind speed $w_b$                 & 2                    & $ft/s$       \\
		Wind speed amplitude $w_A$            & 1                    & $ft/s$       \\
		Base current $I_b$                     & 1500                 & $A$          \\
		Current amplitude $I_A$                & 100                  & $A$         \\ \bottomrule
	\end{tabular}
\end{table}

\begin{figure}[t]
	\centering
	\subfloat[Damage.]{\includegraphics[width=0.25\textwidth]{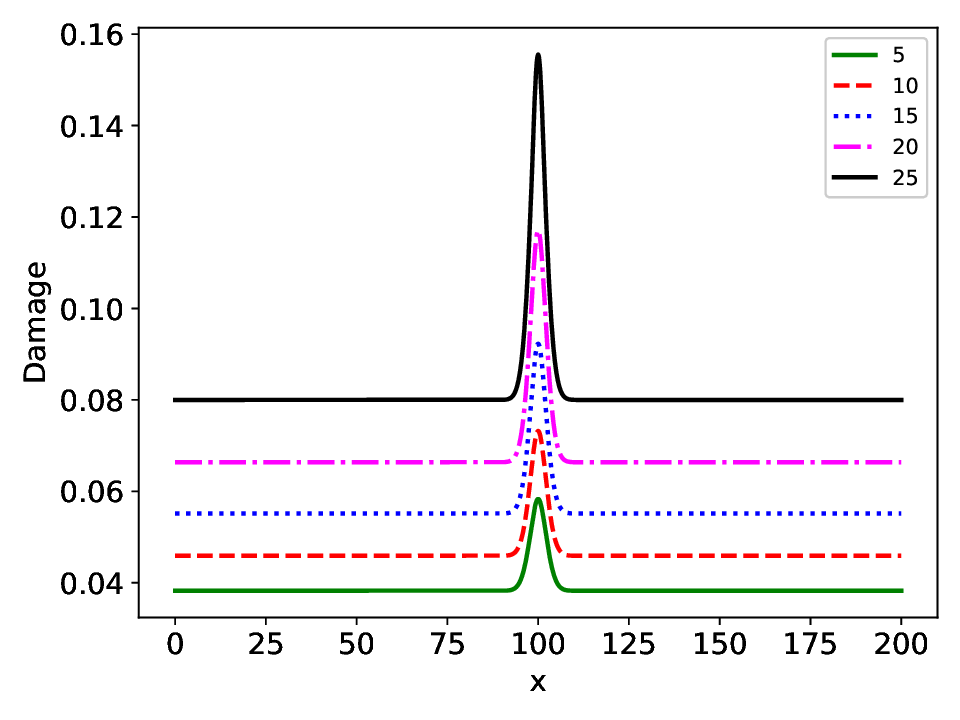}}
	\subfloat[Fatigue.]{\includegraphics[width=0.25\textwidth]{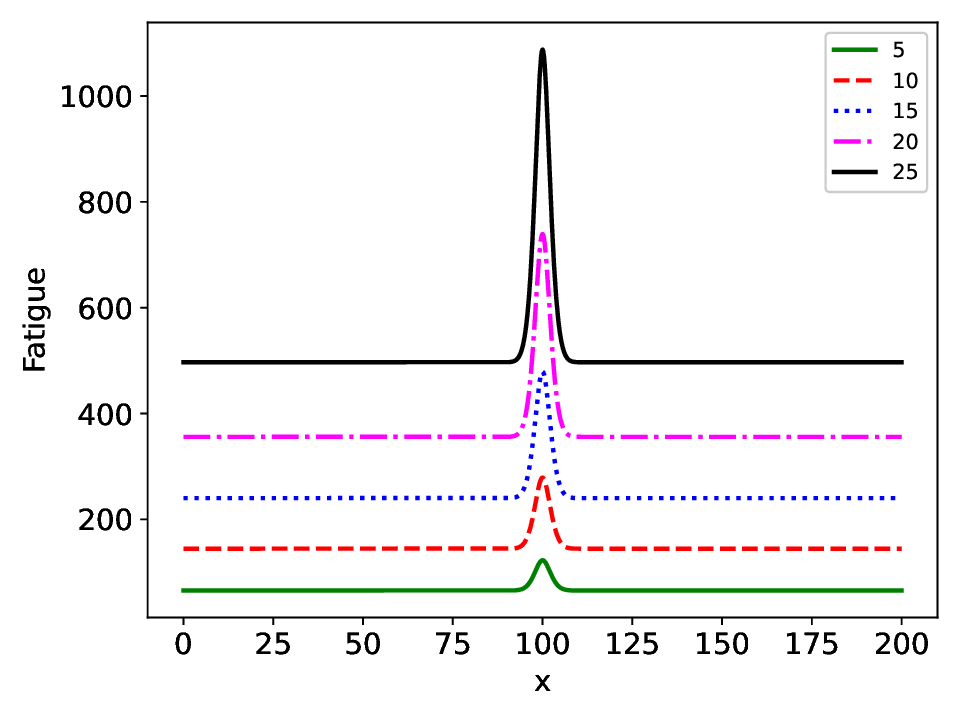}}
	\subfloat[Temperature.]{\includegraphics[width=0.25\textwidth]{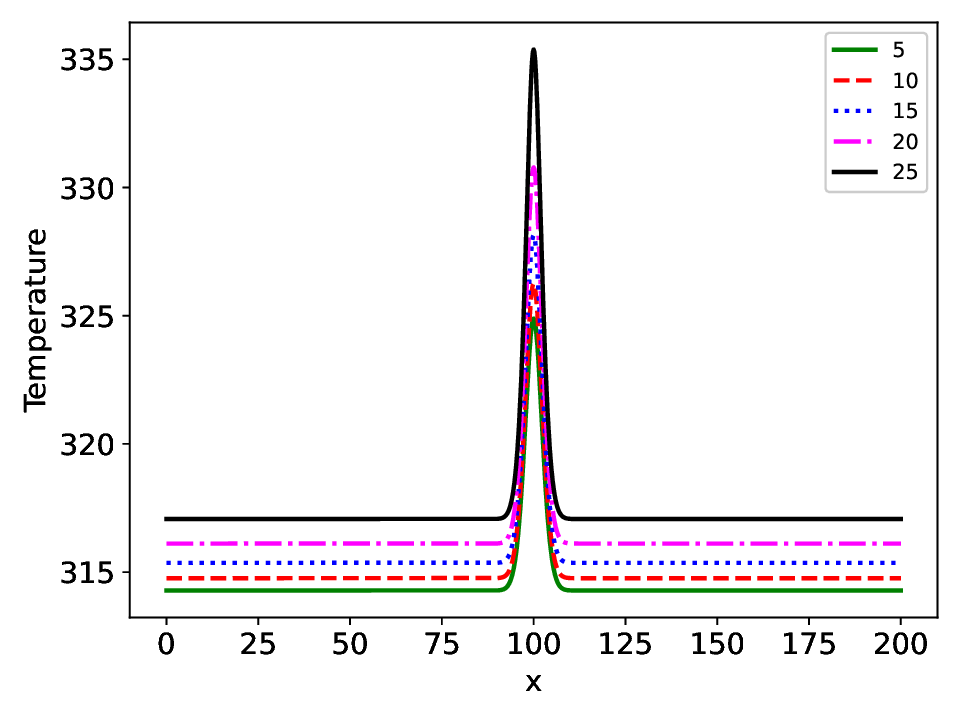}}
	\subfloat[Voltage drop.]{\includegraphics[width=0.25\textwidth]{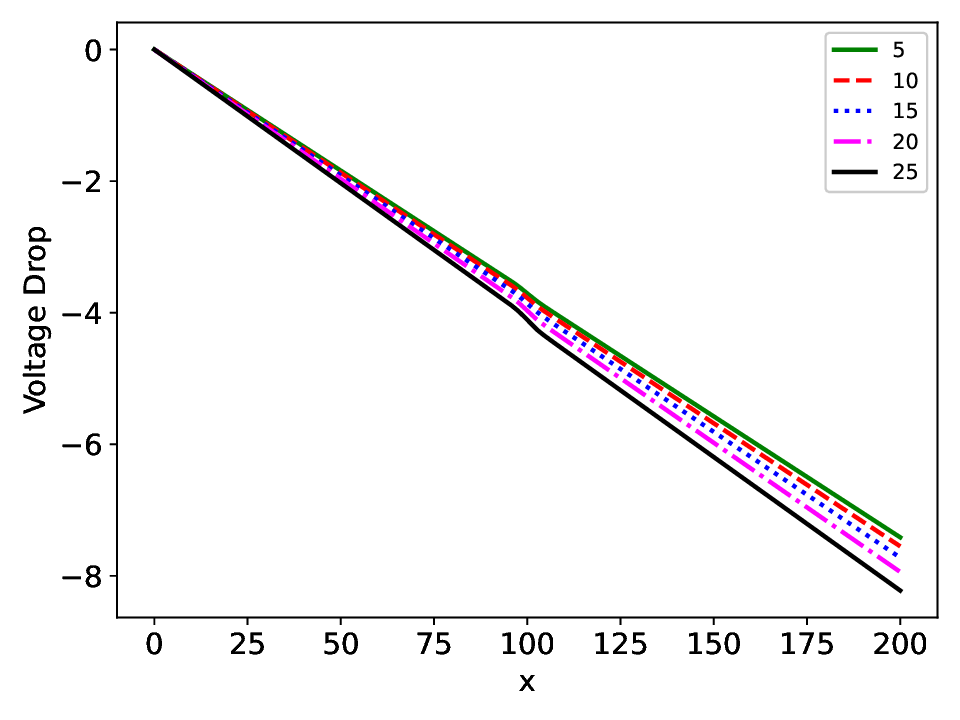}}
	\caption{Evolution of field variables over time at an interval of 5 years.}
	\label{fig:field_1_1}
\end{figure}

We start by investigating the evolution of field quantities in Scenario 1 under the reference parameters from Table~\ref{tab:param}. We plot damage, fatigue, temperature, and voltage fields every 5 years in Fig~\ref{fig:field_1_1}. We observe how damage initiates and concentrates in the region of smaller cross-section area, thus driving an increase in temperature, and a kink in the voltage fields. We also notice that due to damage and temperature increase, the voltage drop along the line increases with time, as the conductor becomes more resistive.

\begin{figure}[t]
	\centering
	\subfloat[Damage.]{\includegraphics[width=0.25\textwidth]{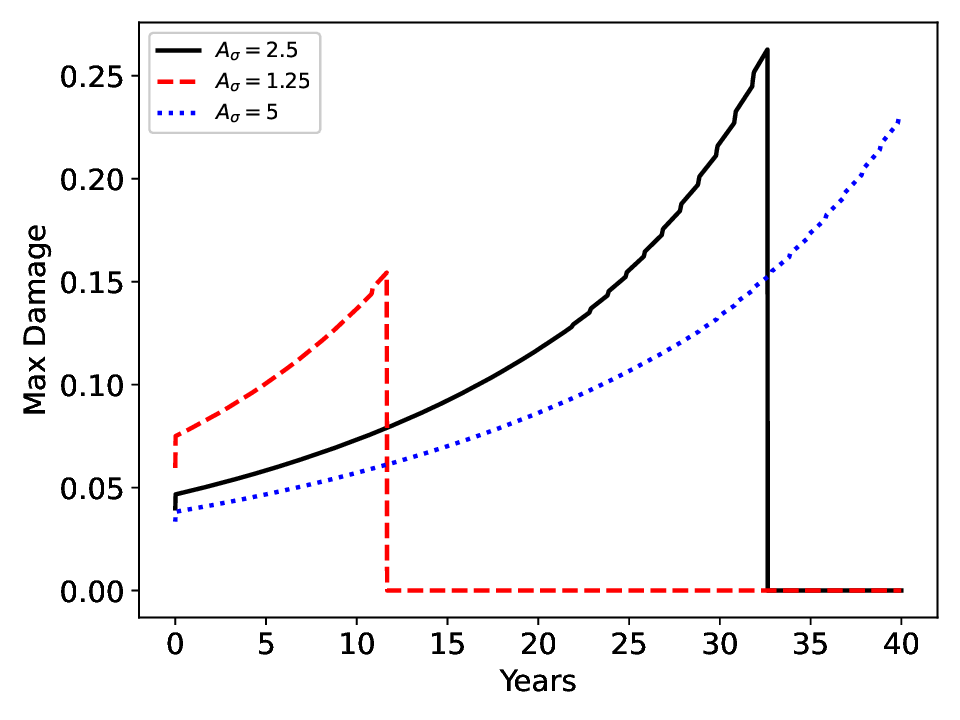}}
	\subfloat[Fatigue.]{\includegraphics[width=0.25\textwidth]{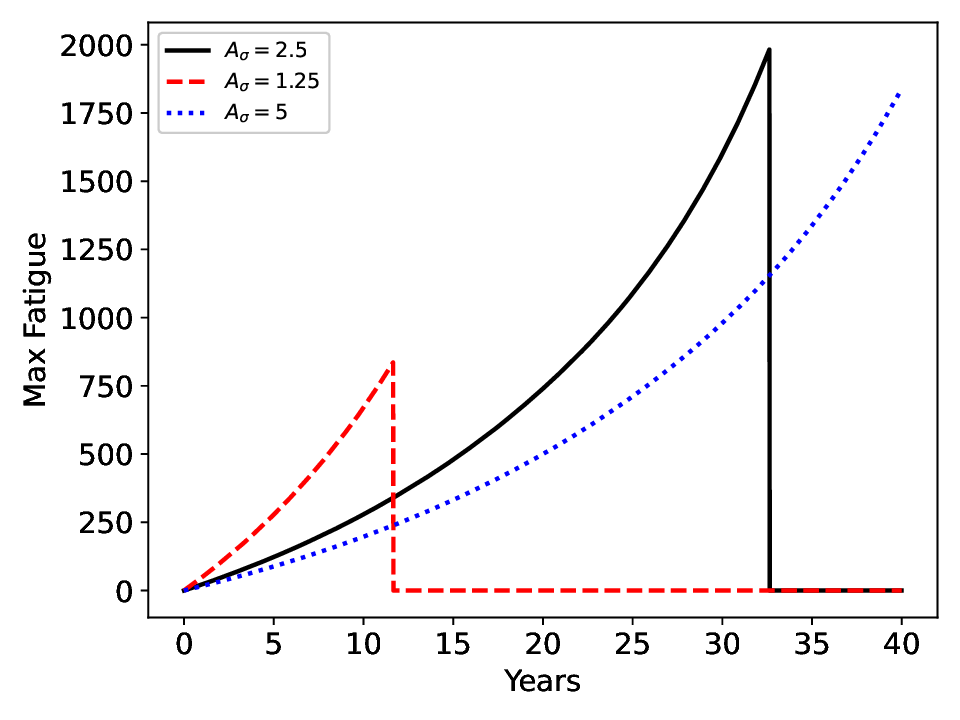}}
	\subfloat[Temperature.]{\includegraphics[width=0.25\textwidth]{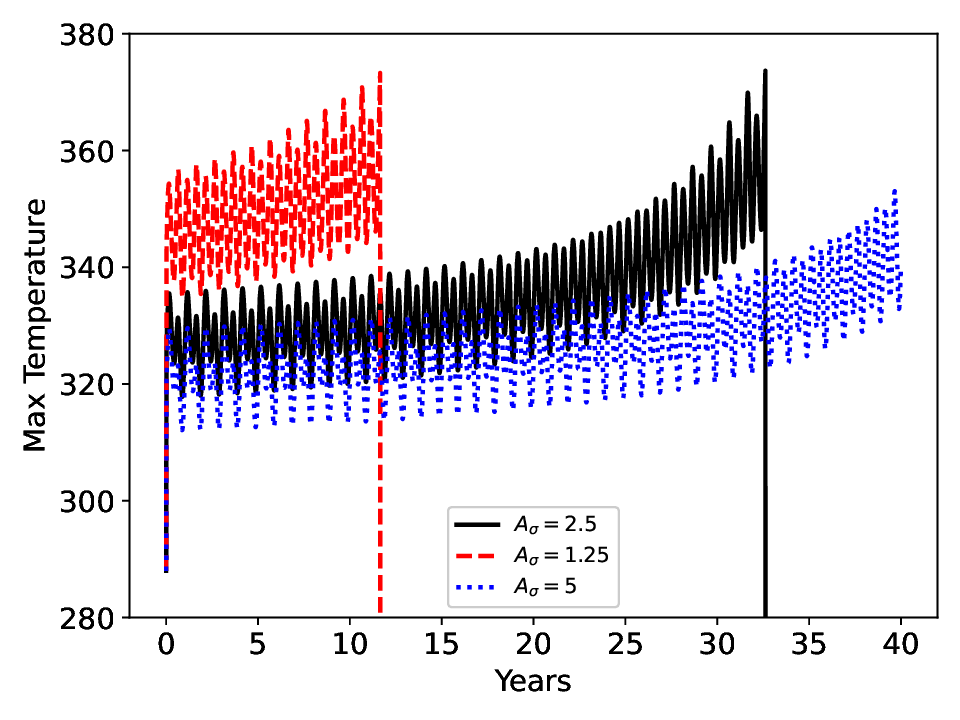}}
	\subfloat[Voltage drop.]{\includegraphics[width=0.25\textwidth]{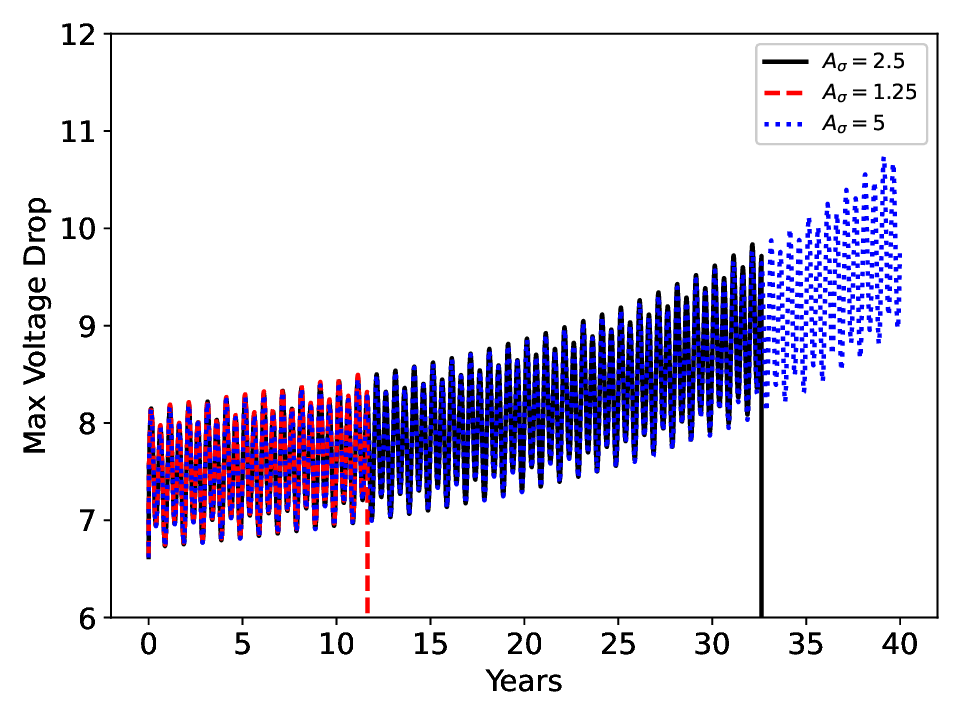}}
	\caption{Effect of initial damage as represented by variable cross-section area $A(x)$ on maximum field values over time.}
	\label{fig:ts_1_1}
\end{figure}

We can study the time-evolution of maximum values in time by tracking the central node in the damage, fatigue, and temperature cases, and track the maximum absolute voltage drop from the end node in the voltage field. Moreover, we compare the material behavior under different operating conditions, still under normal operations of Scenario 1. The most immediate comparison is to see the effect of initial damage represented by $A(x)$, Fig~\ref{fig:ts_1_1}. It is evident that a sharper, intense variation in the area greatly reduces the life, while a smoother variation did not make the material fail. 

\begin{figure}[t]
	\centering
	\subfloat[Damage.]{\includegraphics[width=0.25\textwidth]{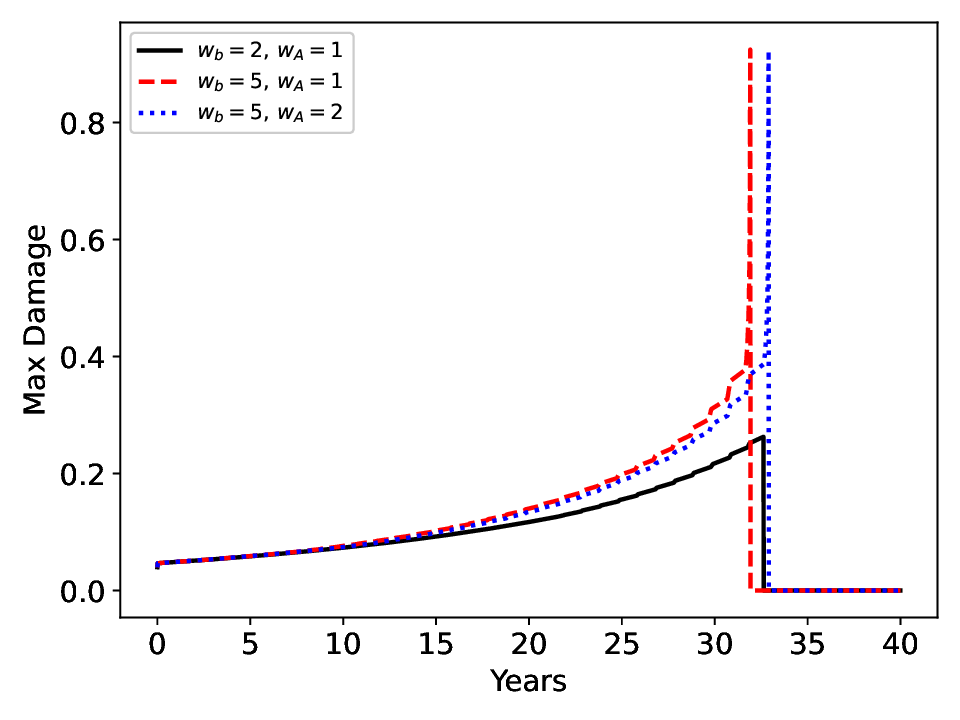}}
	\subfloat[Fatigue.]{\includegraphics[width=0.25\textwidth]{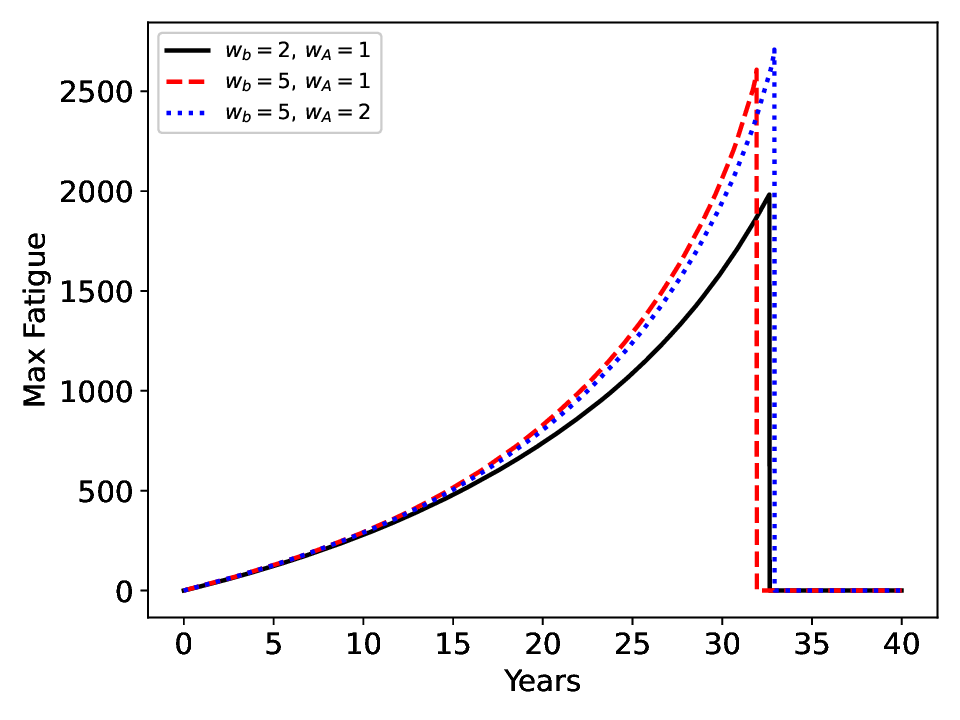}}
	\subfloat[Temperature.]{\includegraphics[width=0.25\textwidth]{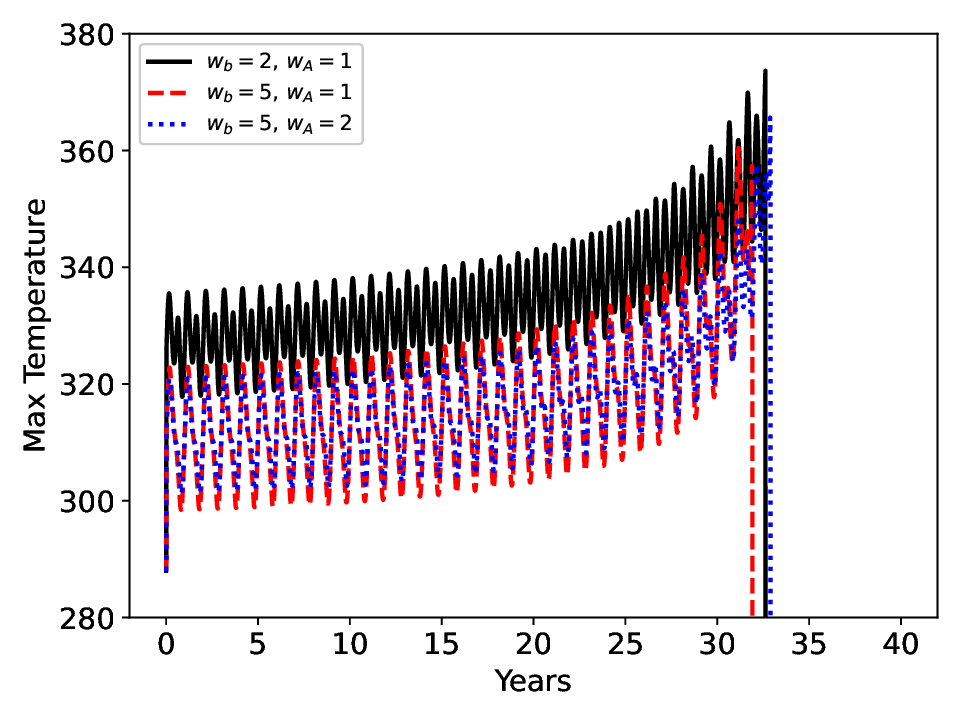}}
	\subfloat[Voltage drop.]{\includegraphics[width=0.25\textwidth]{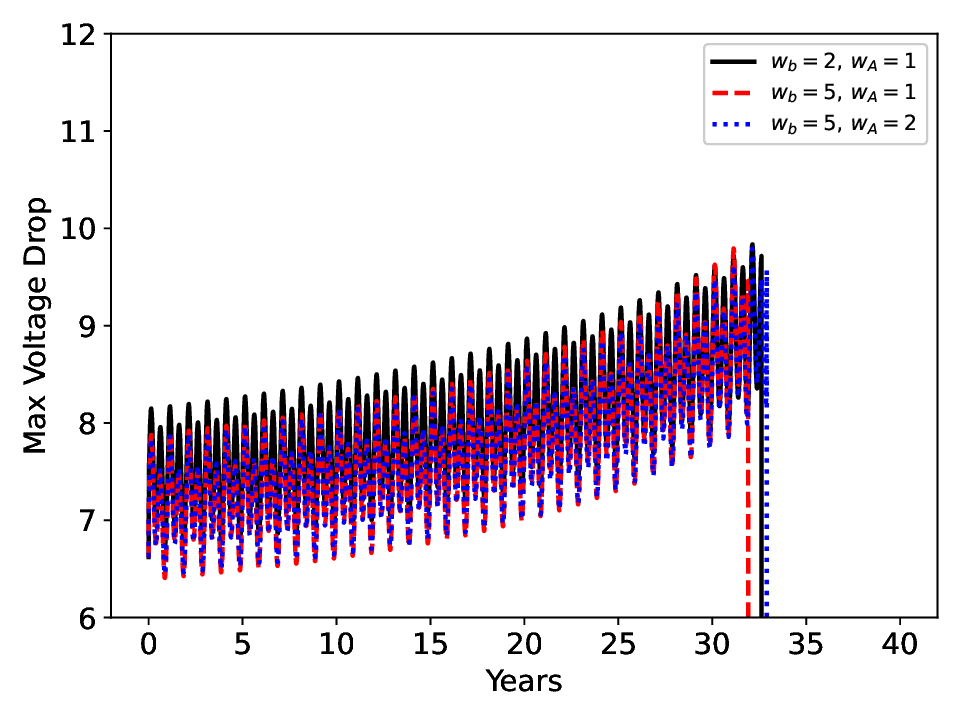}}
	\caption{Effect of wind speed on maximum values over time.}
	\label{fig:ts_1_2}
\end{figure}

\begin{figure}[t]
	\centering
	\subfloat[Damage.]{\includegraphics[width=0.25\textwidth]{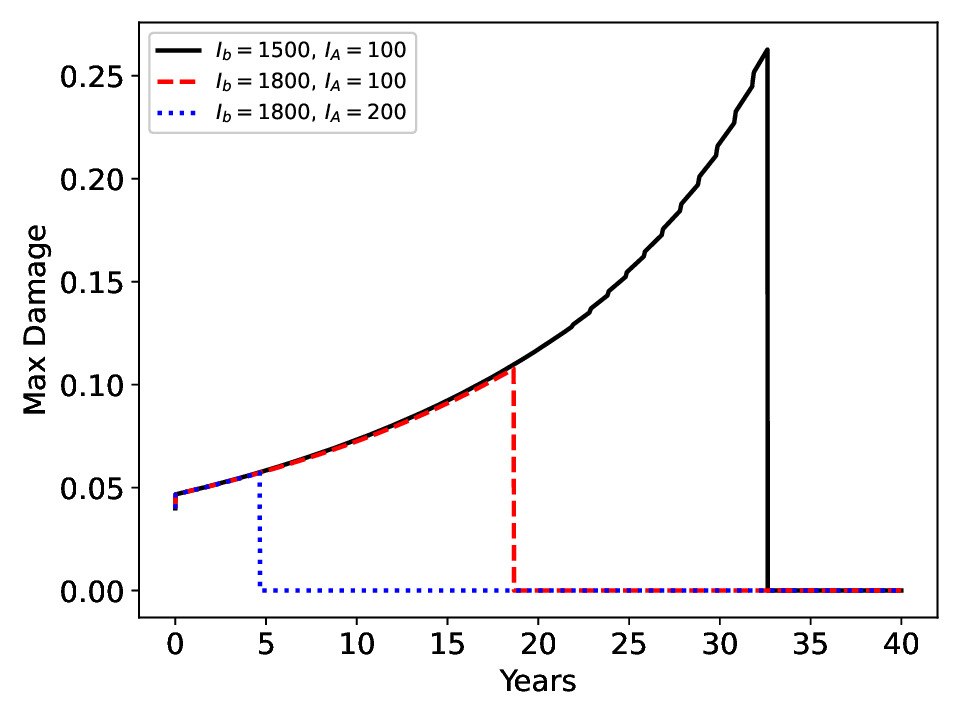}}
	\subfloat[Fatigue.]{\includegraphics[width=0.25\textwidth]{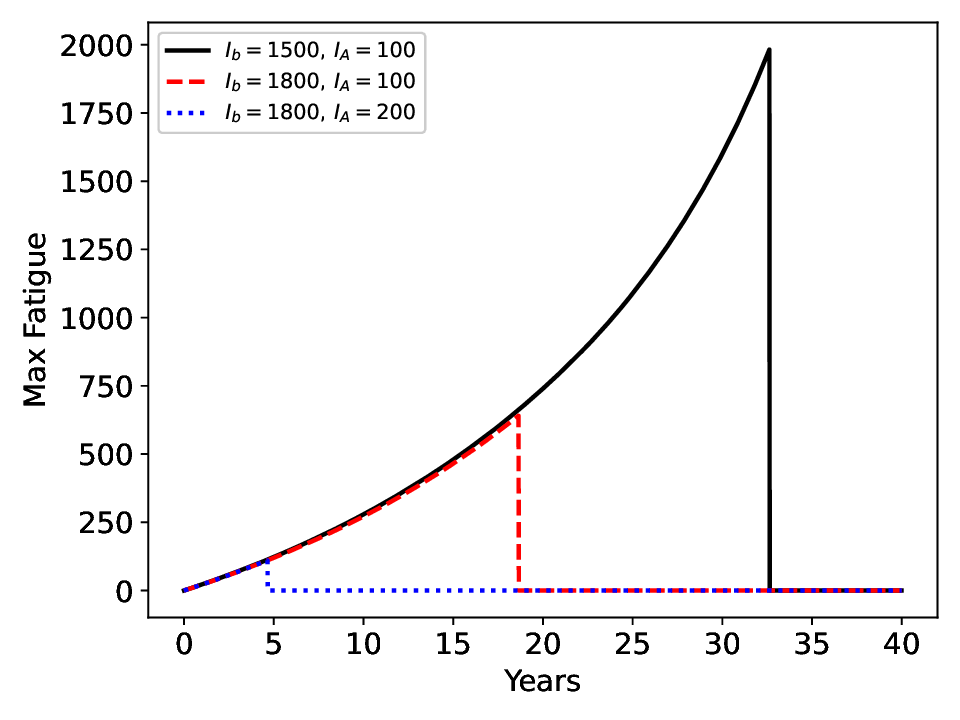}}
	\subfloat[Temperature.]{\includegraphics[width=0.25\textwidth]{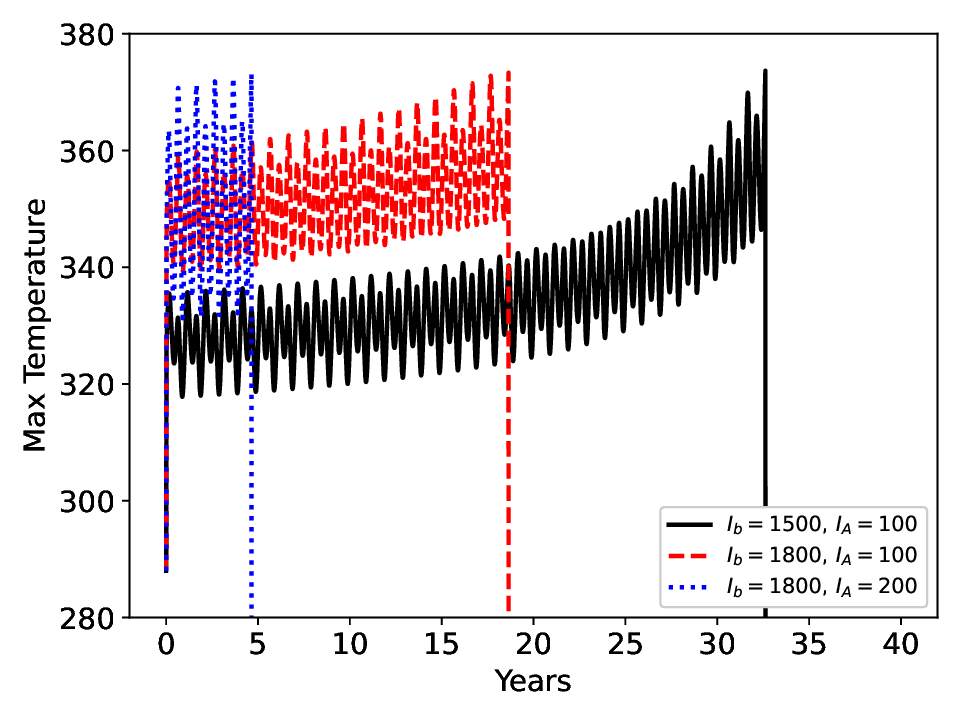}}
	\subfloat[Voltage drop.]{\includegraphics[width=0.25\textwidth]{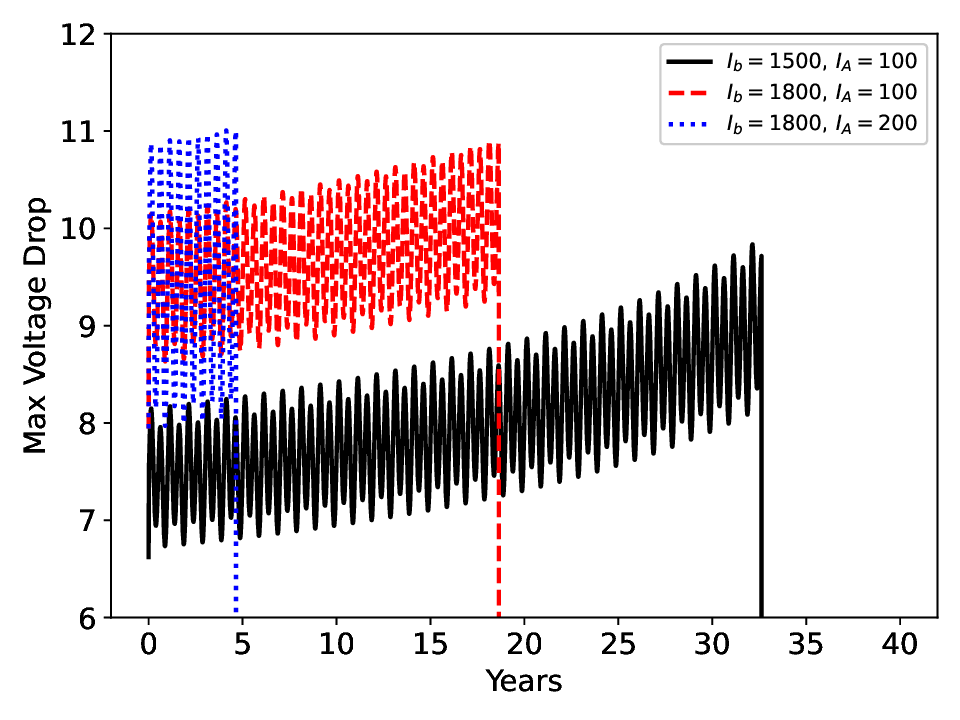}}
	\caption{Effect of electric current on maximum values over time.}
	\label{fig:ts_1_3}
\end{figure}

\begin{figure}[t]
	\centering
	\subfloat[Damage.]{\includegraphics[width=0.25\textwidth]{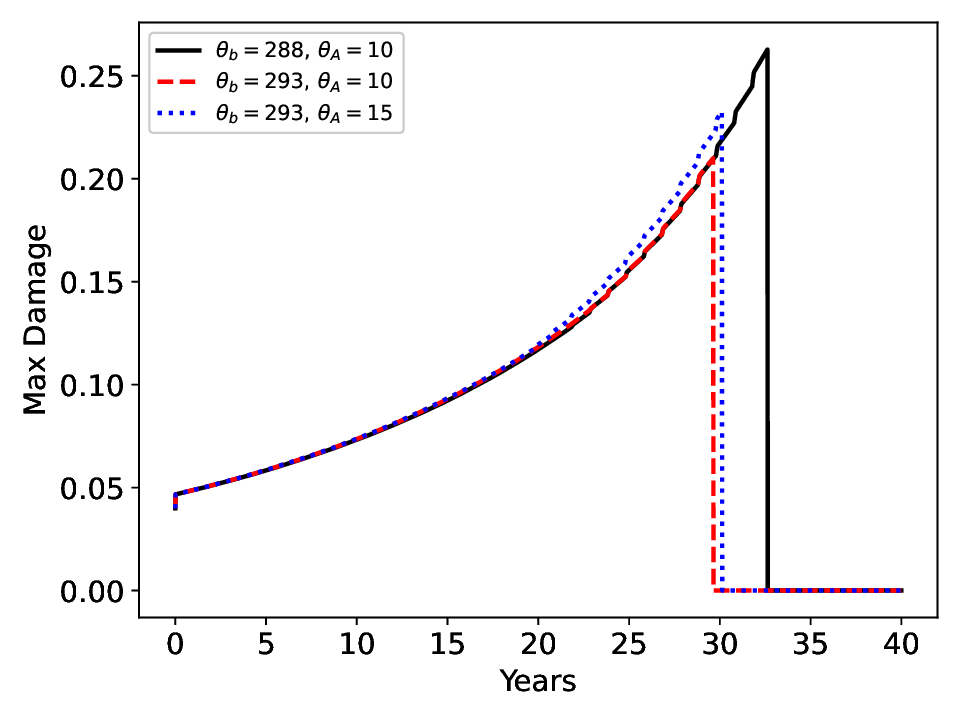}}
	\subfloat[Fatigue.]{\includegraphics[width=0.25\textwidth]{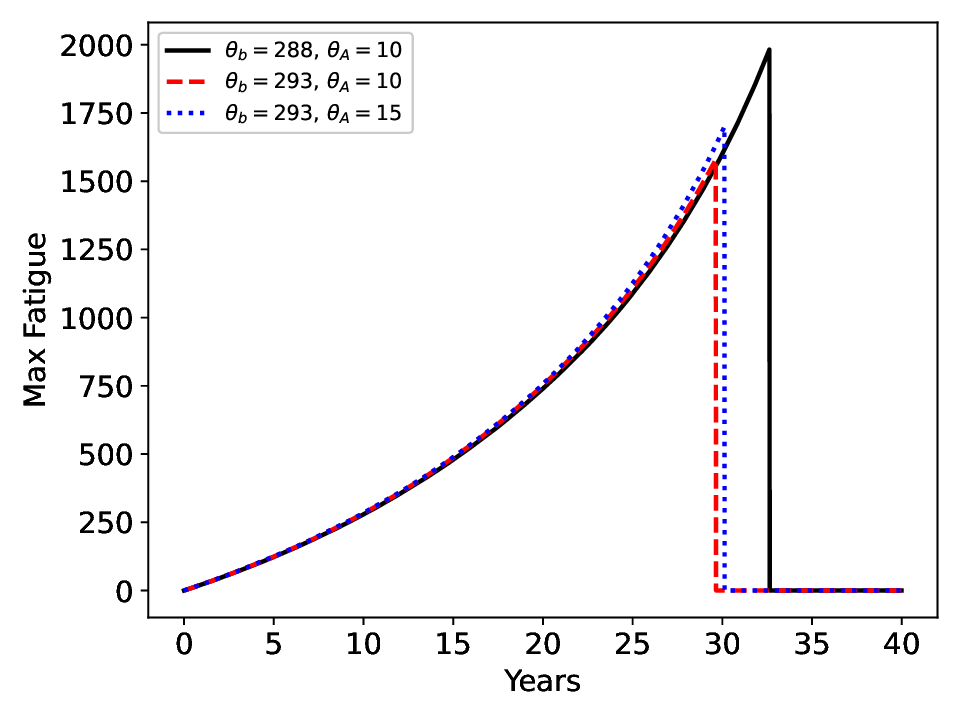}}
	\subfloat[Temperature.]{\includegraphics[width=0.25\textwidth]{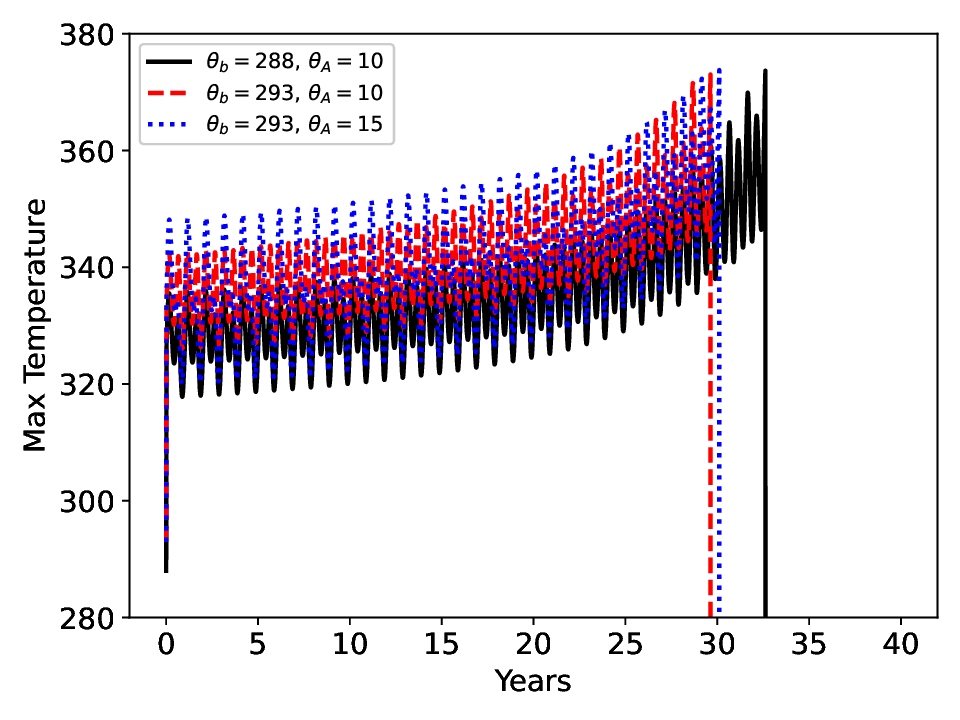}}
	\subfloat[Voltage drop.]{\includegraphics[width=0.25\textwidth]{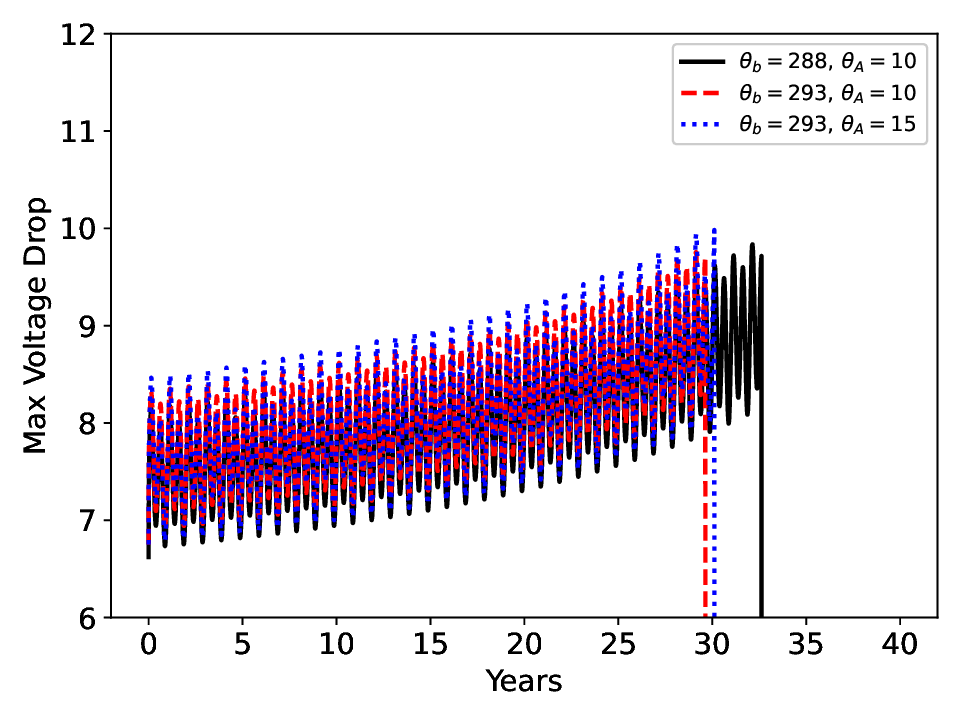}}
	\caption{Effect of air temperature on maximum values over time.}
	\label{fig:ts_1_4}
\end{figure}

Next, we investigate effects of wind, electric current, and air temperature in Figs.~\ref{fig:ts_1_2}, \ref{fig:ts_1_3}, and \ref{fig:ts_1_4}, respectively. Increase in the wind speed slightly advances the aging process, since more wind speed increases the load, accelerating the damaging, at the same time it further cools the conductor, avoiding acceleration of temperature failure. The same competitive effects can be seen with the Increase in temperature (in a warmer region), but the amplitude itself is not as influential, since with more amplitude there is also more cooling. However, the Joule heating from increasing the current base value and amplitude significantly reduce the expected life of the transmission line.

Even in Scenario 1, we see that normal operating conditions under different sets of external loads dramatically change the landscape of transmission line failure. Now, we investigate Scenarios 2, 3, and 4 with a closer focus on the maximum temperature time-series, since it is the quantity of interest related to the failure criterion.

\begin{figure}[t]
	\centering
	\subfloat[Initial damage.]{\includegraphics[width=0.45\textwidth]{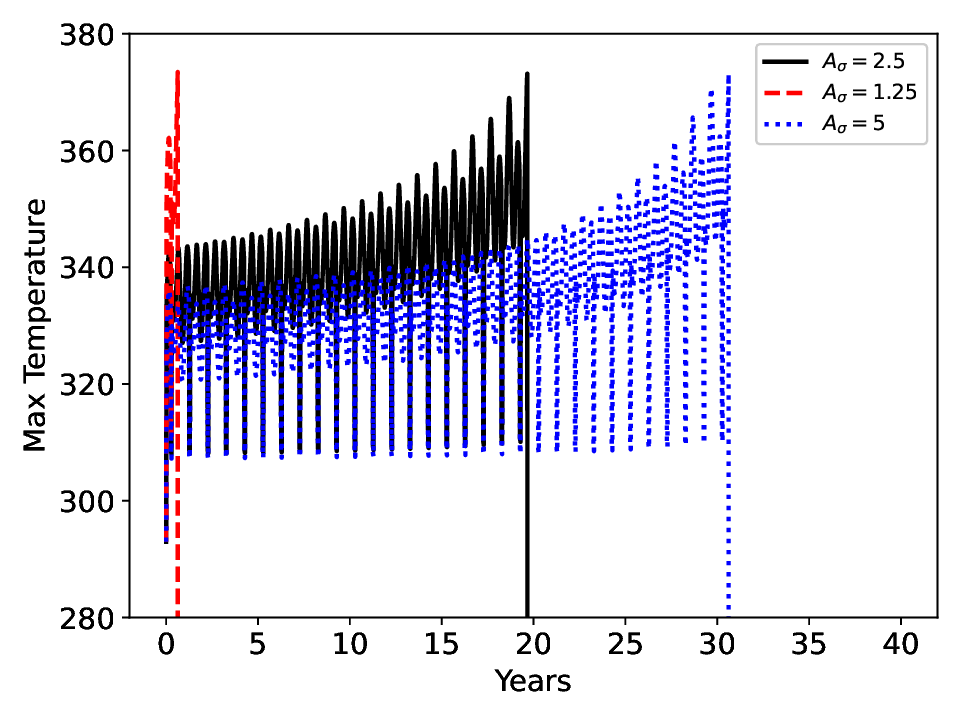}}
	\subfloat[Extreme wind.]{\includegraphics[width=0.45\textwidth]{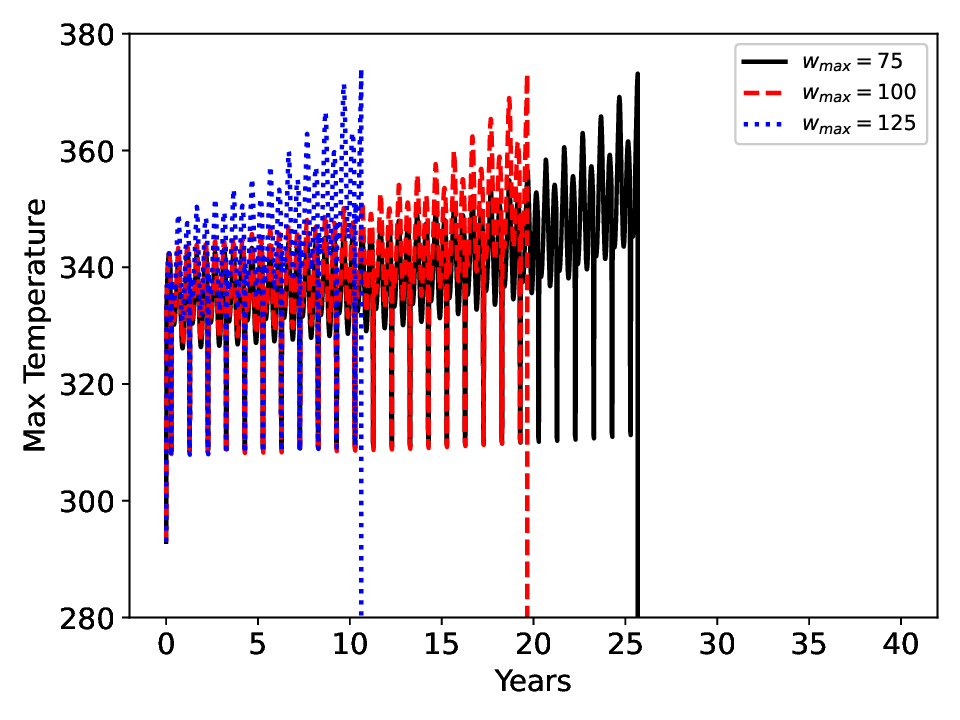}}
	\caption{Effects of initial damage and extreme wind parameter $w_{max}$ on maximum temperature over time in Scenario 2.}
	\label{fig:ts_2}
\end{figure}

In Scenario 2, we simulate the presence of high seasonal wind through the parameter $w_{max}$, applied between the 25th and 30th iterations of each year. The remainder of the year undergoes standard cyclic wind conditions as in Scenario 1. Exclusively for Scenario 2 we choose $\theta_b = 293\ K$. We compare the effects of initial damage at $w_{max}$=100 and different values of $w_{max}$ at reference parameters in Fig~\ref{fig:ts_2}. A more severe initial damage dramatically reduces the useful life of the material. Furthermore, although the high wind speeds enhance convective cooling, over time it progresses the damage to a point that the damage concentration in the center of the cable will overheat and lead to an early failure. It is not a fast effect as a cable snapping during a hurricane, rather an acceleration of aging due to intense winds over several years.

\begin{figure}[t]
	\centering
	\subfloat[Initial damage.]{\includegraphics[width=0.45\textwidth]{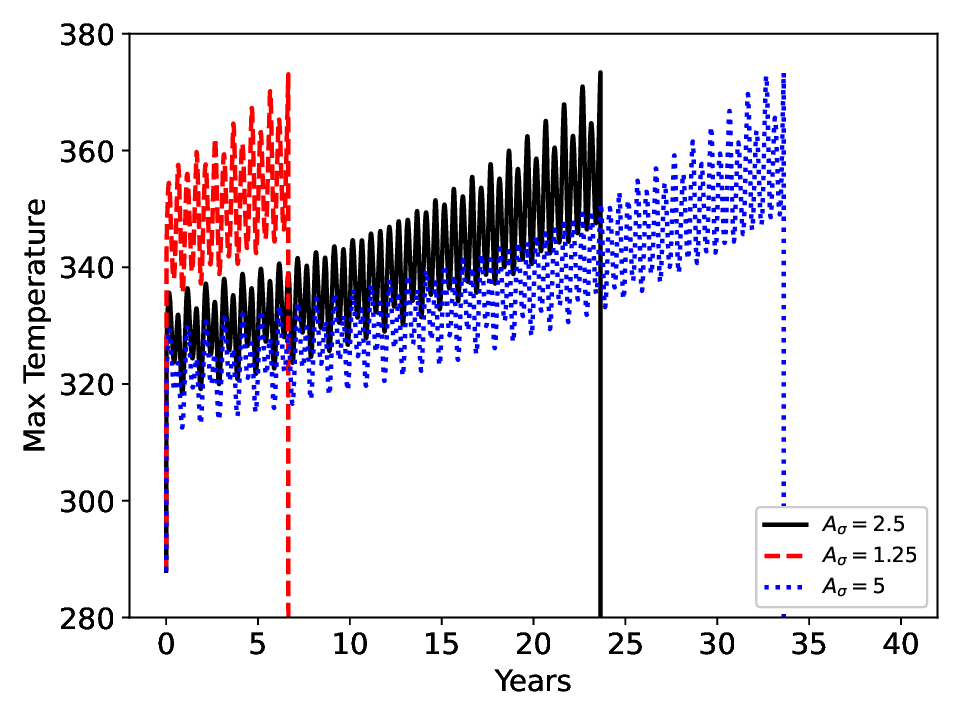}}
	\subfloat[Increase in demand.]{\includegraphics[width=0.45\textwidth]{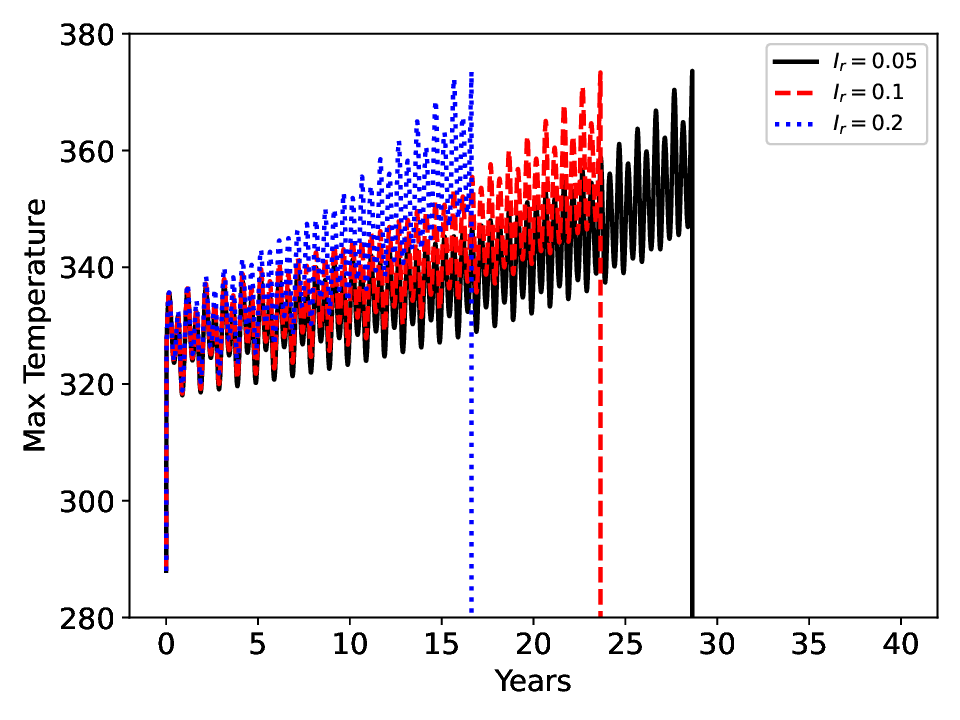}}
	\caption{Effects of initial damage and rate of electric current increase parameter $I_r$ on maximum temperature over time in Scenario 3.}
	\label{fig:ts_3}
\end{figure}

Moving on to Scenario 3, we simulate a steady increase in the base electric current demand, parameterized by the rate of increase of current, $I_r$, at each time-step, such that $I_b(n+1) = I_b(n) + I_r$. This Scenario is representative of a region with increasing population or growing industries, such that $I_r$ is chosen to provide a total increase of $400\ A$ over 40 years when $I_r = 0.1$. Then we observe the effect of initial damage at $I_r$=0.1 and different values of $I_r$ at reference parameters in Fig.~\ref{fig:ts_3}.

\begin{figure}[t]
	\centering
	\subfloat[Initial damage.]{\includegraphics[width=0.45\textwidth]{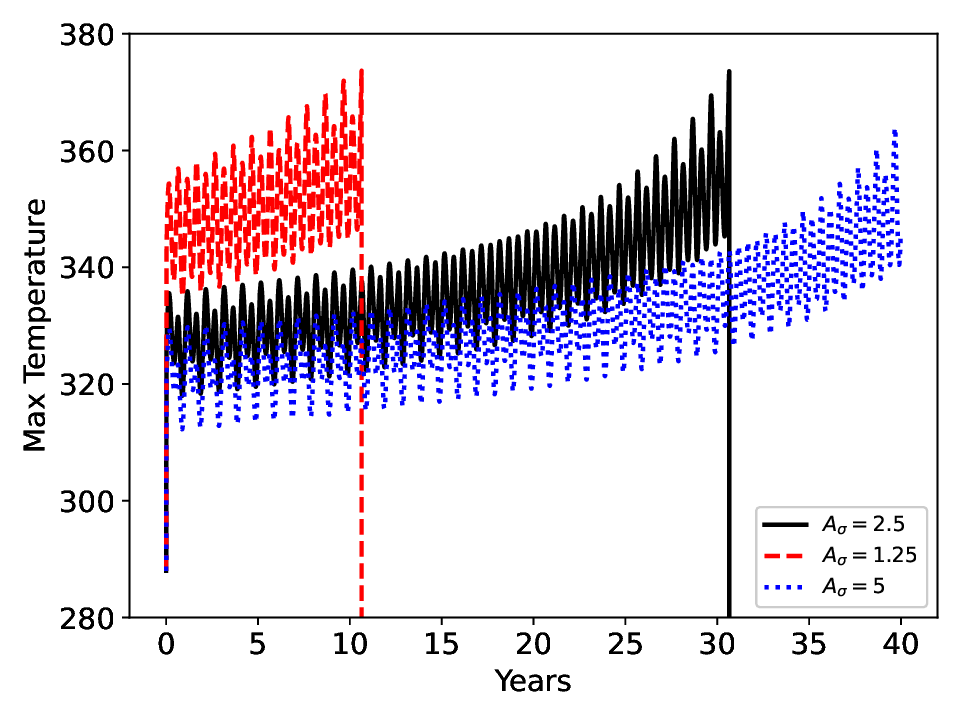}}
	\subfloat[Increase in temperature rate.]{\includegraphics[width=0.45\textwidth]{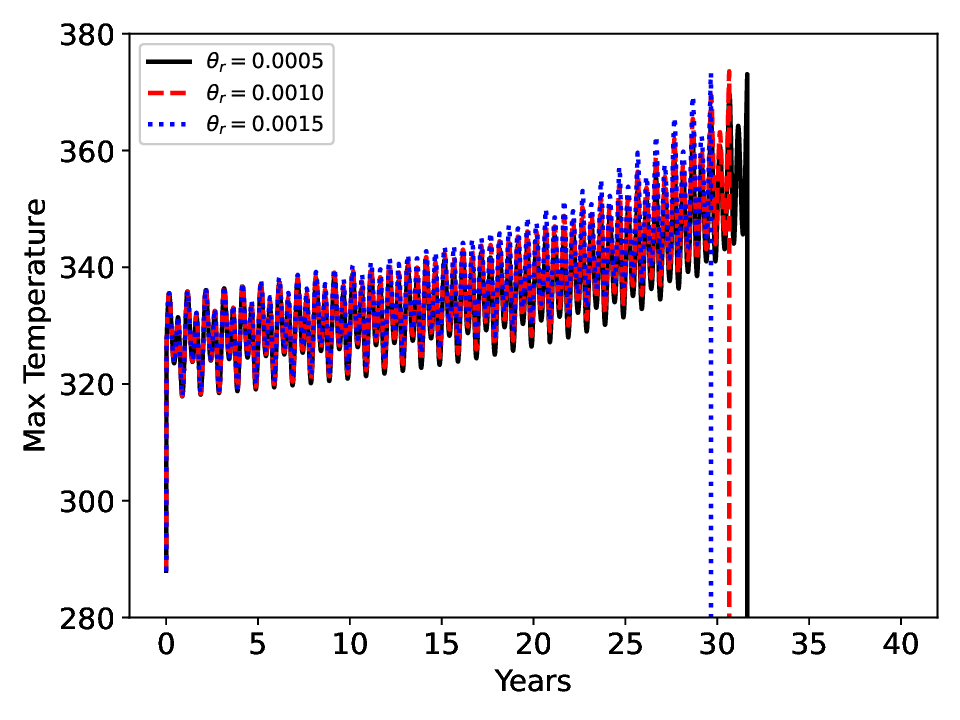}}
	\caption{Effects of initial damage and rate of temperature increase parameter $\theta_r$ on maximum temperature over time in Scenario 4.}
	\label{fig:ts_4}
\end{figure}

Last, we investigate effects increase of base air temperature in Scenario 4, controlled by parameter $\theta_r$. Similarly to Scenario 3, here we design $\theta_r$  to increment $\theta_b$ by $\theta_b(n+1) = \theta_b(n) + \theta_r$. For a $4\ K$ increase in 40 years, $\theta_r = 0.001$. In Fig. ~\ref{fig:ts_4} we see that even a slight increase in average yearly temperatures at reference parameters, is sufficient to accelerate failure in transmission lines by a couple of years.

\section{Stochastic Solution}
\label{sec:stochastic}

We consider the coupled failure model presented as a black-box on which we can perform uncertainty quantification (UQ) and sensitivity analysis (SA). The use of non-intrusive methods is attractive, since we can employ the same procedures used in the deterministic solution to solve each realization of the stochastic problem. We use the Probabilistic Collocation Method (PCM) to compute the moments of our quantity of interest (QoI), which in this work is the conductor temperature. 

Furthermore, the PCM is employed to perform the global sensitivity analysis through the computation of Sobol Sensitivity Index $S_i$ in a computationally efficient way. The sensitivity index measures the importance of input parameters in the total variance of the QoI solution, and the PCm framework performs the computation through a simple post-processing taks from UQ data.

Finally, the PCM building blocks are used to facilitate the computation of $p_f$ directly from the first moment of a Bernoulli random variable defined directly from $g(\theta_{lim},\theta_{max};t)$.

\subsection{Uncertainty Quantification}

To perform the UQ analysis we use the PCM method, which consists in a  polynomial interpolation to approximate the solution in the stochastic space, mapping the points from physical to stochastic space is made through the parametric probability density function (PDF). This is achieved by approximating the solution using orthogonal Lagrange polynomials, which due to orthogonality properties reduces the computation of expectation and variance to evaluation of QoI at the collocation points. This approach reduces computational cost significantly, while improving convergence rates. Here we follow the methodology presented in \cite{barros2021integrated}.

Let $(\Omega_s,\mathcal{G}, \mathbb{P})$ be a complete probability space, where $\Omega_s$ is the space of outcomes $\omega$, $\mathcal{G}$ is the $\sigma-$algebra and $\mathbb{P}$ is a probability measure, $\mathbb{P}: \mathcal{G}\to[0,1]$. We render the transmission line model stochastic by letting material and load parameters to be random variables defined in a set $\xi(\omega)$, which in turn leads to outputs, such as the QoI temperature field also be random variables. We simplify the notation and only explicitly represent the random parameters as $\xi = \xi(\omega)$. 

We denote our quantity of interest as $Q$, and write the mathematical expectation of $Q$, $\mathbb{E}\left[ Q(x, t; \xi) \right]$ in a one-dimensional stochastic space  as

\begin{equation}
\label{eq:pcm_expectation}
\mathbb{E}\left[ Q(x, t; \xi) \right] = \int_{a}^b Q(x, t; \xi) \rho(\xi) d \xi,
\end{equation}

\noindent where $\rho(\xi)$ is the PDF of $\xi$. We evaluate the integration using Gauss quadrature, mapping physical parametric domain to the standard domain $[-1,1]$. The integral should then be written as

\begin{equation}
\label{eq:pcm_expectation2}
\mathbb{E}\left[ Q(x, t; \xi) \right] = \int_{-1}^1 Q(x, t; \xi(\eta)) \rho(\xi(\eta)) J d \xi(\eta),
\end{equation} 

\noindent where $J = d \xi / d \eta$ represents the Jacobian of the transformation. We approximate the expectation by introducing a polynomial interpolation of the exact solution in the stochastic space, $\hat{Q}(x, t; \xi)$:

\begin{equation}
\label{eq:pcm_expectation3}
\mathbb{E}\left[ Q(x, t; \xi) \right] \approx \int_{-1}^1 \hat{Q}(x, yt; \xi(\eta)) \rho(\xi(\eta)) J d \xi(\eta).
\end{equation} 	

We interpolate the solution in the stochastic space using Lagrange polynomials $L_i(\xi)$:

\begin{equation}
\label{eq:poly_approximation}
\hat{Q}(x, t; \xi) = \sum_{i = 1}^{I} Q(x, t; \xi_i) L_i(\xi),
\end{equation}

\noindent which satisfy the Kronecker delta property at the interpolation points:

\begin{equation}
\label{eq:delta}
L_i(\xi_j) = \delta_{ij}.
\end{equation}

Substituting Eq.~(\ref{eq:poly_approximation}) into (\ref{eq:pcm_expectation3}), we approximate the integral using the quadrature rule and evaluate the expectation as

\begin{equation}
\mathbb{E}\left[ Q(x, t; \xi) \right] \approx \sum_{p = 1}^{P} w_p \rho(\xi(\eta)) J \sum_{i = 1}^{I} Q(x, t; \xi(\eta) ) L_i(\xi(\eta) ), \label{eq:pcm_exp_approx2}
\end{equation}

\noindent where we compute the coordinates $\eta_p$ and weights $w_p$ for each integration point $q = 1,\, 2,\, \dots,\, P$. We choose the collocation points to be the same as the integration points $p$ on the paramectric space by the Kronecker property of the Lagrange polynomials Eq.~(\ref{eq:delta}), and simplify the approximation from Eq.~(\ref{eq:pcm_exp_approx2}) to a single summation:

\begin{equation}
\label{eq:pcm_single}
\mathbb{E}\left[ Q(x, t; \xi) \right] = \sum_{p = 1}^{P} w_p \rho(\xi_p(\eta_p))  J Q(x, t; \xi_p(\eta_p)).
\end{equation}

A linear affine mapping from the standard to the real domain $\xi_p(\eta_p) = a + \frac{(b-a)}{2}(\eta_p + 1)$ gives us the Jacobian (for a one-dimensional integration) as $J = (b-a)/2$, and also provides the respective values of the random variable in the physical space.

Finally, we write the integration as a summation over the collocation points, where we assume a uniform distribution for the parameters in the physical space $\xi \sim \mathcal{U}[a,b]$, with $\rho(\xi) = 1/(b-a)$. The expectation becomes 

\begin{equation}
\mathbb{E}\left[ Q(x, y, t; \xi) \right] = \frac{1}{2}\sum_{p = 1}^{P}  w_p Q(x, t; \xi_p).
\end{equation} 

The standard deviation is computed as

\begin{equation}
\sigma \left[ Q(x, t; \xi) \right] =  \sqrt{ \frac{1}{2}\sum_{p = 1}^{P}  w_p \left( Q(x, t; \xi_p) - \mathbb{E}\left[ Q(x, t; \xi) \right] \right)^2}.
\end{equation}

Generalization of PCM to higher dimensions through tensor product is an extension of Eq.~(\ref{eq:pcm_expectation})

\begin{align}
\mathbb{E}\left[ Q(x, t; \xi^1,\, \dots ,\, \xi^k) \right] &= \mathbb{E}_{PCM}\left[ Q(x, t; \xi^1,\, \dots ,\, \xi^k) \right] \notag \\ &\approx \sum_{p = 1}^{P}\dots \sum_{l = 1}^{L} w_p \dots w_l \, \rho(\xi_p) \dots \rho(\xi_l) \, J_p \dots J_l\,  Q(x,  t; \xi^1_p,\,\dots,\, \xi^k_l) \label{eq:pcm_multi}
\end{align}

\noindent where we have $k$ summations, one for each dimension in the random space. In $\xi^k_l$ the superscript indicates the dimension in the random space, and the subscript specifies the collocation point in that dimension. Simplifying the notation using $\mathbb{E}\left[ Q(x, t; \xi^1,\, \dots ,\, \xi^k) \right] = \mathbb{E}\left[ Q\right]$, we write the expression for the standard deviation as

\begin{align}
&\sigma\left[ Q(x, t; \xi^1,\, \dots ,\, \xi^k) \right] = \sigma_{PCM}\left[ Q(x, t; \xi^1,\, \dots ,\, \xi^k) \right] \notag \\&\approx \sqrt{\sum_{p = 1}^{P}\dots \sum_{l = 1}^{L} w_p \dots w_l \, \rho(\xi_p) \dots \rho(\xi_l) \, J_p \dots J_l\, \left(Q(x, t; \xi^1_p,\,\dots,\, \xi^k_l) -  \mathbb{E}\left[ Q\right]\right)^2}. \label{eq:pcm_multi_std}
\end{align}

We note that we assume the random variables to be mutually independent and the discretization in the parametric space is taken to be isotropic. We further remark that a fully tensorial product as in this project is sufficient in terms of computational efficiency, as we deal with 6 dimensions or less. To avoid the curse of dimensionality in high-dimensional stochastic space Smolyak sparse grids\cite{smolyak1963quadrature} is one popular solution to reduce the number of realizations, while still achieving comparable accuracy. Further methods that aid UQ with dimensionality reduction include Principal Component Analysis (PCA) \cite{abdi2010principal}, low-rank approximations \cite{chevreuil2015least}, and active subpace methods \cite{constantine2017global}.

\subsection{Sensitivity Analysis}

We  study the global sensitivity of input parameters through Sobol indices \cite{sobol1993sensitivity}, where we compute the relative importance of each parameters to the variance of our QoI. We refer to Saltelli et al (2010) \cite{saltelli2010variance} for derivation details. Let the $j-$th parameter in the global sensitivity analysis be denoted by $\xi^j$, $j = 1,2,\dots,k$, $k$ being the total dimension of the parametric space. The effect of parameter $\xi^j$ on variance $V$ of QoI is

\begin{equation}
\label{eq:var}
V_{\xi^j}\left(\mathbb{E}_{\mathbf{\xi}^{\sim j}}(Q | \xi^j)\right)
\end{equation}

\noindent where $\mathbf{\xi}^{\sim j}$ denotes the combination of all possible values for random parameters with the exception of $\xi^j$, which is fixed at some value. Eq~(\ref{eq:var}) is equivalent to taking the expected value of $Q$ having fixed a value for $\xi^j$, and then taking the variance over all possible values of $\xi^j$. From the Law of Total Variance, we have

\begin{equation}
\label{eq:law}
V_{\xi^j}\left(\mathbb{E}_{\mathbf{\xi}^{\sim j}}(Q | \xi^j)\right) + \mathbb{E}_{\mathbf{\xi}^{j}}\left(V_{\mathbf{\xi}^{\sim j}}(Q | \xi^j)\right) = V(Q)
\end{equation}

The second term on the left-hand side is called the residual and $V(Q)$ is the total variance. We normalize Eq.~(\ref{eq:law}) to obtain the first-order sensitivity index that measures the effect on total variance by random variable $\xi^j$ as:

\begin{equation}
\label{eq:si}
S_i = \frac{V_{\xi^j}\left(\mathbb{E}_{\mathbf{\xi}^{\sim j}}(U | \xi^j)\right)}{V(U)}
\end{equation}

The sensitivity indices $S_i$ measure the first-order effect on variance from $\xi^j$, not taking into account interactions between $\xi^j$ and other parameters. From the normalization, $\sum S_i < 1$, the remainder consisting of high-order interactions between the parameters, which we do not consider in this paper, but could be obtained in a similar fashion from post-processing of PCM \cite{barros2021integrated}.

The computation of $S_i$ could be quite challenging if the UQ is done through a MC method, yet here the PCM acts as a building block for fast, cheap computations of global sensitivity.

\subsection{Probability of Failure}

In the last step of the stochastic analysis of transmission line failure, we aim to compute the probability of failure $p_f$ as a function of time. Existing methods to compute $p_f$ in reliability literature normally use MC to count the number of failure events \cite{machado2015reliability}. Stochastic collocation methods only provide the moments of a limit state function $g(R,S)$, which then need to be transformed into a PDF to computation of $p_f$ similarly to Eq.~(\ref{eq:pf}). Such transformation from the moments to the PDF yield another level of complexity, and can be done methods of moments \cite{low2013new,dang2019novel}, Polynomial Chaos \cite{lasota2015polynomial,garcia2021polinomial}, Gaussian transformations \cite{he2014sparse} or optimization through entropy methods \cite{winterstein2011extremes}.

In this work, we propose an alternative to once again take advantage of the efficiency of PCM as a building block, and obtain the probability of failure $p_f$ as a typical UQ expectation. Instead of computing moments of $g$, obtaining an approximate PDF, and then computing $P(g<0)$, we first transform $g$ into another random variable $h$, such that $h$ is Bernoulli with coefficient $p_h$.

The definition of $h$ comes directly from $g$:

\begin{equation}
\label{eq:h}
h = \begin{cases}
0, & \text{if } g \geq 0,\\
1, & \text{otherwise}.
\end{cases}
\end{equation}

In practice, for each realization of PCM, we generate a time-series vector $h$ that is $0$ until the point where $\theta_{max} > \theta_{lim}$, at which $h = 1$ until the final time-step. For a single realization this represents a step function at the failure point. When we consider the expectation of $h$ at a fixed time-step, then by the smoothness of the QoI $h$ becomes a real value $0 \leq h \leq 1$. 

Since we use PCM to compute the expectation of $h$, which is a Bernoulli random variable, the expectation of $h$ gives $p_f$, which is equivalent to the Bernoulli parameter $p_h$ itself. Therefore, a single integration using PCM is capable providing an accurate measure of $p_f$ for each time-step.

\subsection{Numerical Results}

With a deterministic simulation, we have a solid understanding of the effects of different parameters on the overall behavior of the system, including how shorter or longer the life-span of the transmission line would be. We now focus on the effects of parametric uncertainty on maximum temperature. The parametric uncertainty is chosen to be a uniform distribution with a range of $10\%$ from the expected (mean) parameter value equal to the values used in the deterministic solution. 

\subsubsection{Preliminary study: Scenario 1}

We initially consider Scenario 1 as a point of reference to perform preliminary analyzes on parametric and input load uncertainty. First, we perform UQ and SA on material parameters to identify the two most influential in the variance of $\theta_{max}$. We focus on the set of material parameters that form the set $\xi_m (\omega) = \{ A_\sigma(\omega), \gamma(\omega), g_c(\omega), a(\omega)\}$ that are either not accurately measurable or are an artifact of mathematical modeling. We assume all the other material parameters to be deterministic.

In a separate stochastic space, we study the effect of loading conditions by looking into the set $\xi_l(\omega) = \{\theta_b(\omega), \theta_A(\omega), w_b(\omega), w_A(\omega), I_b(\omega),I_A(\omega)\}$. We also select the top two most important from the global SA. Finally, we perform a final UQ/SA assessment with the 4 parameters selected in what we denote set $\xi_1 (\omega)$. Later, those 4 parameters will be combined with specific scenario parameter in Scenarios 2, 3, and 4.

For all simulations we consider $n = 5$ PCM points per dimension. We also remark that in order to study the time-series results of maximum temperature among the entire stochastic space, we need to truncate the time-series minimum time of failure from all realizations, since each realization observes failure at different times.

\begin{figure}[t]
	\centering
	\subfloat[Expectation.]{\includegraphics[width=0.45\textwidth]{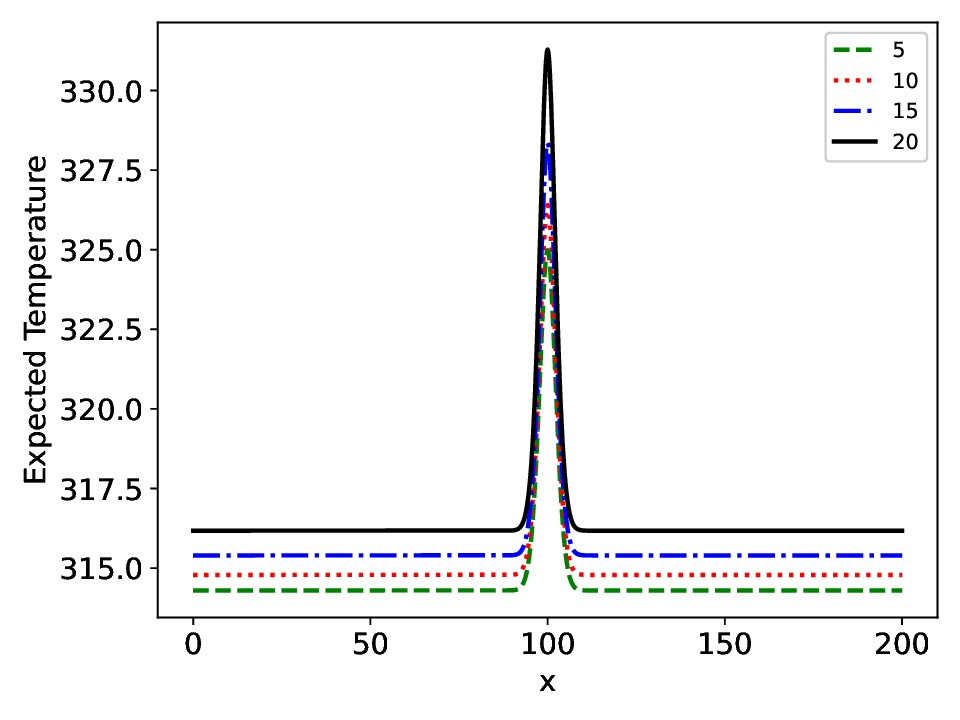}}
	\subfloat[Standard deviation.]{\includegraphics[width=0.45\textwidth]{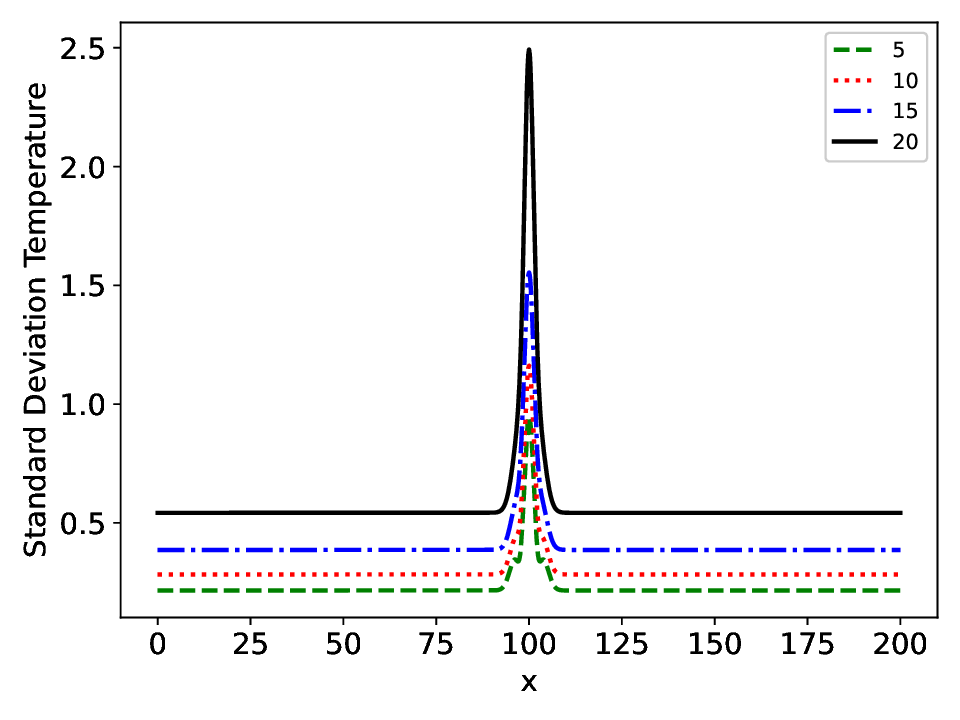}}
	\caption{Expectation and standard deviation temperature fields for material parameter uncertainty, set $\xi_m(\omega)$ in Scenario 1.}
	\label{fig:uq_1_1}
\end{figure}

\begin{figure}[t]
	\centering
	\subfloat[Expectation.]{\includegraphics[width=0.45\textwidth]{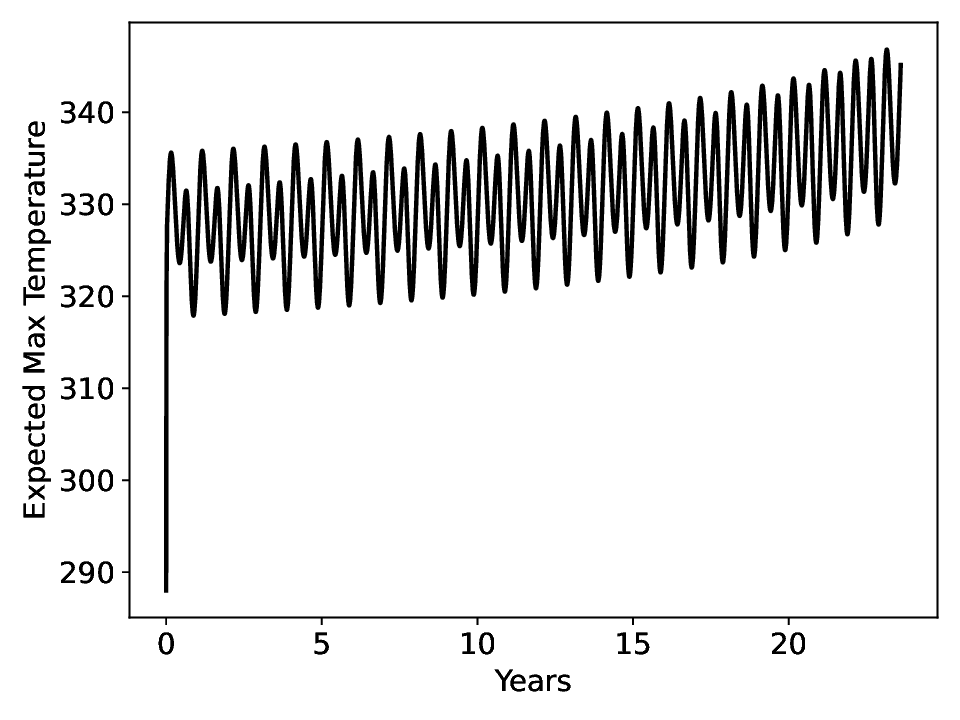}}
	\subfloat[Standard deviation.]{\includegraphics[width=0.45\textwidth]{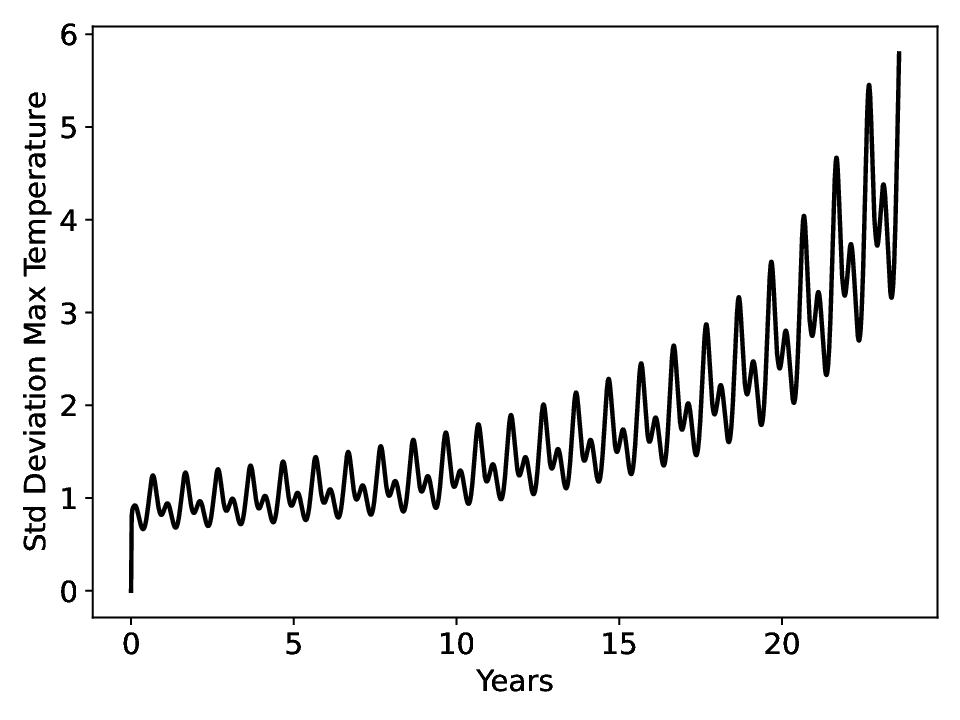}}
	\caption{Time-series of expectation and standard deviation of maximum temperature for material parameter uncertainty, set $\xi_m(\omega)$ in Scenario 1.}
	\label{fig:uq_max_1_1}
\end{figure}

\begin{figure}[t]
	\centering
	\subfloat[Expectation.]{\includegraphics[width=0.45\textwidth]{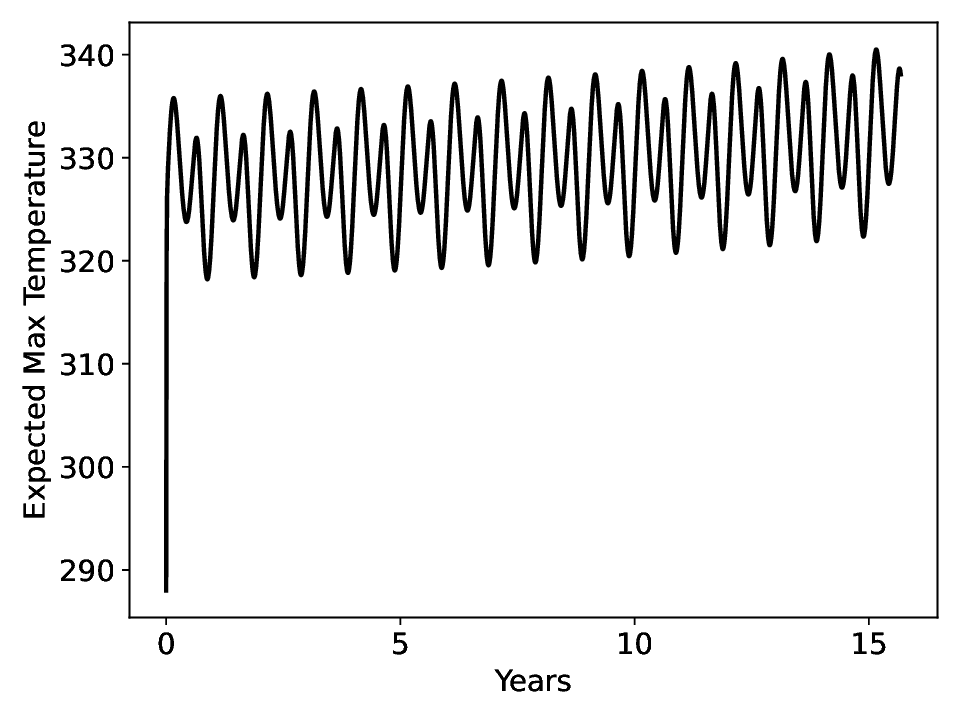}}
	\subfloat[Standard deviation.]{\includegraphics[width=0.45\textwidth]{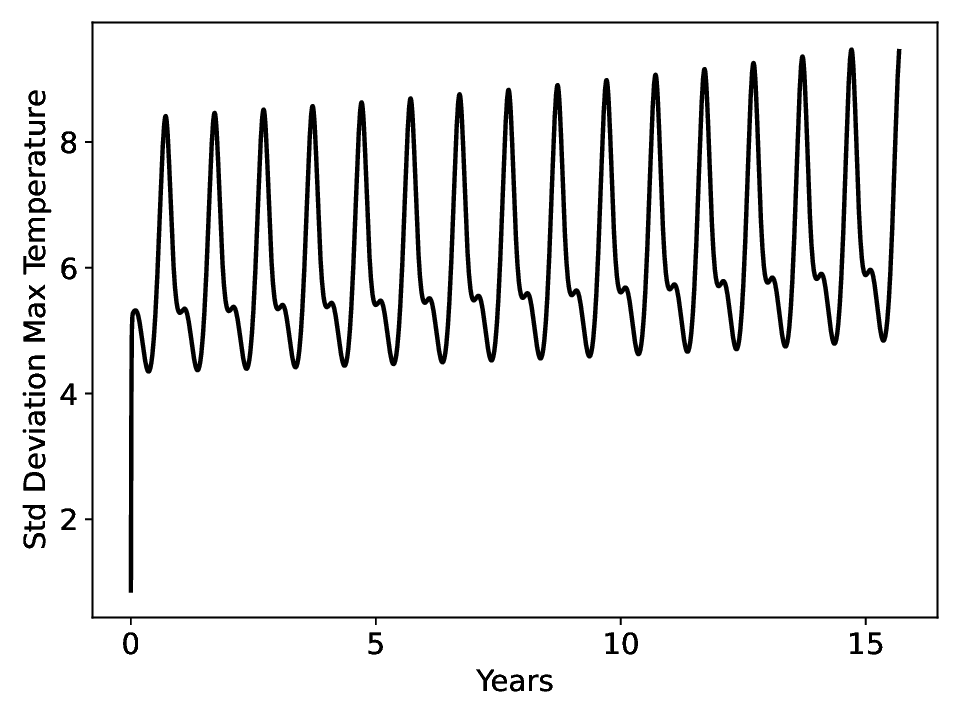}}
	\caption{Time-series of expectation and standard deviation of maximum temperature for external load uncertainty, set $\xi_l(\omega)$ in Scenario 1.}
	\label{fig:uq_max_1_2}
\end{figure}

We start with the material parameter uncertainty from set $\xi_m(\omega)$, and plot the evolution of expected temperature field and its standard deviation over time in 5 year increments, and we plot the results in Fig~\ref{fig:uq_1_1}. We can confirm that indeed the maximum temperature and maximum standard deviation will remain at the center as in the deterministic case. We then check the time-series evolution of the maximum temperature, obtained from the midpoint of the transmission line in Fig.~\ref{fig:uq_max_1_1}, and we note that the standard deviation grows with time. We plot the time-series evolution of maximum temperature for set  $\xi_l(\omega)$ in Fig.~\ref{fig:uq_max_1_2}, and we observe more steady increase in the standard deviation, yet with higher amplitudes.

\begin{figure}[t]
	\centering
	\subfloat[Material parameters.]{\includegraphics[width=0.45\textwidth]{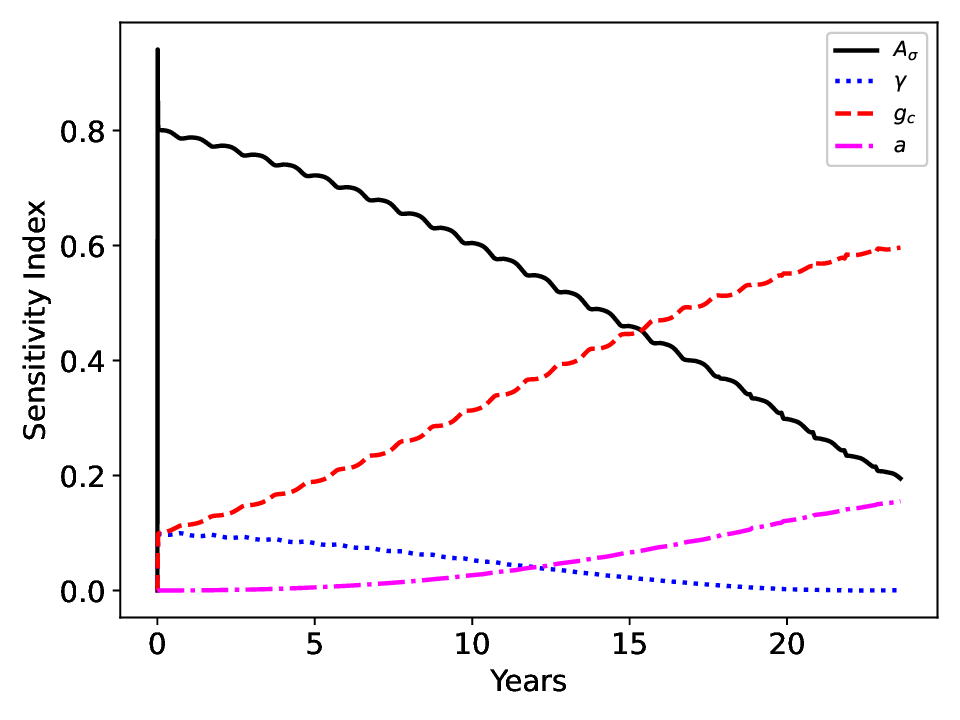}}
	\subfloat[External loading.]{\includegraphics[width=0.45\textwidth]{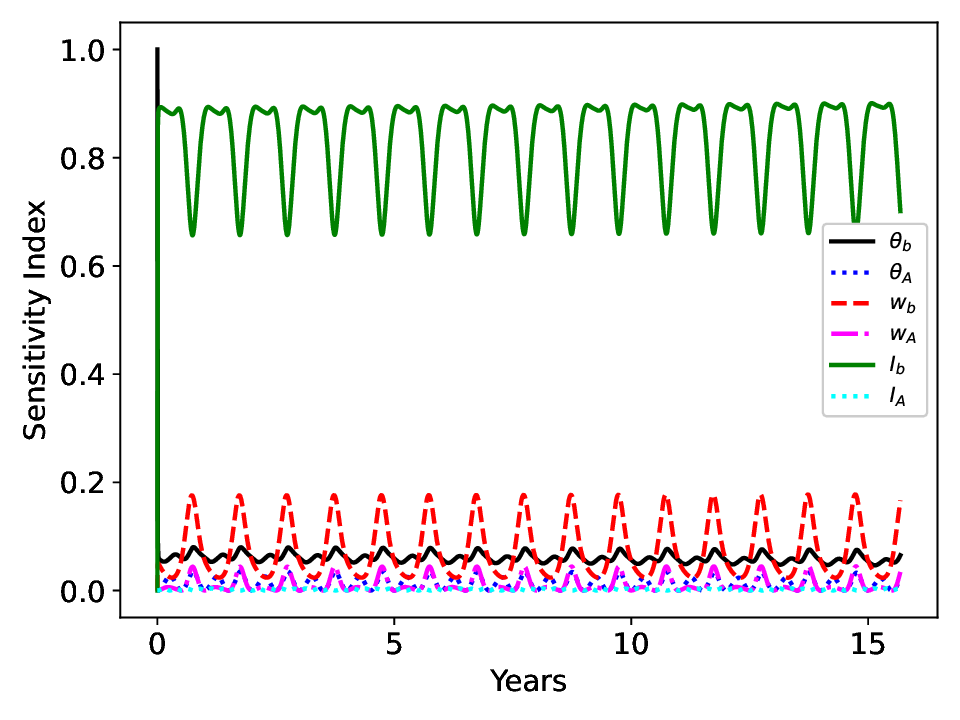}}
	\caption{First-order sensitivity index $S_i$ for Scenario 1 for parameter sets $\xi_m(\omega)$ and $\xi_l(\omega)$.}
	\label{fig:si_1_12}
\end{figure}

From the realizations obtained at the PCM collocation points, we compute the Sobol indices $S_i$ through Eq.~(\ref{eq:si}), and plot the time-series evolution of all parameters from both sets $\xi_m(\omega)$ and $\xi_l(\omega)$ in Fig~\ref{fig:si_1_12}. Within the material parameters, the cross-section area parameter that drives the damage localization is mostly important at the start of the simulation, where it initiates damage, but then becomes unimportant. We see that $g_c$ and $a$ become important in the long-run, as they are associated to the total energetic threshold  for fracture, and the rate of fatigue accumulation, respectively. Among the loading parameters, current base parameter $I_b$ is the most important in the total uncertainty of $\theta_{max}$, as it directly correlates with Joule heating. The competition between Joule heating and convective cooling makes the wind base parameter $w_b$ the second important

\begin{figure}[t]
	\subfloat[Expectation.]{\includegraphics[width=0.33\textwidth]{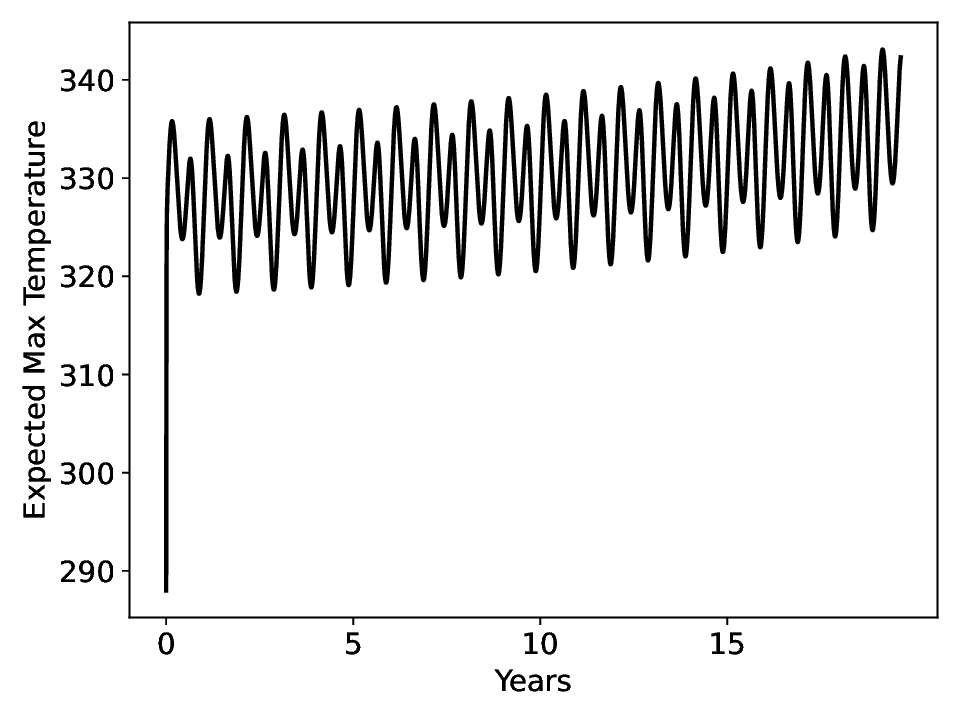}}
	\subfloat[Standard deviation.]{\includegraphics[width=0.33\textwidth]{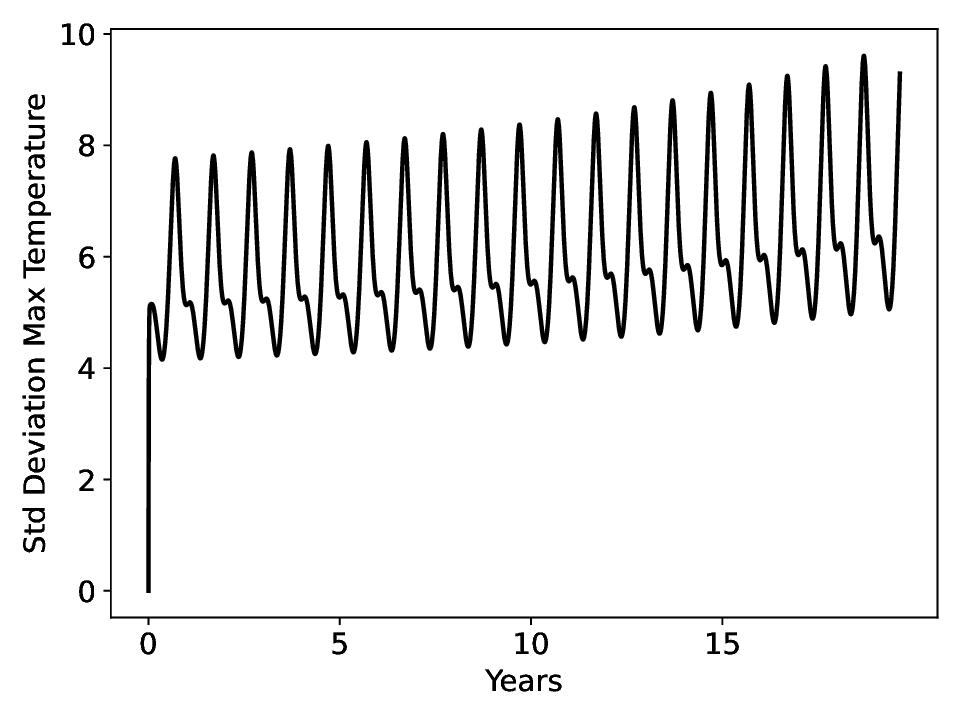}}
	\subfloat[Sensitivity index.]{\includegraphics[width=0.33\textwidth]{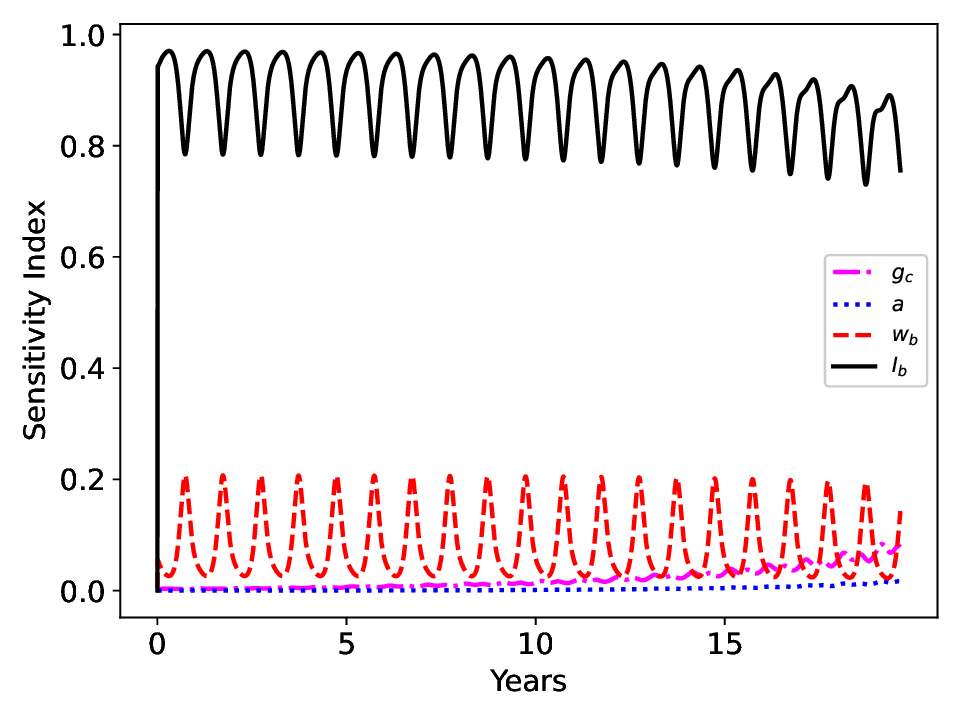}}
	\caption{Time-series of expectation, standard deviation, and sensitivity index of maximum temperature under combined parametric space, set $\xi_1(\omega)$ in Scenario 1.}
	\label{fig:si_1_3}
\end{figure}

We combine the effects of most influential material and loading condition parameters and form the new set of random parameters $\xi_1 (\omega) = \{g_c(\omega), a(\omega),w_b(\omega),I_b(\omega)\}$, and perform a final round of SA for Scenario 1. We plot the results in Fig.~\ref{fig:si_1_3} and observe that loading conditions are more significant than material parameters from the beginning, however the relative influence of material parameters grows with time as aging effects take place.

\subsubsection{Convergence analysis}

Such results present an interesting prospective for the analysis of the remaining scenarios. Before we include the specific scenario parameters in the set of random variables, we check the consistency of preliminary PCM results by a convergence analysis. Due to the coupled, nonlinear nature of the deterministic problem, an analytical solution cannot be found, so we rely on a refined PCM solution as reference.

For the convergence study, we focus solely on the effects of the most influential parameter $I_b$, as seen from the preliminary analysis. This turns the PCM into a 1-D problem, allowing the computation of accurate integration using 100 collocation points to be a reference solution. We compare lower-order PCM solutions with Monte Carlo simulations, and plot the relative error of the norm of temperature field solution (expectation and standard deviation) with respect to the reference PCM at a chosen time as

\begin{equation}
\epsilon = \frac{\Vert \theta -\theta_{ref}\Vert_2 }{\Vert \theta_{ref} \Vert_2}.
\end{equation}

\begin{figure}[t]
	\centering
	\subfloat[PCM.]{\includegraphics[width=0.25\textwidth]{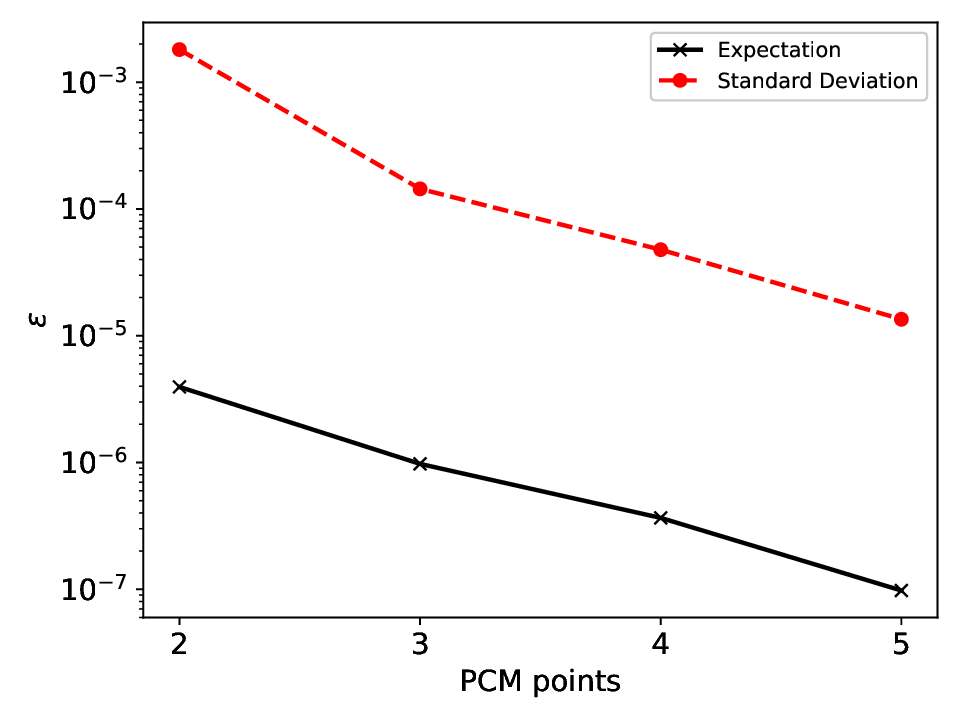}}
	\subfloat[MC.]{\includegraphics[width=0.25\textwidth]{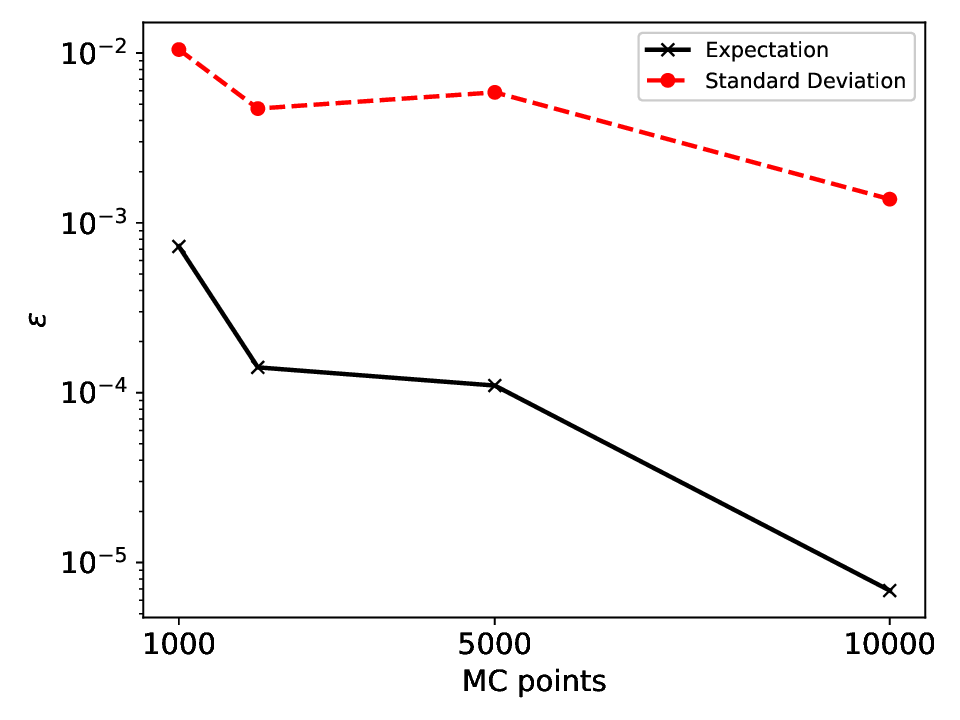}}
	\subfloat[PCM.]{\includegraphics[width=0.25\textwidth]{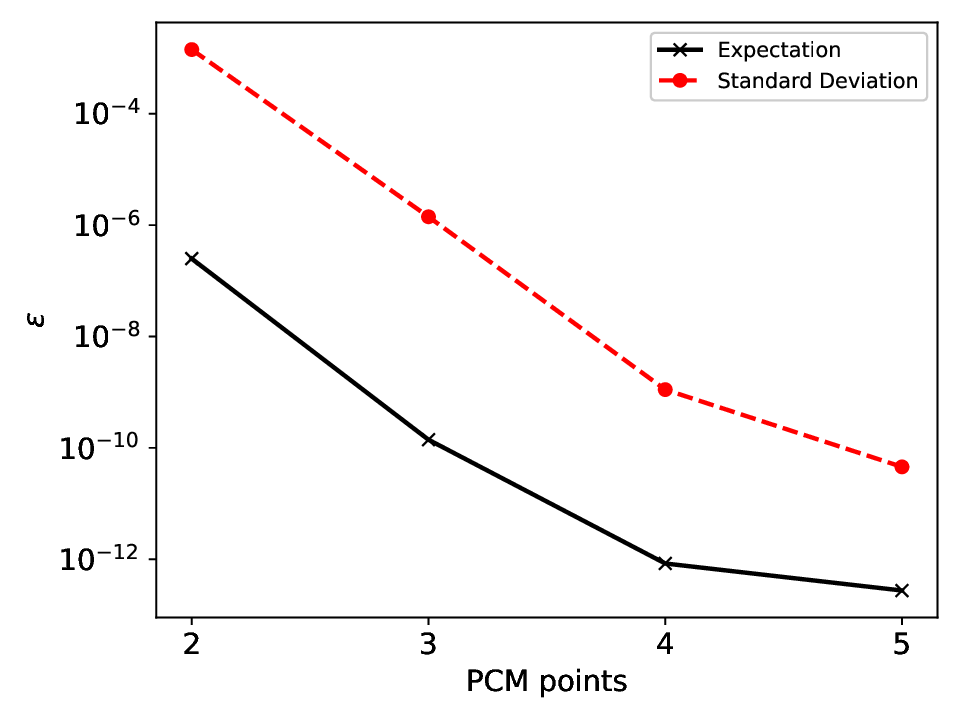}}
	\subfloat[MC.]{\includegraphics[width=0.25\textwidth]{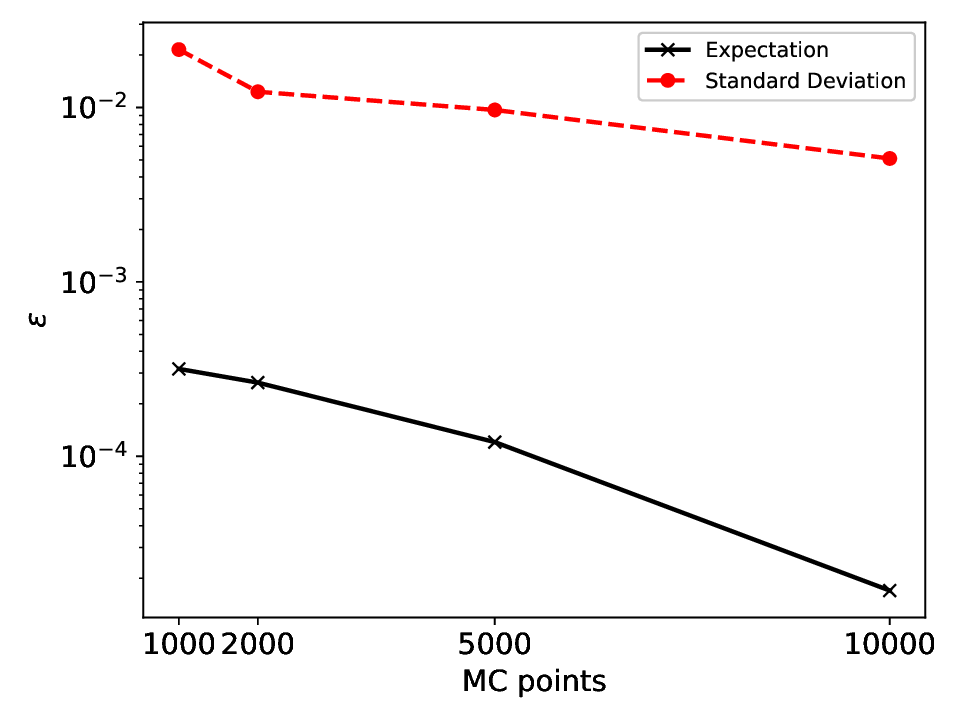}}
	\caption{Relative errors in temperature at 40 years and 10 years for PCM and MC compared to a reference 100-point PCM solution.}
	\label{fig:convergence}
\end{figure}

We compute the all relative errors at year 25 and 10 years. The original time step was reduced by 10 times for 25 years simulation run. The plots of the results are in Fig~\ref{fig:convergence}. We see that PCM achieves a much lower error, around 2 orders of magnitude, than 10000 MC realizations using only 5 points. This feature becomes even more important when we go to higher dimensions, since MC would need number of realizations orders of magnitude higher to properly represent the high-dimensional parametric space. We have sufficient accuracy in PCM with 5 points, and we choose $n=5$ for all subsequent simulations in either 4-D or 5-D.

\subsubsection{Scenarios 2, 3, and 4}

In Scenarios 2, 3, and 4, we combine the 4 influential parameters from Scenario 1, set $\xi_1(\omega)$, with the corresponding parameter for each specific scenario, resulting in sets $\xi_2 (\omega) = \{g_c(\omega), a(\omega),w_b(\omega),I_b(\omega),w_{max}\}$, $\xi_3 (\omega) = \{g_c(\omega), a(\omega),w_b(\omega),I_b(\omega),I_r\}$, and $\xi_4 (\omega) = \{g_c(\omega), a(\omega),w_b(\omega),I_b(\omega),\theta_r\}$, respectively.

\begin{figure}[t]
	\subfloat[Expectation.]{\includegraphics[width=0.33\textwidth]{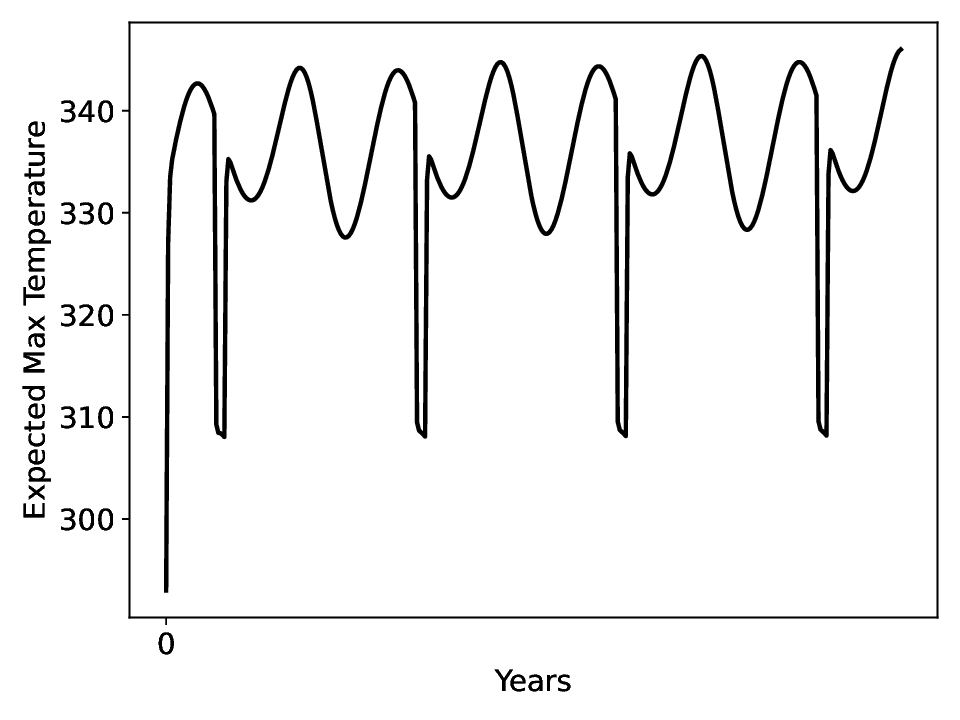}}
	\subfloat[Standard deviation.]{\includegraphics[width=0.33\textwidth]{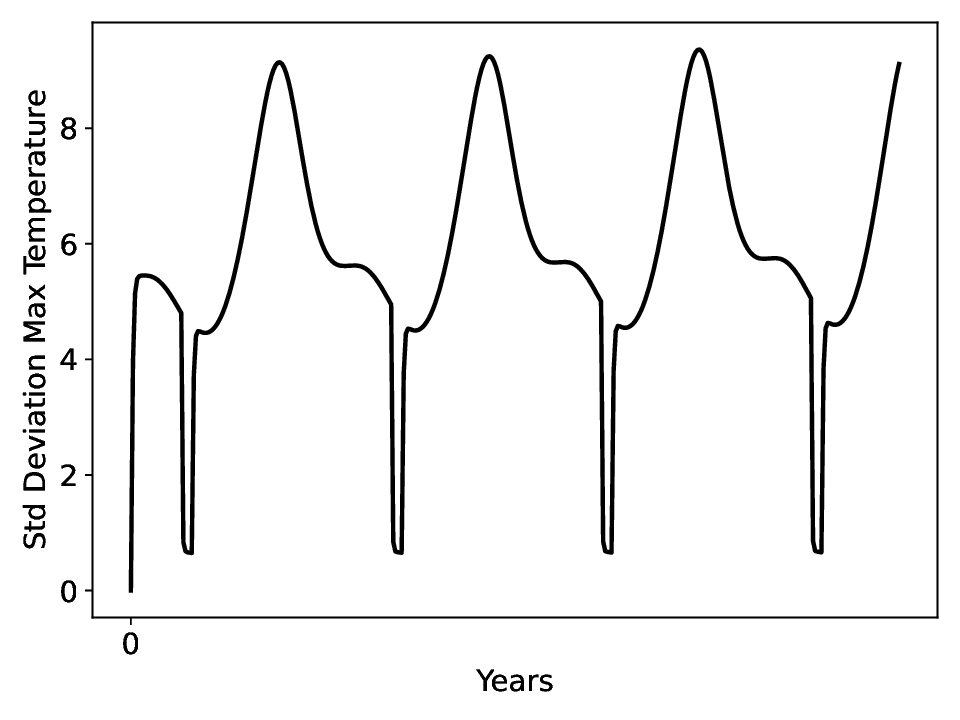}}
	\subfloat[Sensitivity index.]{\includegraphics[width=0.33\textwidth]{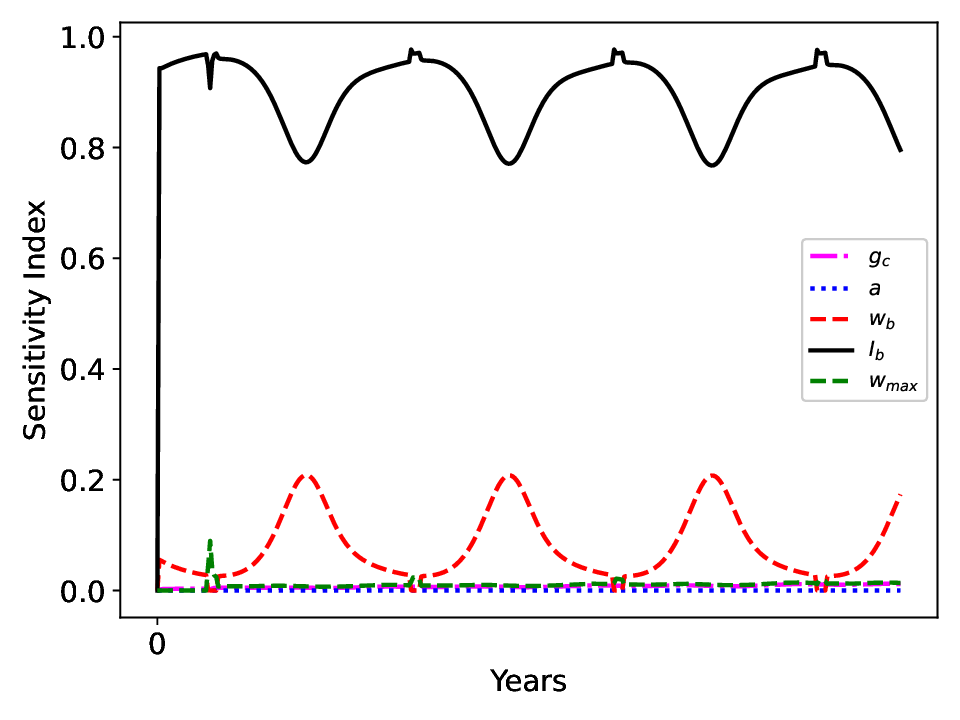}}
	\caption{Time-series of expectation, standard deviation, and sensitivity index of maximum temperature under combined parametric space, set $\xi_2(\omega)$ in Scenario 2.}
	\label{fig:si_2}
\end{figure}

In Scenario 2 the loading parameters are even more significant, since the high winds drive damage faster and therefore do not leave time for aging to take place, as seen in Fig.~\ref{fig:si_2}. The scenario parameter of high wind speed with mean value $w_{max} = 100\ ft/s$ has great importance when it initially hits the line, greatly increasing  damage. Afterwards, it does not have an impact as great as the first hit. Although the extra wind cools the conductor significantly, the expected failure happens before 5 years.

\begin{figure}[t]
	\subfloat[Expectation.]{\includegraphics[width=0.33\textwidth]{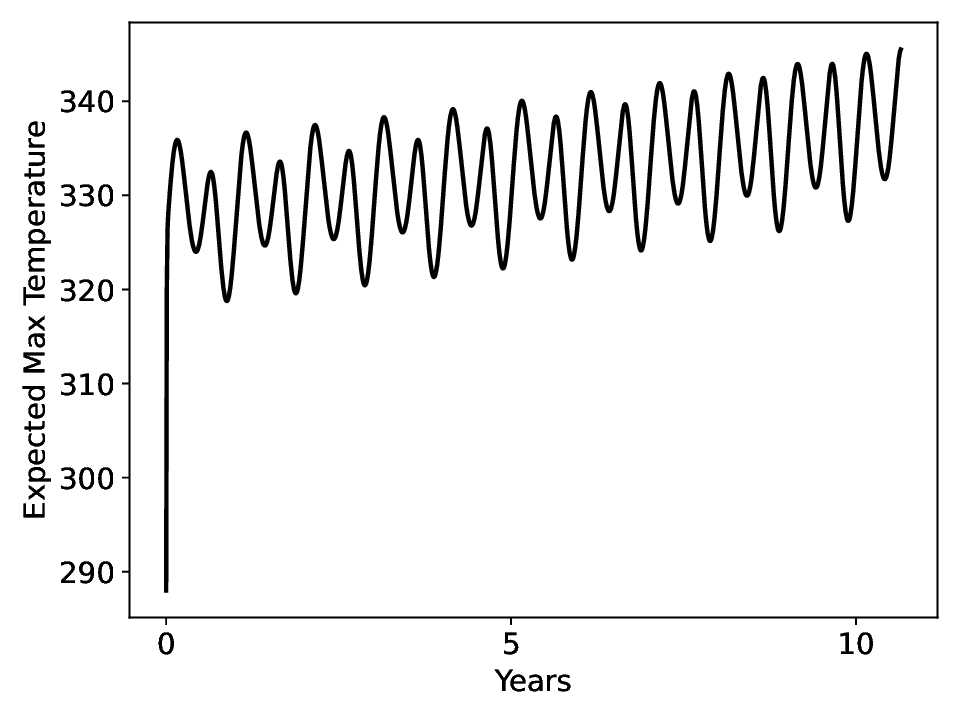}}
	\subfloat[Standard deviation.]{\includegraphics[width=0.33\textwidth]{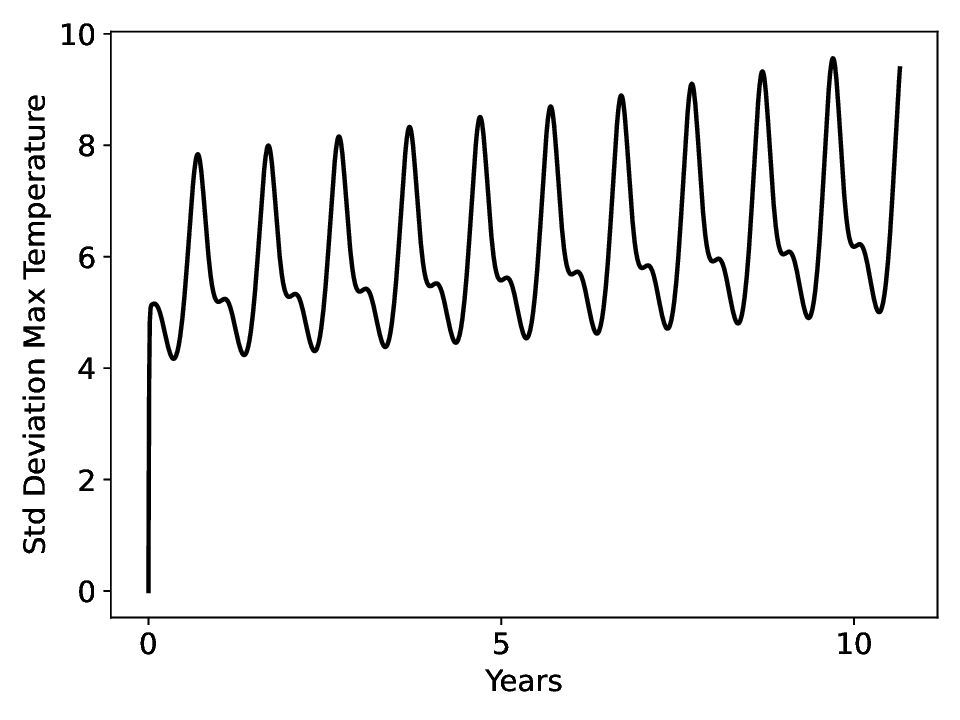}}
	\subfloat[Sensitivity index.]{\includegraphics[width=0.33\textwidth]{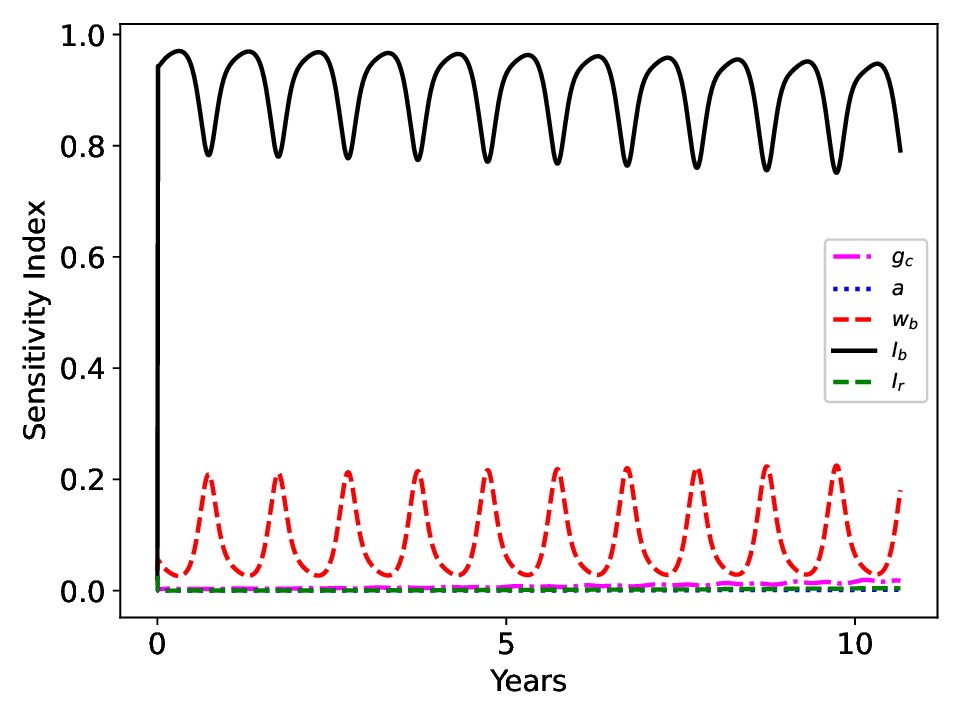}}
	\caption{Time-series of expectation, standard deviation, and sensitivity index of maximum temperature under combined parametric space, set $\xi_3(\omega)$ in Scenario 3.}
	\label{fig:si_3}
\end{figure}

\begin{figure}[t]
	\subfloat[Expectation.]{\includegraphics[width=0.33\textwidth]{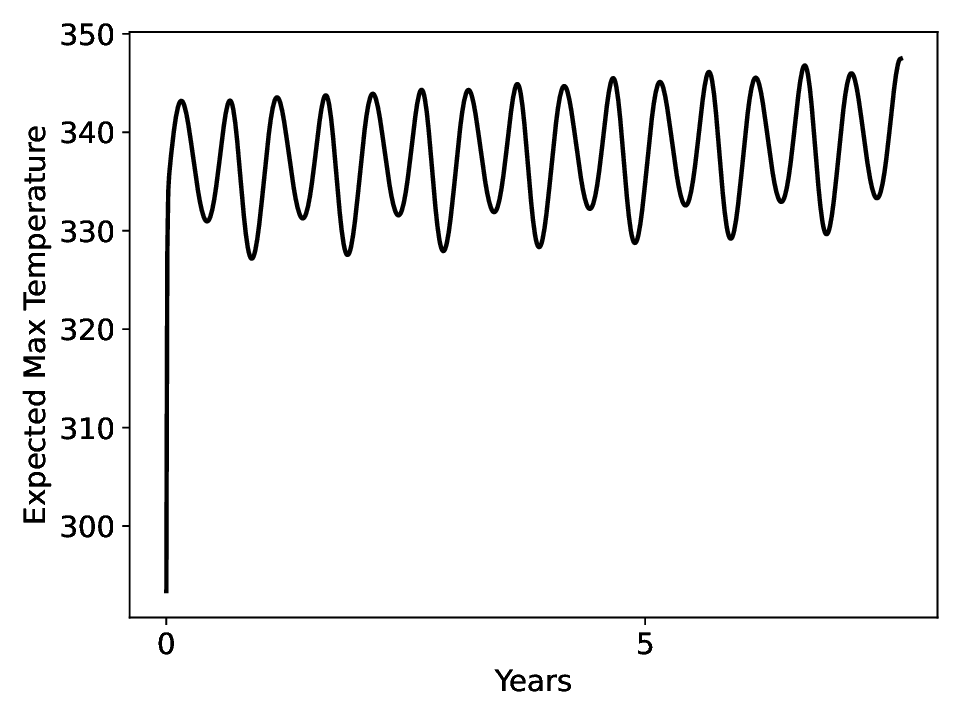}}
	\subfloat[Standard deviation.]{\includegraphics[width=0.33\textwidth]{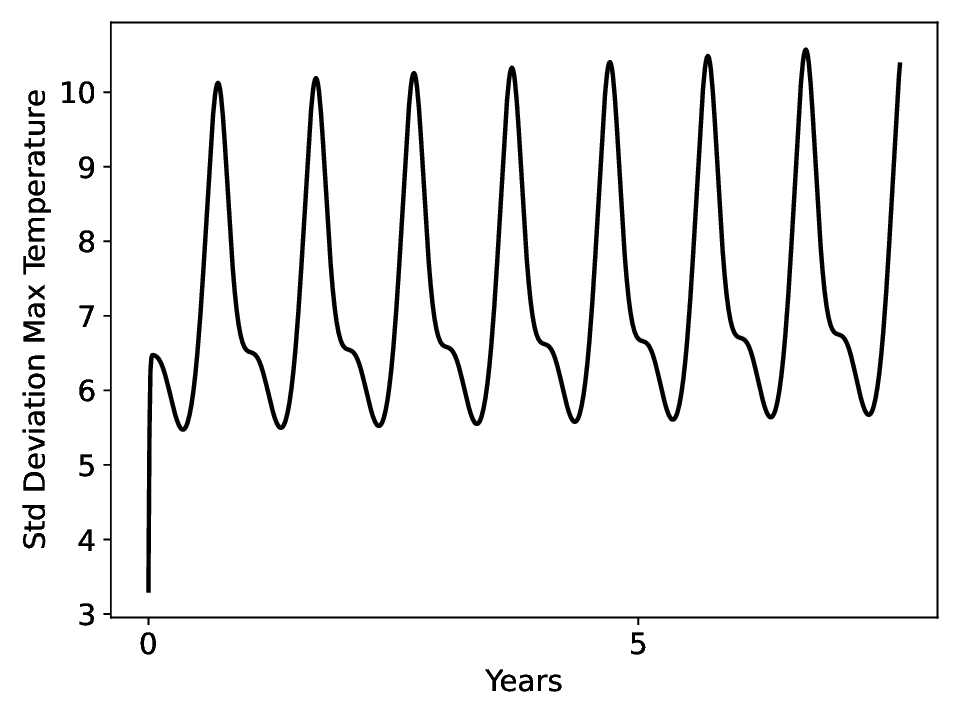}}
	\subfloat[Sensitivity index.]{\includegraphics[width=0.33\textwidth]{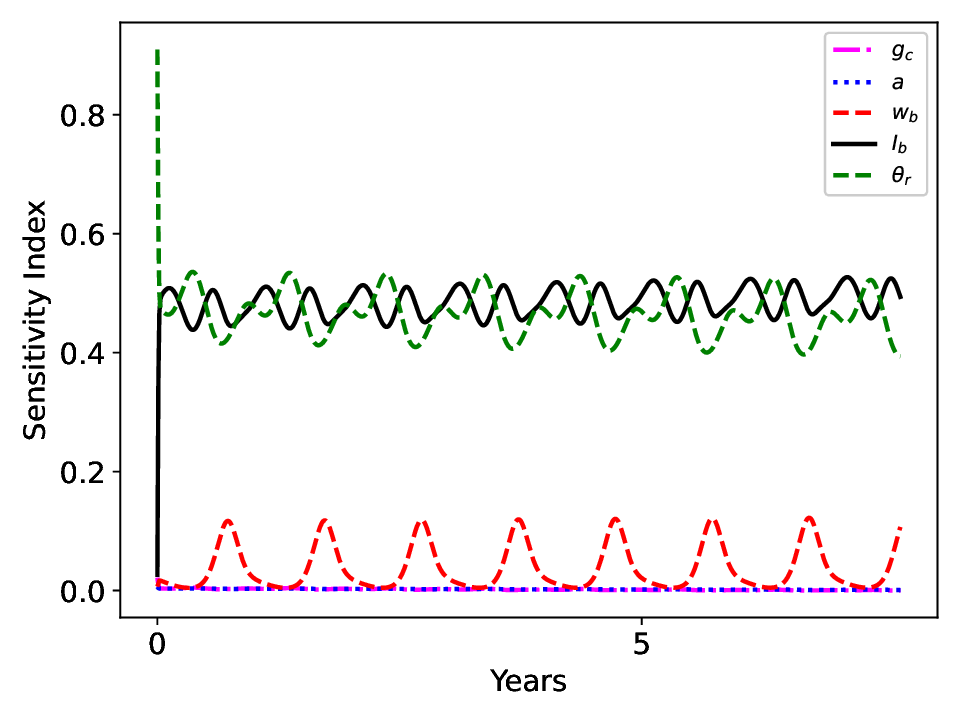}}
	\caption{Time-series of expectation, standard deviation, and sensitivity index of maximum temperature under combined parametric space, set $\xi_4(\omega)$ in Scenario 4.}
	\label{fig:si_4}
\end{figure}

Similarly, we observe that in Scenario 3, the mean value of $I_r = 0.1\ A$ also leads to more sensitivity in the loading conditions, specially the $I_b$ parameter, as seen in Fig.~\ref{fig:si_3}. Scenario 4, however, shows a surprising importance to the parameter that controls the rate of temperature increase, taken with mean value $\theta_r = 0.001\ K$, Fig.~\ref{fig:si_4}. This result does not indicate whether the temperature increase itself leads to premature failure, instead it conveys the message that uncertainties in the prediction of $\theta_r$ significantly affects the uncertainties in predicting transmission line failure.

\subsubsection{Probability of failure}

\begin{figure}[t]
	\centering
	\includegraphics[width=0.45\textwidth]{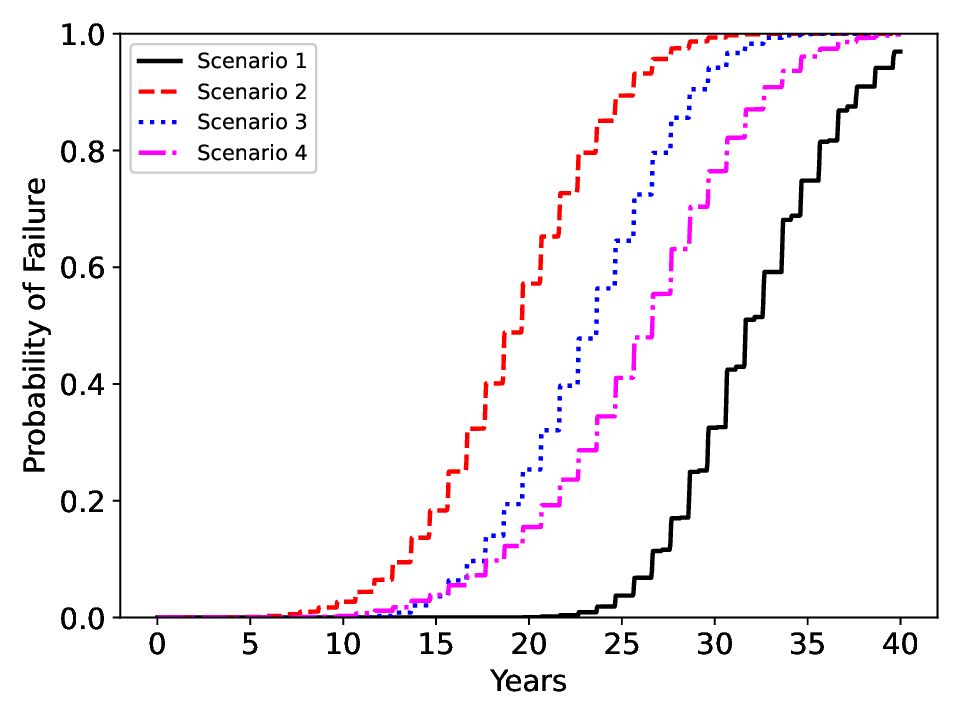}
	\caption{Time-series evolution of probability of failure $p_f$ for each scenario using reference mean parameter values.}
	\label{fig:pf_all}
\end{figure}

In this last part we compute the expectation of Bernoulli variable $h$ as defined in Eq.~(\ref{eq:h}), and using the same PCM building block with $n=5$, through the multi-dimensional expectation Eq.~(\ref{eq:pcm_multi}). We plot the results for reference mean parameter values from sets $\xi_1(\omega)$,  $\xi_2(\omega)$,  $\xi_3(\omega)$, and $\xi_4(\omega)$ in Fig~\ref{fig:pf_all}. We see that normal operating conditions have the curve shifted to the right, while the more intense high wind scenario have the left-most failure curve. The average temperature increase has an interesting effect of increasing the probability of failure early on, followed by a rather smoother increase.

\begin{figure}[t]
	\centering
	\subfloat[Scenario 1.]{\includegraphics[width=0.25\textwidth]{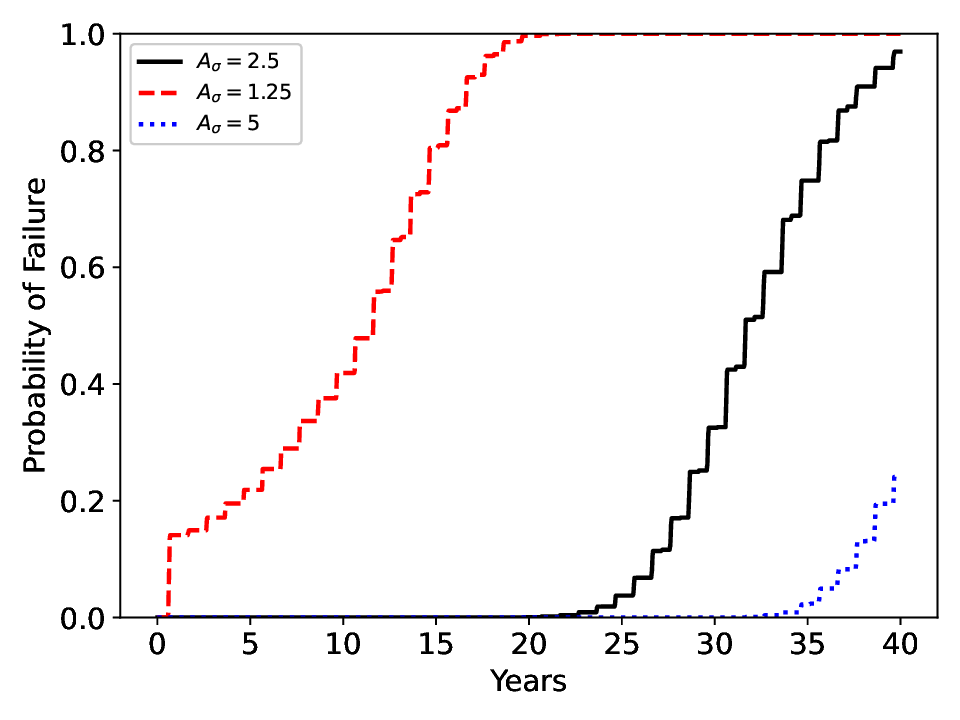}}
	\subfloat[Scenario 2.]{\includegraphics[width=0.25\textwidth]{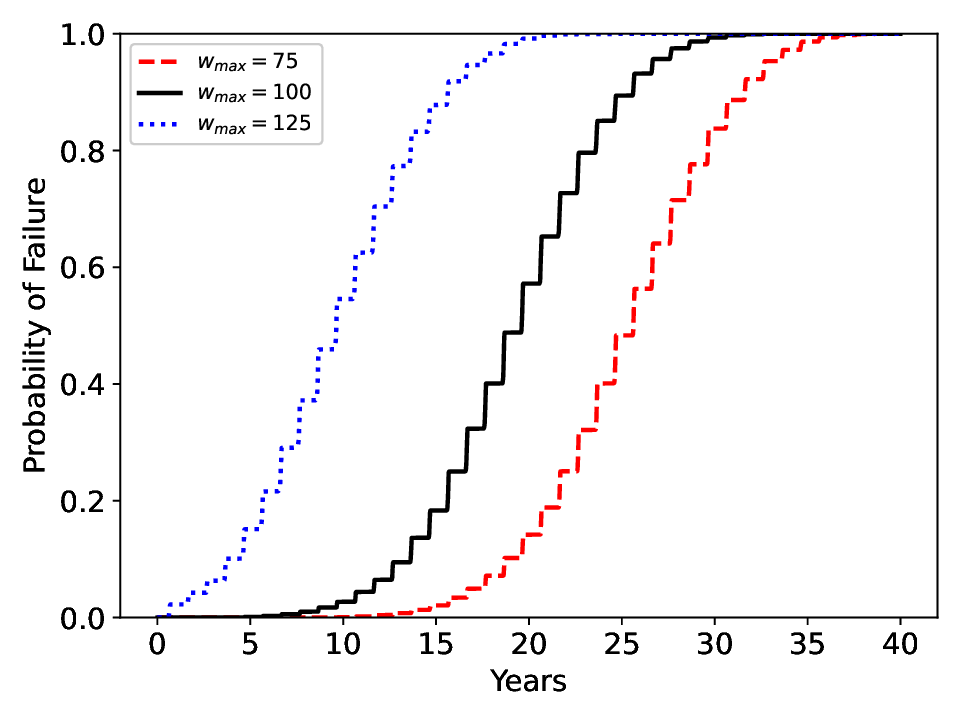}}
	\subfloat[Scenario 3.]{\includegraphics[width=0.25\textwidth]{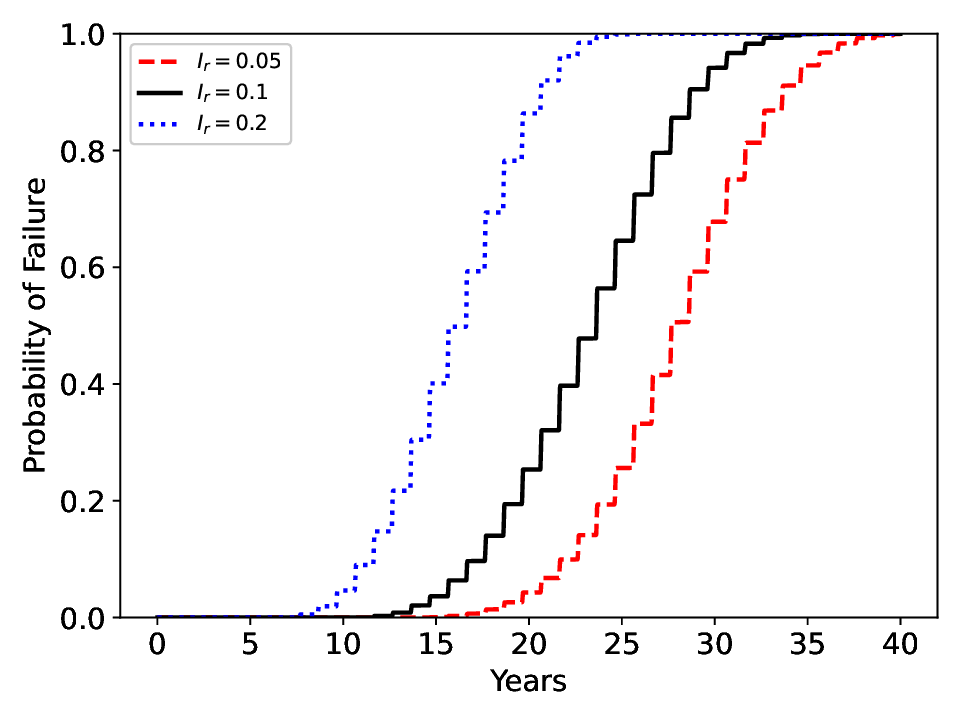}}
	\subfloat[Scenario 4.]{\includegraphics[width=0.25\textwidth]{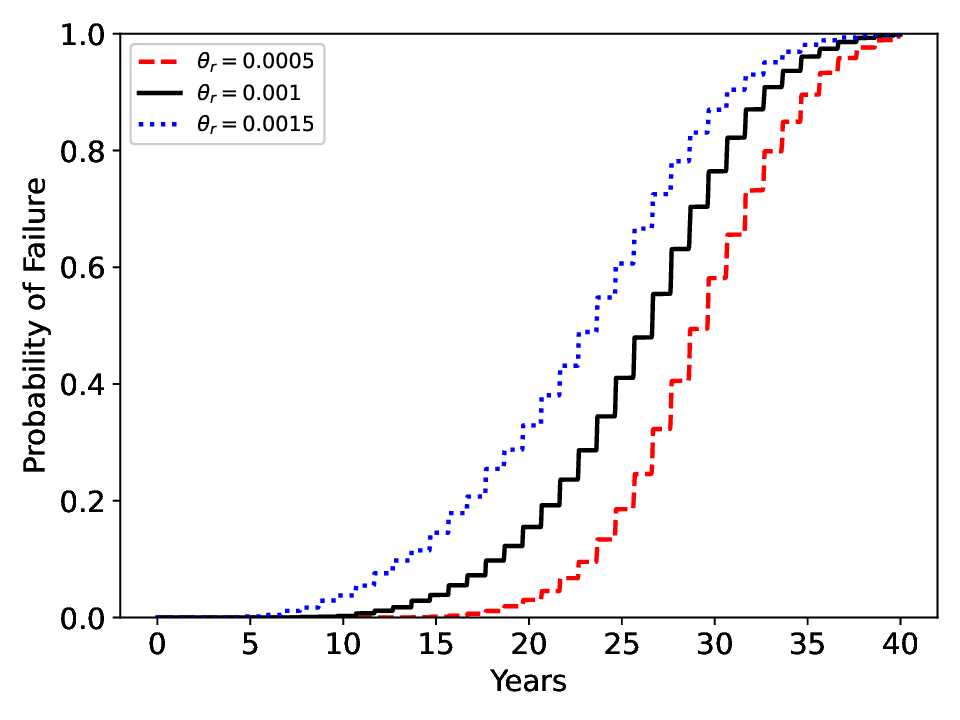}}
	\caption{Comparison of probability of failure time-series under different mean parameters levels for each scenario.}
	\label{fig:pf_comparison}
\end{figure}

Finally, we compare the shape profiles of time-series $p_f$ on each Scenario by changing the mean value of specific parameters.In Scenario 1 we compare normal operation conditions for 3 levels of mean $A_\sigma$ and their respective uncertainty bounds of $\pm 10\%$. Similarly, we compare $p_f$ of Scenarios 2, 3, and 4 under different mean scenario parameters, namely $w_{max}$, $I_r$, and $\theta_r$, respectively,and see how they shift. Unlike in previous SA of material parameters, such as $g_c$ and $a$, or loads (which provide levels of importance with respect to their uncertainty), here we observe the effect of changing the base level of scenario parameters and initial damage and how they affect the $p_f$ curve. We plot the results in Fig.~\ref{fig:pf_comparison}, and 

We observe distinct curve shapes in each scenario. For Scenario 1, the early failure of a sharper initial damage is represented by a left shift of $p_f$ curve, while a smoother damage profile only leads to a $20\%$ chance of failure after 40 years. We observe similar effects for the parameters in Scenarios 2 and 3. For Scenario 4, a slight variation in the rate of base temperature increase is sufficient to increase the probability of failure of this particular line by $10\%$ to $20\%$ in 15 years. Although it is not a high value individually, as opposed to region-specific failure prospectives from Scenarios 2 and 3, increase in average temperatures is a global outcome, and the combined effect of many transmission lines with failure probability as low as $10\%$ have compounding effects and could lead to catastrophic power grid failure.

\section{Conclusions}
\label{sec:conclusions}

We developed a thermo-electro-mechanical model for failure of transmission lines. The mechanical model considers material damage evolution with long-term fatigue-driven aging mechanisms. Heat transfer equation includes Joule heating from an electric current, and convective cooling from the ambient, such that high temperatures accelerate aging and damage, and influences the cable tension, which in turn compromises the conductivity of the material. With this construction, an initially damaged transmission line conductor enters a positive feedback-loop that leads to thermal runaway and failure when it reaches the material's annealing temperature.

We solved the quasi-static set of equations by Finite-Element Method and proposed different scenarios for the long-term behavior of the damaged system. We studied each scenario deterministically to observe the evolution of maximum conductor temperature under different conditions. Then, we employed PCM as a building block for UQ, SA, and probability of failure studies that shed light on critical aspects of the system's operation:

\begin{itemize}
	\item Quasi-static solutions of the thermo-electro-mechanical damage model are capable of simulating the long-term behavior of temperature cycles in the conductor, subject to time-changing conditions in wind, air temperature, and average electric current. We observe that slight changes in the initial damage or in the electric current may significantly reduce the life expectancy of the material.
	\item PCM is an accurate and cheap method that overcomes the main difficulties of MC methods. The tensorial product construction makes the computation of global sensitivity index efficient. In the global SA we observed the significant influence of electric current and rate of change of air temperature in the total variance of the solution.
	\item We further exploited the capabilities of PCM and redefined the limit state function to be a Bernoulli random variable, and we compute its expectation from PCM directly as a measure of the probability of failure. 
\end{itemize}

We acknowledge that the current model relies on simplifying assumptions that could be further explored in future projects. To provide a couple of examples, the assumption of quasi-static equations could be lifted in place of a transient analysis. Also, the simplification of the cable mechanics to one dimension could potentially be expanded to include the solution of the catenary shape as well. These and other assumptions could lead to more complex system of equations better suited for each particular case, yet the overall framework and stochastic analyzes structure would remain the same.

Nevertheless, we emphasize the potential of the proposed framework as a high-fidelity descriptor of probabilistic failure based on physical principles, rather than pure mathematical modeling. Inclusion of real sensory data of wind speed, air temperature, and electric current for specific regions are straight-forward, and could lead to more predictive estimation of failure probability. Through this construction, the reliability of transmission lines depends less on stochastic process modeling, and instead benefits from detailed physics models of aging conductors.

\bibliographystyle{siamplain}
\bibliography{reference}

\begin{thebibliography}{10}

\bibitem{abdi2010principal}
{\sc H.~Abdi and L.~J. Williams}, {\em Principal component analysis}, Wiley
  interdisciplinary reviews: computational statistics, 2 (2010), pp.~433--459.

\bibitem{albert2004structural}
{\sc R.~Albert, I.~Albert, and G.~L. Nakarado}, {\em Structural vulnerability
  of the north american power grid}, Physical review E, 69 (2004), p.~025103.

\bibitem{ambati_phase-field_2015}
{\sc M.~Ambati, T.~Gerasimov, and L.~De~Lorenzis}, {\em Phase-field modeling of
  ductile fracture}, Computational Mechanics, 55 (2015), pp.~1017--1040,
  \url{https://doi.org/10.1007/s00466-015-1151-4}.

\bibitem{ambati_phase-field_2016}
{\sc M.~Ambati, R.~Kruse, and L.~De~Lorenzis}, {\em A phase-field model for
  ductile fracture at finite strains and its experimental verification},
  Computational Mechanics, 57 (2016), pp.~149--167.

\bibitem{babuvska2007stochastic}
{\sc I.~Babu{\v{s}}ka, F.~Nobile, and R.~Tempone}, {\em A stochastic
  collocation method for elliptic partial differential equations with random
  input data}, SIAM Journal on Numerical Analysis, 45 (2007), pp.~1005--1034.

\bibitem{babuska2004galerkin}
{\sc I.~Babuska, R.~Tempone, and G.~E. Zouraris}, {\em Galerkin finite element
  approximations of stochastic elliptic partial differential equations}, SIAM
  Journal on Numerical Analysis, 42 (2004), pp.~800--825.

\bibitem{babuvska2005solving}
{\sc I.~Babu{\v{s}}ka, R.~Tempone, and G.~E. Zouraris}, {\em Solving elliptic
  boundary value problems with uncertain coefficients by the finite element
  method: the stochastic formulation}, Computer methods in applied mechanics
  and engineering, 194 (2005), pp.~1251--1294.

\bibitem{barros2021integrated}
{\sc E.~A. Barros~de Moraes, M.~Zayernouri, and M.~M. Meerschaert}, {\em An
  integrated sensitivity-uncertainty quantification framework for stochastic
  phase-field modeling of material damage}, International Journal for Numerical
  Methods in Engineering, 122 (2021), pp.~1352--1377.

\bibitem{benabou2015continuum}
{\sc L.~Benabou, Z.~Sun, P.~Pougnet, and P.-R. Dahoo}, {\em Continuum damage
  approach for fatigue life prediction of viscoplastic solder joints}, Journal
  of Mechanics, 31 (2015), pp.~525--531.

\bibitem{bockarjova2007transmission}
{\sc M.~Bockarjova and G.~Andersson}, {\em Transmission line conductor
  temperature impact on state estimation accuracy}, in 2007 IEEE Lausanne power
  tech, IEEE, 2007, pp.~701--706.

\bibitem{boldrini_non-isothermal_2016}
{\sc J.~Boldrini, E.~{Barros de Moraes}, L.~Chiarelli, F.~Fumes, and
  M.~Bittencourt}, {\em A non-isothermal thermodynamically consistent phase
  field framework for structural damage and fatigue}, Computer Methods in
  Applied Mechanics and Engineering, 312 (2016), pp.~395--427.

\bibitem{borden_higher-order_2014}
{\sc M.~J. Borden, T.~J. Hughes, C.~M. Landis, and C.~V. Verhoosel}, {\em A
  higher-order phase-field model for brittle fracture: {Formulation} and
  analysis within the isogeometric analysis framework}, Computer Methods in
  Applied Mechanics and Engineering, 273 (2014), pp.~100--118.

\bibitem{borden_phase-field_2012}
{\sc M.~J. Borden, C.~V. Verhoosel, M.~A. Scott, T.~J. Hughes, and C.~M.
  Landis}, {\em A phase-field description of dynamic brittle fracture},
  Computer Methods in Applied Mechanics and Engineering, 217-220 (2012),
  pp.~77--95.

\bibitem{chevreuil2015least}
{\sc M.~Chevreuil, R.~Lebrun, A.~Nouy, and P.~Rai}, {\em A least-squares method
  for sparse low rank approximation of multivariate functions}, SIAM/ASA
  Journal on Uncertainty Quantification, 3 (2015), pp.~897--921.

\bibitem{chhetri2023comparative}
{\sc S.~Chhetri, E.~de~Moraes, M.~Naghibolhosseini, and M.~Zayernouri}, {\em A
  comparative study of dislocation dynamics in ductile and brittle crystalline
  materials}, in 2023 International Conference on Modeling, Simulation \&
  Intelligent Computing (MoSICom), IEEE, 2023, pp.~438--441.

\bibitem{cimini2013temperature}
{\sc C.~A. Cimini~Jr and B.~Q.~A. Fonseca}, {\em Temperature profile of
  progressive damaged overhead electrical conductors}, International Journal of
  Electrical Power \& Energy Systems, 49 (2013), pp.~280--286.

\bibitem{constantine2017global}
{\sc P.~G. Constantine and P.~Diaz}, {\em Global sensitivity metrics from
  active subspaces}, Reliability Engineering \& System Safety, 162 (2017),
  pp.~1--13.

\bibitem{constantine2014active}
{\sc P.~G. Constantine, E.~Dow, and Q.~Wang}, {\em Active subspace methods in
  theory and practice: applications to kriging surfaces}, SIAM Journal on
  Scientific Computing, 36 (2014), pp.~A1500--A1524.

\bibitem{constantine2015exploiting}
{\sc P.~G. Constantine, M.~Emory, J.~Larsson, and G.~Iaccarino}, {\em
  Exploiting active subspaces to quantify uncertainty in the numerical
  simulation of the hyshot ii scramjet}, Journal of Computational Physics, 302
  (2015), pp.~1--20.

\bibitem{crucitti2004model}
{\sc P.~Crucitti, V.~Latora, and M.~Marchiori}, {\em Model for cascading
  failures in complex networks}, Physical Review E, 69 (2004), p.~045104.

\bibitem{dang2019novel}
{\sc C.~Dang and J.~Xu}, {\em Novel algorithm for reconstruction of a
  distribution by fitting its first-four statistical moments}, Applied
  Mathematical Modelling, 71 (2019), pp.~505--524.

\bibitem{de2023machine}
{\sc E.~A.~B. de~Moraes, M.~D’Elia, and M.~Zayernouri}, {\em Machine learning
  of nonlocal micro-structural defect evolutions in crystalline materials},
  Computer Methods in Applied Mechanics and Engineering, 403 (2023), p.~115743.

\bibitem{de2021data}
{\sc E.~A.~B. de~Moraes, H.~Salehi, and M.~Zayernouri}, {\em Data-driven
  failure prediction in brittle materials: A phase field-based machine learning
  framework}, Journal of Machine Learning for Modeling and Computing, 2 (2021).

\bibitem{de2021atomistic}
{\sc E.~A.~B. de~Moraes, J.~L. Suzuki, and M.~Zayernouri}, {\em
  Atomistic-to-meso multi-scale data-driven graph surrogate modeling of
  dislocation glide}, Computational Materials Science, 197 (2021), p.~110569.

\bibitem{fishman2013monte}
{\sc G.~Fishman}, {\em Monte Carlo: concepts, algorithms, and applications},
  Springer Science \& Business Media, 2013.

\bibitem{gao2018potential}
{\sc X.~Gao, C.~A. Schlosser, and E.~R. Morgan}, {\em Potential impacts of
  climate warming and increased summer heat stress on the electric grid: a case
  study for a large power transformer (lpt) in the northeast united states},
  Climatic change, 147 (2018), pp.~107--118.

\bibitem{garcia2021polinomial}
{\sc N.~A. Garc{\'\i}a, G.~G. G{\'o}mez, and C.~N. Serna}, {\em Polinomial
  chaos expansion applied to limit state functions}, in Journal of Physics:
  Conference Series, vol.~1981, IOP Publishing, 2021, p.~012010.

\bibitem{grigsby2006electric}
{\sc L.~L. Grigsby}, {\em Electric power engineering handbook}, CRC Press LLC,
  London, 2006.

\bibitem{guo2018determination}
{\sc Y.~Guo, R.~Chen, J.~Shi, J.~Wan, H.~Yi, and J.~Zhong}, {\em Determination
  of the power transmission line ageing failure probability due to the impact
  of forest fire}, IET Generation, Transmission \& Distribution, 12 (2018),
  pp.~3812--3819.

\bibitem{hathout2018impact}
{\sc I.~Hathout, K.~Callery, J.~Trac, and T.~Hathout}, {\em Impact of thermal
  stresses on the end of life of overhead transmission conductors}, in 2018
  IEEE Power \& Energy Society General Meeting (PESGM), IEEE, 2018, pp.~1--5.

\bibitem{he2014sparse}
{\sc J.~He, S.~Gao, and J.~Gong}, {\em A sparse grid stochastic collocation
  method for structural reliability analysis}, Structural Safety, 51 (2014),
  pp.~29--34.

\bibitem{hockenberry2004evaluation}
{\sc J.~R. Hockenberry and B.~C. Lesieutre}, {\em Evaluation of uncertainty in
  dynamic simulations of power system models: The probabilistic collocation
  method}, IEEE Transactions on Power Systems, 19 (2004), pp.~1483--1491.

\bibitem{hofacker_phase_2013}
{\sc M.~Hofacker and C.~Miehe}, {\em A phase field model of dynamic fracture:
  Robust field updates for the analysis of complex crack patterns:},
  International Journal for Numerical Methods in Engineering, 93 (2013),
  pp.~276--301, \url{https://doi.org/10.1002/nme.4387}.

\bibitem{hou2021damage}
{\sc H.~Hou, Z.~Zhang, S.~Yu, Y.~Huang, Y.~Zhang, and Z.~Dong}, {\em Damage
  prediction of transmission lines under typhoon disasters considering
  multi-effect}, Journal of Smart Environments and Green Computing, 1 (2021),
  pp.~90--102.

\bibitem{kaiser2021fundamentals}
{\sc T.~Kaiser and A.~Menzel}, {\em Fundamentals of electro-mechanically
  coupled cohesive zone formulations for electrical conductors}, Computational
  Mechanics, 68 (2021), pp.~51--67.

\bibitem{karoumi1999some}
{\sc R.~Karoumi}, {\em Some modeling aspects in the nonlinear finite element
  analysis of cable supported bridges}, Computers \& Structures, 71 (1999),
  pp.~397--412.

\bibitem{kharazmi2019operator}
{\sc E.~Kharazmi and M.~Zayernouri}, {\em Operator-based uncertainty
  quantification of stochastic fractional partial differential equations},
  Journal of Verification, Validation and Uncertainty Quantification, 4 (2019).

\bibitem{kinney2005modeling}
{\sc R.~Kinney, P.~Crucitti, R.~Albert, and V.~Latora}, {\em Modeling cascading
  failures in the north american power grid}, The European Physical Journal
  B-Condensed Matter and Complex Systems, 46 (2005), pp.~101--107.

\bibitem{knio2006uncertainty}
{\sc O.~Knio and O.~Le~Maitre}, {\em Uncertainty propagation in cfd using
  polynomial chaos decomposition}, Fluid dynamics research, 38 (2006), p.~616.

\bibitem{koufakis2010wildfire}
{\sc E.~I. Koufakis, P.~T. Tsarabaris, J.~S. Katsanis, C.~G. Karagiannopoulos,
  and P.~D. Bourkas}, {\em A wildfire model for the estimation of the
  temperature rise of an overhead line conductor}, IEEE transactions on power
  delivery, 25 (2010), pp.~1077--1082.

\bibitem{lasota2015polynomial}
{\sc R.~Lasota, R.~Stocki, P.~Tauzowski, and T.~Szolc}, {\em Polynomial chaos
  expansion method in estimating probability distribution of rotor-shaft
  dynamic responses}, Bulletin of the Polish Academy of Sciences. Technical
  Sciences, 63 (2015), pp.~413--422.

\bibitem{lin2014uncertainty}
{\sc G.~Lin, M.~Elizondo, S.~Lu, and X.~Wan}, {\em Uncertainty quantification
  in dynamic simulations of large-scale power system models using the
  high-order probabilistic collocation method on sparse grids}, International
  Journal for Uncertainty Quantification, 4 (2014).

\bibitem{low2013new}
{\sc Y.~M. Low}, {\em A new distribution for fitting four moments and its
  applications to reliability analysis}, Structural Safety, 42 (2013),
  pp.~12--25.

\bibitem{machado2015reliability}
{\sc M.~Machado and J.~Dos~Santos}, {\em Reliability analysis of damaged beam
  spectral element with parameter uncertainties}, Shock and Vibration, 2015
  (2015).

\bibitem{miehe_phase_2010}
{\sc C.~Miehe, M.~Hofacker, and F.~Welschinger}, {\em A phase field model for
  rate-independent crack propagation: {Robust} algorithmic implementation based
  on operator splits}, Computer Methods in Applied Mechanics and Engineering,
  199 (2010), pp.~2765--2778.

\bibitem{miehe2010phase_el}
{\sc C.~Miehe, F.~Welschinger, and M.~Hofacker}, {\em A phase field model of
  electromechanical fracture}, Journal of the Mechanics and Physics of Solids,
  58 (2010), pp.~1716--1740.

\bibitem{miehe_thermodynamically_2010}
{\sc C.~Miehe, F.~Welschinger, and M.~Hofacker}, {\em Thermodynamically
  consistent phase-field models of fracture: {Variational} principles and
  multi-field {FE} implementations}, International Journal for Numerical
  Methods in Engineering, 83 (2010), pp.~1273--1311,
  \url{https://doi.org/10.1002/nme.2861}.

\bibitem{nayak2019microstructure}
{\sc S.~Nayak and S.~Das}, {\em A microstructure-guided numerical approach to
  evaluate strain sensing and damage detection ability of random heterogeneous
  self-sensing structural materials}, Computational Materials Science, 156
  (2019), pp.~195--205.

\bibitem{rezaei2016analysis}
{\sc S.~N. Rezaei, L.~Chouinard, S.~Langlois, and F.~L{\'e}geron}, {\em
  Analysis of the effect of climate change on the reliability of overhead
  transmission lines}, Sustainable cities and society, 27 (2016), pp.~137--144.

\bibitem{saltelli2010variance}
{\sc A.~Saltelli, P.~Annoni, I.~Azzini, F.~Campolongo, M.~Ratto, and
  S.~Tarantola}, {\em Variance based sensitivity analysis of model output.
  design and estimator for the total sensitivity index}, Computer Physics
  Communications, 181 (2010), pp.~259--270.

\bibitem{sarajlic2018identification}
{\sc M.~Sarajli{\'c}, J.~Pihler, N.~Sarajli{\'c}, and G.~{\v{S}}tumberger},
  {\em Identification of the heat equation parameters for estimation of a bare
  overhead conductor’s temperature by the differential evolution algorithm},
  Energies, 11 (2018), p.~2061.

\bibitem{shen2021numerical}
{\sc F.~Shen and L.-L. Ke}, {\em Numerical study of coupled
  electrical-thermal-mechanical-wear behavior in electrical contacts}, Metals,
  11 (2021), p.~955.

\bibitem{smith2013uncertainty}
{\sc R.~C. Smith}, {\em Uncertainty quantification: theory, implementation, and
  applications}, vol.~12, Siam, 2013.

\bibitem{smolyak1963quadrature}
{\sc S.~A. Smolyak}, {\em Quadrature and interpolation formulas for tensor
  products of certain classes of functions}, in Doklady Akademii Nauk,
  vol.~148, Russian Academy of Sciences, 1963, pp.~1042--1045.

\bibitem{sobol1993sensitivity}
{\sc I.~M. Sobol}, {\em Sensitivity estimates for nonlinear mathematical
  models}, Mathematical modelling and computational experiments, 1 (1993),
  pp.~407--414.

\bibitem{stefanou2009stochastic}
{\sc G.~Stefanou}, {\em The stochastic finite element method: past, present and
  future}, Computer methods in applied mechanics and engineering, 198 (2009),
  pp.~1031--1051.

\bibitem{stengel2014finite}
{\sc D.~Stengel and M.~Mehdianpour}, {\em Finite element modelling of
  electrical overhead line cables under turbulent wind load}, Journal of
  structures, 2014 (2014).

\bibitem{suzuki2016fractional}
{\sc J.~Suzuki, M.~Zayernouri, M.~Bittencourt, and G.~Karniadakis}, {\em
  Fractional-order uniaxial visco-elasto-plastic models for structural
  analysis}, Computer Methods in Applied Mechanics and Engineering, 308 (2016),
  pp.~443--467.

\bibitem{suzuki2021thermodynamically}
{\sc J.~Suzuki, Y.~Zhou, M.~D’Elia, and M.~Zayernouri}, {\em A
  thermodynamically consistent fractional visco-elasto-plastic model with
  memory-dependent damage for anomalous materials}, Computer Methods in Applied
  Mechanics and Engineering, 373 (2021), p.~113494.

\bibitem{suzuki2023fractional}
{\sc J.~L. Suzuki, M.~Gulian, M.~Zayernouri, and M.~D’Elia}, {\em Fractional
  modeling in action: A survey of nonlocal models for subsurface transport,
  turbulent flows, and anomalous materials}, Journal of Peridynamics and
  Nonlocal modeling, 5 (2023), pp.~392--459.

\bibitem{suzuki2022general}
{\sc J.~L. Suzuki, M.~Naghibolhosseini, and M.~Zayernouri}, {\em A general
  return-mapping framework for fractional visco-elasto-plasticity}, Fractal and
  fractional, 6 (2022), p.~715.

\bibitem{tan2022phase}
{\sc Y.~Tan, Y.~He, C.~Liu, and X.~Li}, {\em Phase field fracture model of
  transversely isotropic piezoelectric materials with thermal effect},
  Engineering Fracture Mechanics, 268 (2022), p.~108479.

\bibitem{tartakovsky2019physics}
{\sc A.~Tartakovsky and R.~Tipireddy}, {\em Physics-informed machine learning
  method for forecasting and uncertainty quantification of partially observed
  and unobserved states in power grids}, in Proceedings of the 52nd Hawaii
  International Conference on System Sciences, 2019.

\bibitem{vasquez2017end}
{\sc W.~A. Vasquez, D.~Jayaweera, and J.~J{\'a}tiva-Ibarra}, {\em End-of-life
  failure modelling of overhead lines considering loading and weather effects},
  in 2019 IEEE International Conference on Power, Electrical, and Electronics
  and Industrial Applications (PEEIACON), 2017.

\bibitem{ward2013effect}
{\sc D.~M. Ward}, {\em The effect of weather on grid systems and the
  reliability of electricity supply}, Climatic Change, 121 (2013),
  pp.~103--113.

\bibitem{winterstein2011extremes}
{\sc S.~R. Winterstein and C.~A. MacKenzie}, {\em Extremes of nonlinear
  vibration: models based on moments, l-moments, and maximum entropy}, in
  International Conference on Offshore Mechanics and Arctic Engineering,
  vol.~44342, 2011, pp.~617--626.

\bibitem{xiu2005high}
{\sc D.~Xiu and J.~S. Hesthaven}, {\em High-order collocation methods for
  differential equations with random inputs}, SIAM Journal on Scientific
  Computing, 27 (2005), pp.~1118--1139.

\bibitem{xiu2002modeling}
{\sc D.~Xiu and G.~E. Karniadakis}, {\em Modeling uncertainty in steady state
  diffusion problems via generalized polynomial chaos}, Computer methods in
  applied mechanics and engineering, 191 (2002), pp.~4927--4948.

\bibitem{xiu2002wiener}
{\sc D.~Xiu and G.~E. Karniadakis}, {\em The wiener--askey polynomial chaos for
  stochastic differential equations}, SIAM journal on scientific computing, 24
  (2002), pp.~619--644.

\bibitem{yang2013probability}
{\sc H.~Yang, C.~Chung, J.~Zhao, and Z.~Dong}, {\em A probability model of ice
  storm damages to transmission facilities}, IEEE Transactions on power
  delivery, 28 (2013), pp.~557--565.

\bibitem{yang2019failure}
{\sc S.~Yang, W.~Zhou, S.~Zhu, L.~Wang, L.~Ye, X.~Xia, and H.~Li}, {\em Failure
  probability estimation of overhead transmission lines considering the spatial
  and temporal variation in severe weather}, Journal of Modern Power Systems
  and Clean Energy, 7 (2019), pp.~131--138.

\end{thebibliography}

\end{document}